\documentclass[manuscript]{aastex62}
\usepackage[encapsulated]{CJK}
\usepackage{subfigure}
\usepackage{mathrsfs}
\usepackage{amsmath}
\usepackage{epstopdf}
\usepackage{longtable}
\usepackage{booktabs}
\usepackage{hyperref} %\usepackage{threeparttable}
\def\etal {et al.~}

\newbox\grsign \setbox\grsign=\hbox{$>$} \newdimen\grdimen \grdimen=\ht\grsign
\newbox\laxbox \newbox\gaxbox
\setbox\gaxbox=\hbox{\raise.5ex\hbox{$>$}\llap
     {\lower.5ex\hbox{$\sim$}}}\ht1=\grdimen\dp1=0pt
\setbox\laxbox=\hbox{\raise.5ex\hbox{$<$}\llap
     {\lower.5ex\hbox{$\sim$}}}\ht2=\grdimen\dp2=0pt
\shorttitle{CO Outflow Feedback}
\shortauthors{Li \etal}

\newcommand{\co}{$^{12}$CO }                             % 12co
\newcommand{\xco}{$^{13}$CO }                            % 13co
\begin{document}
\begin{CJK*}{UTF8}{gbsn}

\title{CO Outflow Candidates Toward the W3/4/5 Complex II: Feedback from Candidate Outflows}

\correspondingauthor{Yingjie Li}
\email{liyj@pmo.ac.cn, xuye@pmo.ac.cn}

\author{Yingjie Li}\affiliation{Purple Mountain Observatory, Chinese Academy of Sciences, Nanjing 210023, China}
\affiliation{University of Science and Technology of China, Hefei, Anhui 230026, China}

\author{Ye Xu}
\affiliation{Purple Mountain Observatory, Chinese Academy of Sciences, Nanjing 210023, China}

\author{Yan Sun}
\affiliation{Purple Mountain Observatory, Chinese Academy of Sciences, Nanjing 210023, China}

\author{Ji Yang}
\affiliation{Purple Mountain Observatory, Chinese Academy of Sciences, Nanjing 210023, China}

%\author{Xiuhui Chen}
%\affiliation{Hunan University of Arts and Science, Changde 415000, China}

\begin{abstract}
To date, few studies have focused on protostellar outflow feedback at scales larger than several parsecs. 
To address this paucity of research, we investigate the effects of feedback from CO outflow candidates on 
their parent clouds over $\sim$ 110 deg$^2$ toward the W3/4/5 complex and its surroundings. 
Our search identified 265 \xco clouds with radii being $\sim$ 0.04 -- 17.12 pc. We estimate the turbulent support and potential disruptive effect of the outflow activities through analyzing physical properties of outflow candidates and their host clouds in terms of turbulence and gravitational binding energy. We find:
(1) clouds of larger size might be less affected by feedback; 
(2) the possible scale break is $\gtrsim 4.7$ pc for both turbulent support and potential disruptive effect; 
(3) if outflows couple to dense gas where stars are forming, for clouds in the Perseus arm, a scale $\lesssim$ 0.2 -- 0.4 pc is sufficient for the energy and momentum injected by outflow activity to maintain turbulence, while for clouds in the Local arm, the scale is $\lesssim$ 0.1 -- 0.2 pc; 
and (4) for clouds in the Perseus arm, a scale $\lesssim$ 0.3 -- 1.0 pc is needed for outflow activity to potentially disperse material away from the natal clouds, while for clouds in the Local arm, the scale is $\lesssim$ 0.2 -- 0.6 pc.
The strength of outflow activity may affect the values in the points (3) and (4).
Finally, we find that outflow feedback probably possesses the power to alter the linewidth-size relation. 
\end{abstract}


\section{Introduction}

Protostellar outflows are an intrinsic process during the early stages of star formation.
They can impact clustered star formation in two feedback scenarios \citep{B2016} which are related to eruptive star formation \citep[e.g., rapid cloud dispersal following a burst of star formation;][]{E2007, HB2007} and slow, quasi-equilibrium star formation regulated by the replenishment of turbulence via feedback \citep{TKM2006, NL2007, NL2014}.
As a protostar accretes mass from its host core, the subsequent outflow likely represents the first rung of the feedback ladder arising from ever more powerful momentum and energy injection mechanisms \citep[such as ejection by wide angle winds and soft-UV radiation from moderate-mass stars, or by ionizing radiation, stellar winds, and the explosions of massive stars;][]{FRC2014, B2016}. Feedback from these mechanisms can impact the ecology of star-forming regions \citep{KKM2011, MKK2014, B2016}. Overall, outflow feedback is related to two critical star formation issues: the relatively low efficiency of star formation and the driving source of turbulence in clouds \citep{ES2004, MO2007, FRC2014}.

Two perspectives are frequently explored to investigate the feedback of outflows: their turbulent support and potential disruptive effect. The first one compares the total outflow kinetic energy, luminosity, and conserved momentum with the corresponding properties of cloud turbulence \citep{FRC2014}. For instance, \citet{GRB2010} found that the total outflow kinetic energy is $\sim$ 70\% of the total turbulent energy in the Serpens cloud. Similar conclusions have been also drawn based on other regions such as NGC2264C \citep{MAL2009}, $\rho$ Ophiuchi \citep{NKK2011}, Serpens South \citep{NSS2011}, L1641-N \citep{NMK2012}, etc. A few studies have investigated areas contained several star-forming regions, e.g., six sub-regions in the Perseus cloud complex \citep{ABG2010} and several nearby Gould Belt clouds \citep{DHB2016}. In these two studies, conclusions were drawn whereby the total outflow kinetic energy accounted for at least $\sim$ 14\% of the natal cloud's turbulent energy. All of these studied regions were confined to distances $\lesssim$ 800 pc and cloud radii $\lesssim$ 2.0 pc. Moreover, in other studies, it was found that when the cloud's radius is $\gtrsim$ 9 pc, the outflow kinetic energy only makes up $\lesssim$ 1\% of the host cloud's turbulent energy \citep{LLQ2015, LLX2018, WYX2017}.

The potential for outflows to disrupt their parent clouds has been previously traced by comparing the total outflow kinetic energy and the cloud's gravitational binding energy \citep[e.g., $E_{\mathrm{flow}}/E_{\mathrm{grav}}$ or the quantity $\eta_\mathrm{out} = 2 E_{\mathrm{flow}}/E_{\mathrm{grav}}$; see the definition in][]{NL2014}. In filaments or isolated small clouds, outflows can blow out of their parent cores \citep{B2016}. \citet{NL2014} investigated $\eta_\mathrm{out}$ towards eight nearby clumps (distances $\lesssim$ 415 pc and cloud radii $\lesssim 2$ pc) and found $0.01 < \eta_\mathrm{out} < 0.10$ except for one clump that had $\eta_\mathrm{out} = 5$, implying that outflows cannot disrupt most clumps. Similar conclusions have been drawn based on studies toward the Perseus molecular cloud complex \citep{ABG2010} and several  nearby Gould Belt clouds \citep{DHB2016}. It was also shown that $E_{\mathrm{flow}}/E_{\mathrm{grav}}$ decreases to $<$ 1\% when the cloud radius is $\gtrsim$ 9 pc \citep[e.g.,][]{LLQ2015, LLX2018}.

Theoretical studies have shown that outflows modify the environments in which stars form by injecting energy and momentum into their surroundings \citep{NS1980, Mc1989, KBA2014}. Dynamic equilibrium between momentum input and turbulent dissipation is achieved for a young cluster once star formation and its outflows have commenced \citep{FRC2014}. Numerical simulations have demonstrated that a cluster-forming clump can remain in quasi-equilibrium when subjected to turbulence driven by outflows \citep{LN2006}. Jets and collimated outflows are more efficient in driving turbulence (\citealp{M2007, CFC2009, CFB2009}; for more detail see the review by \citealp{KBA2014}). So far, though it has widely been believed that outflows are sources of energy and momentum input, a fundamental question, i.e., whether the momentum injected by protostellar outflows is enough to counteract turbulence decay in clouds \citep{FRC2014}, is still outstanding. A coexistent and unanswered question is how protostellar outflows limit the collapse rate or disrupt the parent clouds \citep[e.g.,][]{M2007, MJ2015, B2016}.

Whether there is a clear break of the scale of turbulence driven by outflow feedback is still an open question \citep{B2016}. Almost all studies have been confined to scales below several parsecs except those by \citet{LLQ2015, LLX2018} in which the scales were $\sim$ 4, $\sim$ 9 and $\sim$ 21 -- 32 pc, although only eight clouds were investigated in these two studies. Fortunately, we are now able to systematically investigate the feedback of outflow activities for a much larger sample thanks to the study of the structures and physical properties of the molecular (\co and its other two isotopic molecules) gas toward the W3/4/5 complex and its surroundings \citep{SYX2020}, and the related outflow survey \citep[][hereafter, Paper I]{LXS2019}. 

The remainder of the paper is organized as follows. In Section \ref{data}, we describe the data used in this work. In Section \ref{result}, we present the cloud detection process, the physical properties of the clouds and the effect of feedback of the outflow activities on their parent clouds. In Section \ref{discussion}, we discuss the outflow feedback in terms of turbulent support and potential disruptive effects caused by the outflow activities and the potential effect of the outflow activities on the linewidth-size relation. Finally, we summarize our conclusions in Section \ref{summary}.

\section{Data}\label{data}

As stated in Paper I, the data used in our analysis covers $\sim$ 110 deg$^2$ ($129\fdg75\leq l\leq140\fdg25$, $-5\fdg25\leq b\leq5\fdg25$), which were obtained by the Milky Way Imaging Scroll Painting Project \citep[MWISP,][]{SYZ2018}. The data of \co (J = 1 $\rightarrow$ 0) (115.271 GHz) and \xco (J = 1 $\rightarrow$ 0) (110.201 GHz) were observed from November 2011 to November 2017 using the Purple Mountain Observatory Delingha (PMODLH) 13.7-m telescope with the 9-beam superconducting array receiver (SSAR) which works in the sideband separation mode and employs a fast Fourier transform spectrometer \citep{ZLS2011, SYS2012}. The velocity resolution is $\sim0.16$ km s$^{-1}$ for \co and $\sim0.17$ km s$^{-1}$ for $^{13}$CO. The main beam root mean squared noise (RMS) after main beam efficiency correction at these velocity resolutions is $\sim0.45$ K for $^{12}$CO and $\sim0.25$ K for $^{13}$CO. The half-power beam-width (HPBW) is $\sim 49\arcsec$ for $^{12}$CO and $\sim 51\arcsec$ for $^{13}$CO, and the data were gridded to 30$\arcsec$ pixels for both transitions.
%\footnote{It is a huge ongoing project to survey CO and its Isotopic Transitions toward the northern Galactic plane within $-10\degr.25\leq l\leq250\degr.25$ and $-5\degr.25\leq b\leq5\degr.25$, and other regions of interest, see details in \url{http://www.radioast.nsdc.cn/english/}.}

\section{Data Analysis and Results}\label{result}

\subsection{Summary of Outflow Candidates}

In this work, the outflow candidates in the Perseus arm, the Local arm and interarm 1, as presented in Paper I, are considered. Those candidates were searched for based on the longitude-latitude-velocity \co data, where the cores were traced by three-dimensional \xco data. 
The outflow candidate sample was finally obtained after searching for the \co velocity bulges based on 
the \xco peak velocity distribution maps and conducting line diagnoses of each candidate (see more details in Paper 
I and \citealt{LLX2018}). The proportion of bipolar outflow candidates is $\sim$ 22\% (see table 3 in Paper I). 
The median value of dynamical timescale of outflow candidates, $t_{\mathrm{flow}}$, before correction for inclination 
is $\sim$ 0.4 Myr, which is consistent with the estimated duration of the Class 0 and likely Class I phases of the evolution of young stellar object  \citep{EDJ2009, B2016}. 
Thus, these candidates were more likely driven by protostars. To further distinguish individual protostellar outflow candidate from high-velocity outflowing \co gas driven by stellar wind, new observations with improved spatial resolution are required.

The momentum, $P_{\mathrm{flow}}$, of an outflow candidate was calculated by multiplying the outflow candidate's mass, $M_\mathrm{flow}$, with its velocity relative to the central cloud weighted by the main beam brightness temperature of each channel in the line wing (outflow candidate's velocity, $\langle\Delta v_{\mathrm{flow}}\rangle$). For kinetic energy, $E_{\mathrm{flow}}$, of an outflow candidate, $\langle\Delta v_{\mathrm{flow}}\rangle$ was substituted with  $\langle\Delta v^2_{\mathrm{flow}}\rangle$.  The luminosity of an outflow candidate was $L_{\mathrm{flow}}=E_{\mathrm{flow}}/t_{\mathrm{flow}}$ with $t_\mathrm{flow} = l_\mathrm{flow}/ \Delta v_\mathrm{max}$, where $\Delta v_\mathrm{max}$ and $l_\mathrm{flow}$ are respectively the maximum velocity and the length of an outflow candidate (see more details in Paper I). We list the outflow candidates' physical properties after corrections for optical depth, average inclination and blending effect in Table~\ref{Table:relation}.

\subsection{Cloud Identification}\label{Sec:indentify}

Similar to \citet{ABG2010} and \citet{LLQ2015, LLX2018}, the \xco emission was used to evaluate the physical quantities of each cloud. \xco clouds near/covering the outflow candidates were identified with the following steps.
First, the integrated intensity map of \xco was mapped with the velocity range of integration being the full velocities range characterized by zero intensity of an outflow candidate. The boundary of a cloud was determined by pixels where the main beam brightness temperatures were larger than 3 $\times$ RMS in at least three successive channels.

Second, we searched for a cloud which covered the position of the outflow candidate. If we did not obtain such a cloud, we searched for a cloud where the boundary was $<$ 30$\arcsec$ away from the outflow candidate. For the latter case, if two or more clouds satisfied the criterion, we chose the nearest one.

Third, we searched for other outflow candidates which were covered by or close to\footnote{The distance to the boundary of the acquired cloud was $<$ 30$\arcsec$.} the acquired cloud. We then adjusted the velocity range accordingly, and then mapped the intensity map over the adjusted velocity range. This step was repeated until the size of the \xco cloud and the number of outflow candidates associated with the cloud reached their maximum values.
This step provides the maximal velocity range of \xco cloud, because the full velocity range of \co outflow candidates is larger than that of \xco outflow or cloud, at least at current detection limits. Similar to \citet{ABG2010} and \citet{LLQ2015, LLX2018}, we are not attempting to differentiate the ambient cloud and components of outflowing gas for the properties of clouds.
 
Finally, in order to try to diminish impact of uncorrelated components, the velocity range of \xco cloud was further adjusted referring to the mean spectrum of the \xco cloud. Following the above steps, the associations between the \xco clouds and \co outflow candidates were obtained.

\begin{figure}[!ht]
 %\figurenum{1-cloud 1}
 \centering
  \subfigure[Intensity]{\includegraphics[width=0.45\textwidth]{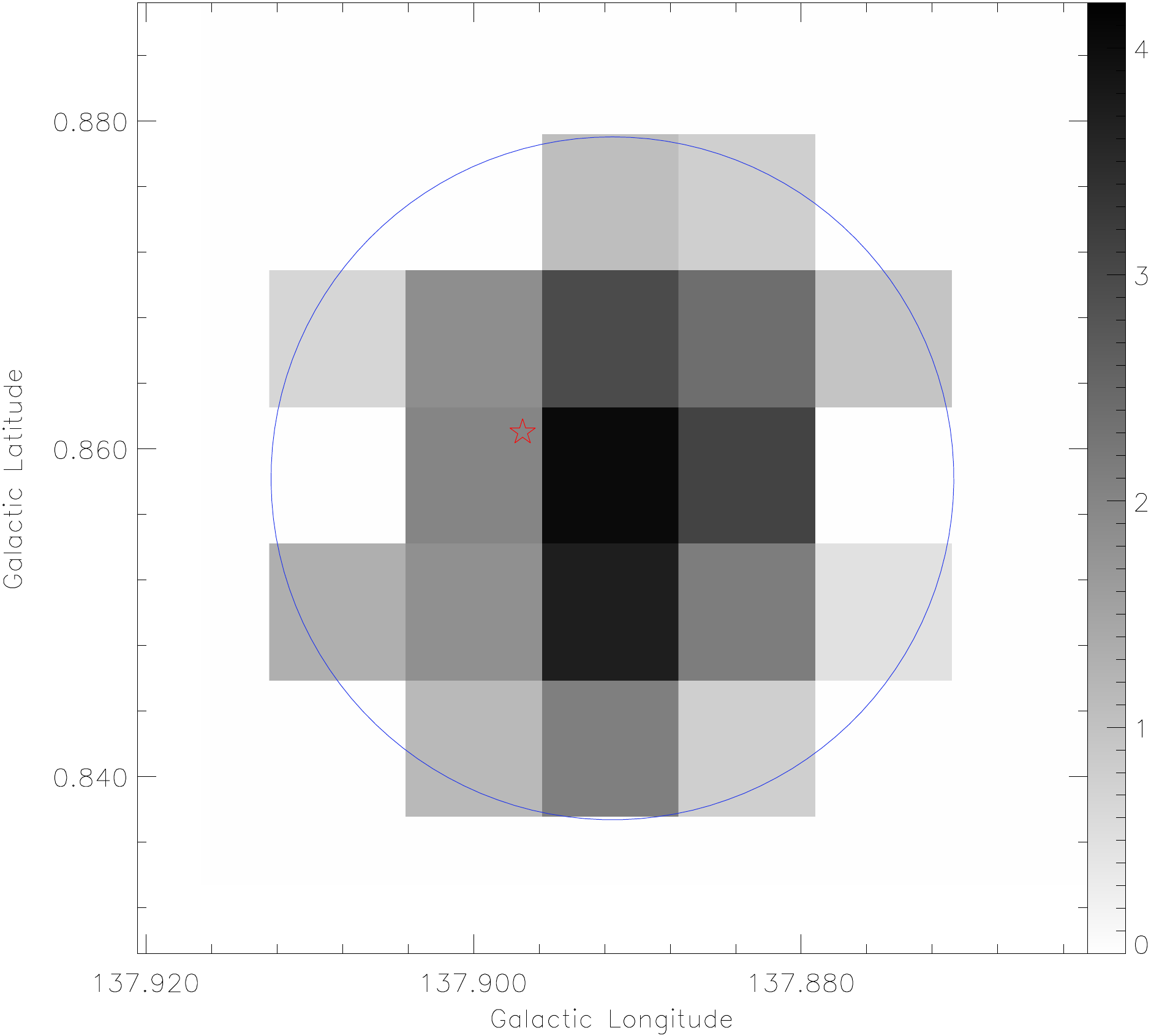}}
  \subfigure[Spectra]{ \includegraphics[width=0.45\textwidth,trim=100 50 50 60]{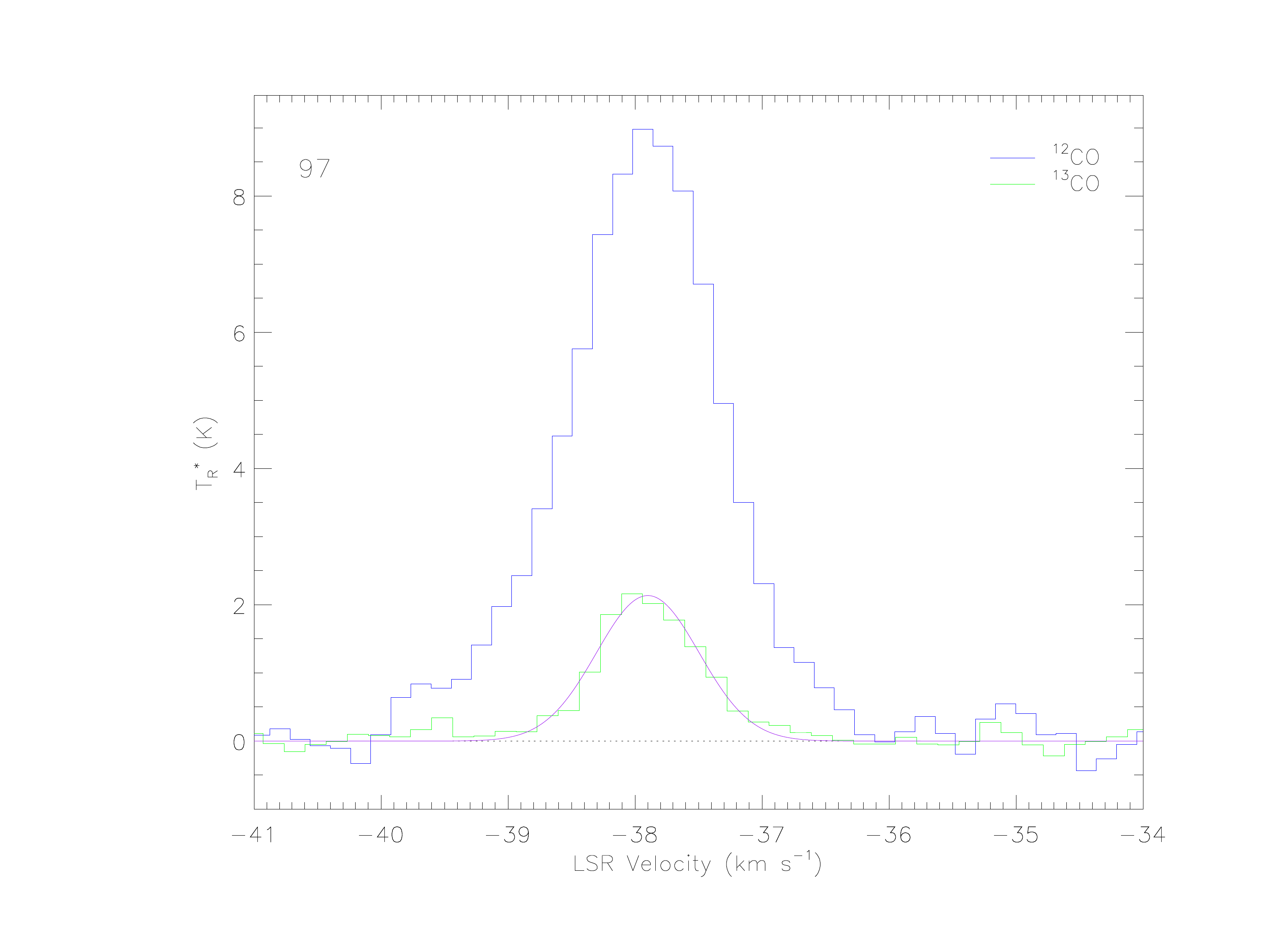}}
\caption{The integrated intensity map and spectrum for cloud 97. The color bar is in units of K km s$^{-1}$. Panel (a) shows the \xco integrated intensity over the velocity range shown in Table \ref{Table:relation}. The red pentagram indicates the outflow candidate, and the ellipse reports the geometric figure used to locate the cloud. The pixel size is 30$\arcsec$ (HPBW is $\sim 51\arcsec$), corresponding to $\sim$ 0.28 pc at a distance of $\sim$ 1950 pc (see Paper I). Panel (b) shows the averaged spectra of the $^{12}$CO and \xco emission over the entire cloud by the blue and green lines, respectively, and the cloud index is reported in the upper left corner.}
\label{fig:cloud}
\end{figure}

We detected 265 such clouds, including 136 in the Perseus arm, \footnote{They were correlated with 284 outflow candidates in the Perseus arm, $\sim 21\%$ of which were associated with at least two outflow candidates.} 124 in the Local arm, \footnote{They were correlated with 159 outflow candidates in the Local arm, of which $\sim 17\%$ were associated with at least two outflow candidates.} and five in interarm 1 (correlated with five outflow candidates). The details of the correlations between the clouds and outflow candidates are catalogued in Table \ref{Table:relation}, and the integrated intensity maps and mean spectra of the clouds are mapped in Figure \ref{fig:cloud}.

\begin{deluxetable*}{lrrrrrrccl}
\tablecolumns{10}
\tabletypesize{\scriptsize}
\setlength{\tabcolsep}{0.1in}
\tablewidth{0pt}
\tablecaption{Correlation between Clouds and Outflow Candidates. \label{Table:relation}}
%\begin{tabular}
\tablehead{
\\[-0.2cm]
 \colhead{Cloud Index} & \colhead{$l$} & \colhead{$b$} & \colhead{$V_\mathrm{low}$} & \colhead{$V_\mathrm{high}$} & \colhead{$M_\mathrm{flow}$} & \colhead{$P_\mathrm{flow}$} & \colhead{$E_\mathrm{flow}$} & \colhead{$L_\mathrm{flow}$} & \colhead{Outflow Index} \\
 \colhead{} & \colhead{($\degr$)} & \colhead{($\degr$)} & \colhead{(km\,s$^{-1}$)} & \colhead{(km\,s$^{-1}$)} & \colhead{($M_{\sun}$)} & \colhead{($M_{\sun}$\,km\,s$^{-1}$)} & \colhead{(erg)} & \colhead{(erg\,s$^{-1}$)} & \colhead{} \\
 \colhead{(1)} & \colhead{(2)} & \colhead{(3)} & \colhead{(4)} & \colhead{(5)} & \colhead{(6)} & \colhead{(7)} & \colhead{(8)} & \colhead{(9)} & \colhead{(10)}
}
\startdata
\multicolumn{10}{l}{Perseus Arm}\\
\hline
1	&	130.103 	&	-4.471 	&	-46.0 	&	-37.0 	&	1.59 	&	5.12 	&	1.59E+44	&	1.14E+31	&	1, 2	\\
2	&	130.380 	&	-0.784 	&	-38.0 	&	-26.0 	&	27.41 	&	126.20 	&	5.44E+45	&	2.10E+32	&	3, 5	\\
3	&	130.392 	&	1.638 	&	-47.0 	&	-38.0 	&	0.14 	&	0.46 	&	1.46E+43	&	2.05E+30	&	4	\\
4	&	130.578 	&	1.953 	&	-49.0 	&	-40.0 	&	5.14 	&	21.69 	&	8.68E+44	&	7.98E+31	&	6 -- 9	\\
5	&	130.854 	&	-0.929 	&	-38.0 	&	-29.0 	&	0.24 	&	0.97 	&	3.66E+43	&	3.80E+30	&	10	\\
...
\enddata
\tablecomments{Column (1): the index for each molecular cloud. Columns (2) -- (3): 
central position of each molecular cloud in Galactic 
coordinates (see Figure \ref{fig:cloud}). Columns (4) -- (5): the integral interval used to calculate 
the cloud mass (see Section \ref{section:calculation}). Columns (6) -- (9): the total outflow mass, 
momentum, kinetic energy and luminosity corresponding to each cloud. They are from Paper I, but here we newly apply the corrections for inclination angle, 
blending effect and optical depth. All values in columns (6) -- (9) also have been multiplied by 1.36 to take into consideration the 
mean molecular weight \citep{B2010} instead of the weight of only molecular H$_2$ given in \citet{LXS2019}. Column (10):
the outflow candidate indexes corresponding to each cloud.}
\end{deluxetable*}

\subsection{Physical Properties of Turbulent Support and Potential Disruptive Effect}\label{section:calculation}

The turbulent support and potential disruptive effect of outflow candidates are considered here to evaluate their feedback effect. Because the W3/4/5 complex in the Perseus arm is a massive star-forming region \citep{W1958, HT1998}, the outflow candidates in Paper I are probably similar to those presented by \citet{BSS2002} and \citet{TM2002}, and the physical quantities of each are most likely dominated by a single massive flow. Therefore the effects due to multiple flows are neglected in our analysis.

\subsubsection{Physical Properties of the Clouds}\label{section:pclouds}

Since the absent of accurate distance to each cloud, following our previous works on this region, we adopt a constant distance to each cloud within individual spiral arm or interarm \citep[i.e., the Perseus arm $\sim1950$~pc, the Local arm $\sim 600$~pc, and interarm 1 $\sim 1280$ pc, see][and references therein]{SYX2020}.
The cloud radius, $R_\mathrm{cloud}$, was calculated  by $R_\mathrm{cloud}=d\sqrt{4A/\pi-\mathrm{HPBW}^2}/2$ \citep{LMG1994}, where $A$ is the area of the cloud, and $d$ is the distance. The excitation temperature of a cloud, $T_\mathrm{cloud}$, was estimated from the peak intensity of the mean spectrum of $^{12}$CO \citep{NMO1998}. The line width, $\Delta V_\mathrm{cloud}$, and its error, $\Delta V_\mathrm{err}$, of a cloud was determined from a Gaussian fit of the mean spectra of $^{13}$CO. If there were two or more peaks, we took the mean line width weighted by the mass.\footnote{The weight is given by $\tau\cdot\Delta V\cdot\tau/(1-e^{-\tau})$, where $\tau$ and $\Delta V$ are the optical depth \citep[calculated from $T_\mathrm{cloud}$ and the mean spectrum of $^{13}$CO,][]{KOY1998, GL1999} and the line width of a component, respectively, and $\tau/(1-e^{-\tau})$ is a correction factor for the optical depth. In the weight, we assumed that all components shared the same excitation temperature.} Next, the cloud mass, $M_{\mathrm{cloud}}$, was evaluated by multiplying the mean mass \citep[computed from $T_\mathrm{cloud}$ and the mean spectrum of $^{13}$CO; ][]{DXY2017} by the area enclosed by the boundary of the \xco emission (see Section \ref{Sec:indentify}), where the optical depth correction, $\tau_\mathrm{cloud}/(1-e^{-\tau_\mathrm{cloud}})$, and $N\left({\mathrm{H_2}}\right)/N\left(\mathrm{^{13}CO}\right)=7\times10^5$ \citep{FLW1982} are included. All these quantities are cataloged in Table \ref{Table:parameter}.

Although no attempt is made to differentiate the ambient cloud and components of outflowing gas for the properties of clouds, the influence of outflowing gas on cloud properties (such as mass, line width and other properties discussed below) is limited. The reasons include: 
%1) only a small fraction of the \xco spectra show high-velocity wing emission, thus the emission is dominated by cloud; 
1)  high-velocity wing emission (if any) contributes only a small fraction of the \xco emission, i.e., the emission is dominated by cloud; 2) the line width was estimated from Gaussian fit where the contribution from the line wing of \xco emission is negligible or very small.

Fixing the distance, the uncertainty of  mass of most clouds ($\sim$ 95\%) was estimated to be less than 4\%, by assuming a cloud diameter of one beam size, and an integrated intensity of 3 $\times$ RMS. 
However, if considering the intrinsic width of a spiral arm, the distance error is $\sim$ 20\% -- 67\% \citep[see][]{SYX2020}, leading to the uncertainty of the estimated mass larger than 40\%. This indicates that the uncertainty of cloud mass is dominated by distance, and this conclusion also applies to other properties (e.g., $M_\mathrm{flow}$, 
$P_\mathrm{flow}$, $E_\mathrm{flow}$, $L_\mathrm{flow}$, $R_\mathrm{cloud}$, $E_\mathrm{turb}$, $P_\mathrm{turb}$, $L_\mathrm{turb}$, $E_\mathrm{grav}$, and $M_\mathrm{esc}$ presented below). 
It is difficult to estimate the error contributed by distance due to absence of accurate distance to each cloud. Therefore, we do not present the errors of those physical parameters here, 
but the impact of uncertainty of distance would be discussed in detail in Sections \ref{sec:effect} and \ref{sec:velo-dis}.

\begin{deluxetable}{lrrrrrrrrrrrrrl}
\centering
\setlength\tabcolsep{3pt}
\tablecolumns{15}
\tabletypesize{\tiny}
\tablewidth{0pt}
\tablecaption{Global Physical Properties of the Clouds \label{Table:parameter}}
%\begin{tabular}
\tablehead{
\\
   \noalign{\smallskip} 
 \colhead{Cloud Index} & \colhead{$M_{\mathrm{cloud}}$} & \colhead{$R_{\mathrm{cloud}}$} & \colhead{$\Delta V_\mathrm{cloud}$} & \colhead{$\Delta V_\mathrm{err}$} & \colhead{$T_\mathrm{cloud}$} & \colhead{$\tau_\mathrm{cloud}$} & \colhead{$E_{\mathrm{turb}}$} & \colhead{$P_\mathrm{turb}$} & \colhead{$t_{\mathrm{ff}}$} & \colhead{$t_{\mathrm{diss}}$} & \colhead{$L_{\mathrm{turb}}$} & \colhead{$v_{\mathrm{esc}}$} & \colhead{$M_{\mathrm{esc}}$} & \colhead{$E_{\mathrm{grav}}$} \\
 \colhead{} & \colhead{($M_{\sun}$)} & \colhead{(pc)} & \colhead{(km s$^{-1}$)} & \colhead{(km s$^{-1}$)} & \colhead{(K)} & \colhead{} & \colhead{(erg)} & \colhead{($M_{\sun}$ km s$^{-1}$)} & \colhead{($10^6$ yr)} & \colhead{($10^6$ yr)} & \colhead{(erg s$^{-1}$)} & \colhead{(km s$^{-1}$)} & \colhead {($M_{\sun}$)} & \colhead{(erg)}
}
\startdata
\multicolumn{13}{l}{Perseus Arm}\\
\hline
1	&	299.5 	&	1.96 	&	2.0 	&	0.05 	&	8.4 	&	0.3 	&	6.2E+45	&	430.8 	&	2.6 	&	1.7 	&	1.1E+32	&	1.1 	&	4.5 	&	3.9E+45	\\
2	&	649.6 	&	2.39 	&	1.8 	&	0.05 	&	10.4 	&	0.2 	&	1.1E+46	&	845.8 	&	2.4 	&	3.2 	&	1.1E+32	&	1.5 	&	82.5 	&	1.5E+46	\\
3	&	46.5 	&	1.08 	&	1.2 	&	0.05 	&	6.2 	&	0.5 	&	3.4E+44	&	40.0 	&	2.7 	&	1.5 	&	7.0E+30	&	0.6 	&	0.8 	&	1.7E+44	\\
4	&	686.8 	&	3.37 	&	1.8 	&	0.04 	&	7.6 	&	0.3 	&	1.2E+46	&	921.2 	&	3.9 	&	1.7 	&	2.3E+32	&	1.3 	&	16.4 	&	1.2E+46	\\
5	&	16.3 	&	0.42 	&	2.9 	&	0.14 	&	10.0 	&	0.2 	&	7.6E+44	&	35.2 	&	1.1 	&	0.8 	&	2.9E+31	&	0.6 	&	1.7 	&	5.5E+43	\\
...
\enddata
\end{deluxetable}

\subsubsection{Turbulent Support}\label{section:TSR}

The turbulent energy, $E_{\mathrm{turb}}$, and momentum, $P_{\mathrm{turb}}$, of a cloud are given approximately by
\begin{equation}
   E_{\mathrm{turb}} = 0.5M_{\mathrm{cloud}}\sigma^2_{\mathrm{3d}},
\end{equation}
and
\begin{equation}\label{equ:2}
   P_{\mathrm{turb}} = M_{\mathrm{cloud}}\sigma_{\mathrm{3d}},
\end{equation}
respectively, where $M_{\mathrm{cloud}}$ and $\sigma_{\mathrm{3d}}$ are its mass and three-dimensional velocity dispersion ($\sigma_{\mathrm{3d}}=\sqrt{3}\Delta V_\mathrm{cloud}/2\sqrt{2\ln 2}$), respectively. The turbulent dissipation rate (turbulent luminosity), $L_{\mathrm{turb}}$, of a cloud can be calculated as
\begin{equation}\label{equ:lturb}
 L_{\mathrm{turb}}=\frac{E_{\mathrm{turb}}}{t_{\mathrm{diss}}},
\end{equation}
where $t_{\mathrm{diss}}$ is the turbulent dissipation timescale. The value of $t_{\mathrm{diss}}$, which arises from the energy dissipation of uniformly driven magnetohydrodynamic turbulence, is approximately given by \citep[see numerical study in][]{M1999}
\begin{equation}\label{equ:tdiss}
 t_{\mathrm{diss}}=\left(\frac{3.9\kappa}{M_{\mathrm{rms}}}\right)t_{\mathrm{ff}},
\end{equation}
where $\kappa$ is the ratio of the driving wavelength \citep[was approximately equal to the length of the outflow lobe of a continuous outflow; ][]{NL2007, CFC2009} over the Jean's length of the cloud, $\lambda_{\mathrm{J}}$. $t_{\mathrm{ff}}$ is the free-fall timescale of the cloud, and $M_{\mathrm{rms}} = \sigma_{\mathrm{3d}}/c_{\mathrm{s}}$ where $c_{\mathrm{s}}=(3kT_{\mathrm{cloud}}/m_\mathrm{H}\mu)^{1/2}$ is the Mach number of the turbulence \citep{M1999}.
$k$ is the Boltzmann's constant, $m_\mathrm{H}$ is the mass of atomic hydrogen, and $\mu=2.72$ is the mean molecular weight \citep{B2010}. For more details regarding the calculation of $L_{\mathrm{turb}}$ see \citet{LLX2018}. The derived turbulent properties of each cloud are listed in Table~\ref{Table:parameter}.

Three indicators were provided to estimate the turbulent support of the outflow candidates, i.e., the ratios of the total kinetic energy, momentum and luminosity (kinetic energy injection rate) of the outflow candidates to their respective cloud turbulence values (see Table \ref{Table:ratio}). These three ratios are denoted by $E_{\mathrm{flow}}/E_{\mathrm{turb}}$, $P_{\mathrm{flow}}/P_{\mathrm{turb}}$ and $L_{\mathrm{flow}}/L_{\mathrm{turb}}$, respectively.

\begin{deluxetable}{lrrrrrr}
\centering
\tablecolumns{7}
\tabletypesize{\footnotesize}
\tablewidth{0pt}
\tablecaption{The Effect of Feedback of Outflow Candidates on Their Parent Cloud\label{Table:ratio}}
%\begin{tabular}
\tablehead{
 \colhead{Cloud Index} & \colhead{$E_{\mathrm{flow}}/E_{\mathrm{turb}}$}  & \colhead{$P_{\mathrm{flow}}/P_{\mathrm{turb}}$} & \colhead{$L_{\mathrm{flow}}/L_{\mathrm{turb}}$} & \colhead{$E_{\mathrm{flow}}/E_{\mathrm{grav}}$} & \colhead{$M_{\mathrm{esc}}/M_{\mathrm{cloud}}$} & \colhead{$M_{\mathrm{esc}}/M_{\mathrm{flow}}$}
}
\startdata
\multicolumn{7}{l}{Perseus Arm} \\
\hline
1	&	2.6E-02	&	1.2E-02	&	1.0E-01	&	4.1E-02	&	1.5E-02	&	2.8 	\\
2	&	5.0E-01	&	1.5E-01	&	2.0E+00	&	3.6E-01	&	1.3E-01	&	3.0 	\\
3	&	4.3E-02	&	1.2E-02	&	2.9E-01	&	8.6E-02	&	1.6E-02	&	5.6 	\\
4	&	7.1E-02	&	2.4E-02	&	3.5E-01	&	7.2E-02	&	2.4E-02	&	3.2 	\\
5	&	4.8E-02	&	2.8E-02	&	1.3E-01	&	6.7E-01	&	1.0E-01	&	6.9 	\\
...
\enddata
%\vspace{-11pt}
\end{deluxetable}

The driving wavelength and $t_{\mathrm{diss}}$ from Equation (\ref{equ:tdiss}) may be overestimated if we misinterpret a clustered outflow as a single one (see Paper I). Therefore, $L_{\mathrm{turb}}$ might be underestimated from Equation (\ref{equ:lturb}) and $L_{\mathrm{flow}}/L_{\mathrm{turb}}$ might be overestimated. In addition, $t_\mathrm{diss}$ and $L_{\mathrm{turb}}$, as stated by \citet{ABG2010}, can only be roughly estimated owing to the differences of outflow activity between numerical simulations and reality. Therefore, $L_{\mathrm{flow}}/L_{\mathrm{turb}}$ may be highly uncertain due to the uncertainties both in $L_{\mathrm{turb}}$ and the dynamical timescale of the outflow lobe candidates, which are related to $L_{\mathrm{flow}}$ (see Paper I).

Approximately $\sim 18\%$ (24/136) and $\sim 3\%$ (4/136) of the clouds have values of $E_{\mathrm{flow}}/E_{\mathrm{turb}}$ and $P_{\mathrm{flow}}/P_{\mathrm{turb}}$ that are greater than unity in the Perseus arm. The proportions are $\sim 29\%$ (36/124) and $\sim 6\%$ (8/124) for the Local arm, and 0\% (0/5) and 0\% (0/5) for interarm 1, respectively. These proportions indicate that: in a minority of clouds, outflow activity is enough to maintain turbulence. Note that we did not consider the proportion of $L_{\mathrm{flow}}/L_{\mathrm{turb}}$ because it might be highly uncertain.

\subsubsection{Potential Disruptive Effect}\label{section:DER}

Outflow candidates may potentially disrupt their parent clouds \citep{AG2002}. One method to evaluate this effect is to compare the total kinetic energy of the outflow candidates to the cloud's gravitational binding energy,  $E_{\mathrm{flow}}/E_{\mathrm{grav}}$, or $\eta_\mathrm{out} = 2E_{\mathrm{flow}}/E_{\mathrm{grav}}$. $E_{\mathrm{grav}}$ read
\begin{equation}
 E_{\mathrm{grav}}=\frac{GM^2_{\mathrm{cloud}}}{R_{\mathrm{cloud}}}.
\end{equation}
Another method to estimate the potential disruptive effect is to determine the ratio of the escape mass, $M_\mathrm{esc}$, to the cloud mass, $M_\mathrm{cloud}$, where $M_\mathrm{esc}$ is defined following \citet{ABG2010}
\begin{equation}
 M_{\mathrm{esc}}=\frac{P_{\mathrm{flow}}}{v_{\mathrm{esc}}}=
 \frac{P_{\mathrm{flow}}}{\sqrt{2GM_{\mathrm{cloud}}/R_{\mathrm{cloud}}}}.
\end{equation}
For this equation, $P_\mathrm{flow}$ is catalogued in Table \ref{Table:relation}, and $v_{\mathrm{esc}}$ is the cloud's escape velocity. The ratio of $M_\mathrm{esc}$ to the total mass of the outflow candidates in the cloud, $M_{\mathrm{esc}}/M_{\mathrm{flow}}$, can be used to estimate the impact of the outflow candidates on the environment in their immediate vicinity \citep[e.g., ][]{LLX2018}. These quantities are listed in Table \ref{Table:parameter}, and the ratios are listed in Table \ref{Table:ratio}.
%The scale of such impact is less than the value derived by $v_{\mathrm{esc, m}}\cdot t_\mathrm{lobe, m}/2 \sim 0.3$ pc with $v_{\mathrm{esc, m}} \sim 1$ km s$^{-1}$ and $t_\mathrm{lobe, m} \sim 5 \times 10^5$ yr (the dynamical timescale of outflow lobe candidates, see Paper II) in the Perseus arm, where subscript ``m'' means mean value.

Roughly $\sim 54\%$ (74/136) and $\sim 5\%$ (7/136) of the clouds have values of $\eta_\mathrm{out} = 2E_{\mathrm{flow}}/E_{\mathrm{grav}}$ and $M_{\mathrm{esc}}/M_{\mathrm{cloud}}$ that are greater than unity in the Perseus arm. The proportions are $\sim 66\%$ (82/124) and $\sim 26\%$ (32/124) for the Local arm, and 100\% (5/5) and 0\% (0/5) for interarm 1, respectively. It is possible that outflow activities could potentially disrupt the entire cloud, at least for some clouds. $M_{\mathrm{esc}}/M_{\mathrm{flow}}$ reached $\sim 12.2$, $\sim 34.0$ and $\sim 17.6$ for the clouds in the Perseus arm, the Local arm and interarm 1, respectively, indicating that outflow candidates could disperse the gas in their immediate vicinities.

\section{Discussion}{\label{discussion}}

In this work, 265 clouds with radii ranging from $\sim 0.04$ to $\sim 17.12$ pc were detected. This large sample enabled us to calculate the feedback properties of the outflow candidates as functions of the cloud radius. In the following, we further investigate the obtained fitted functions.

\subsection{Correlations among Cloud Properties}

\begin{figure}[!ht]
%\figurenum{A. 5}
\vspace{0.2cm}
\centering
 \subfigure[$M_\mathrm{cloud}$]{\includegraphics[width=0.32\textwidth]{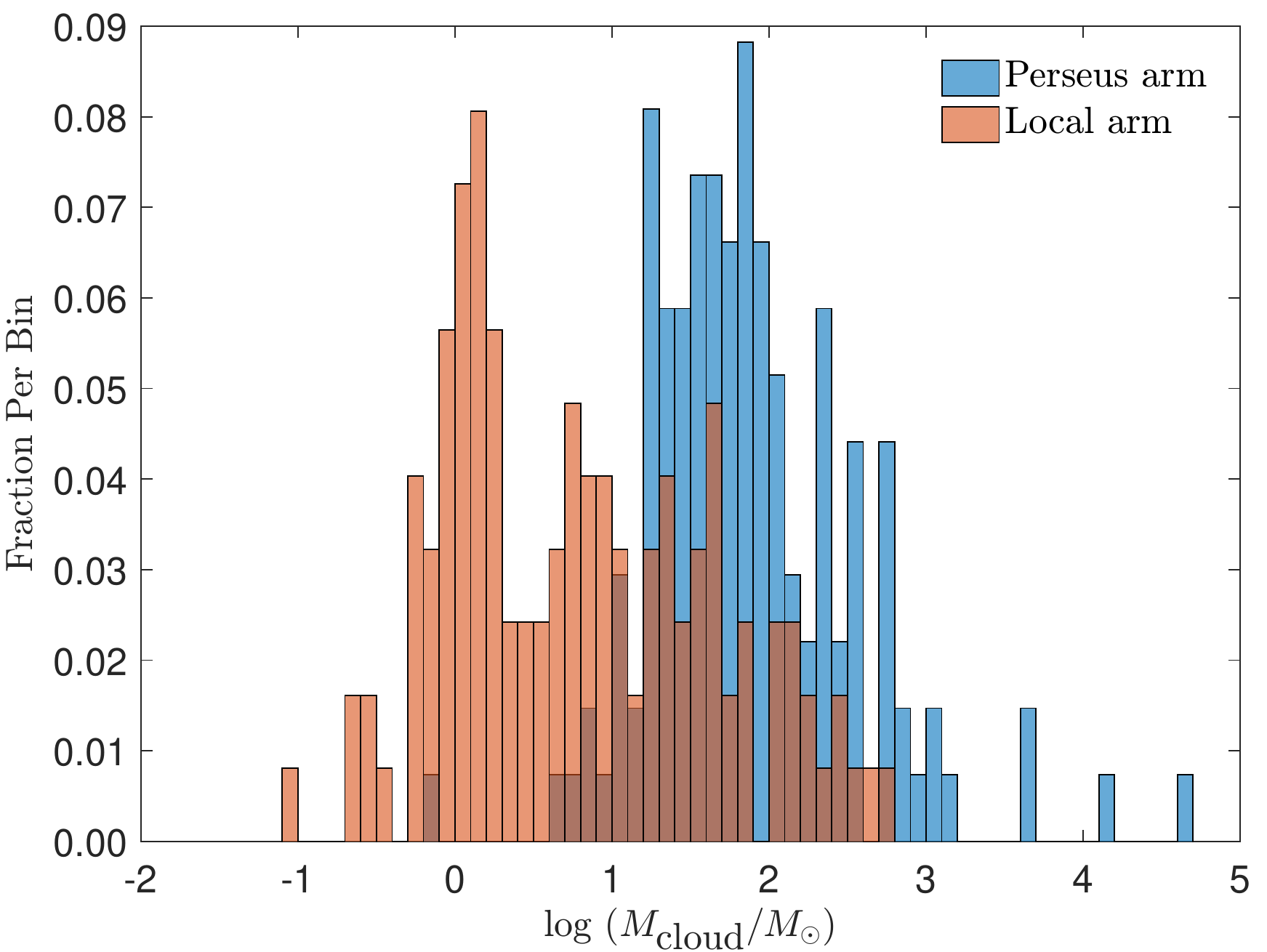}}
 \subfigure[$R_\mathrm{cloud}$]{\includegraphics[width=0.32\textwidth]{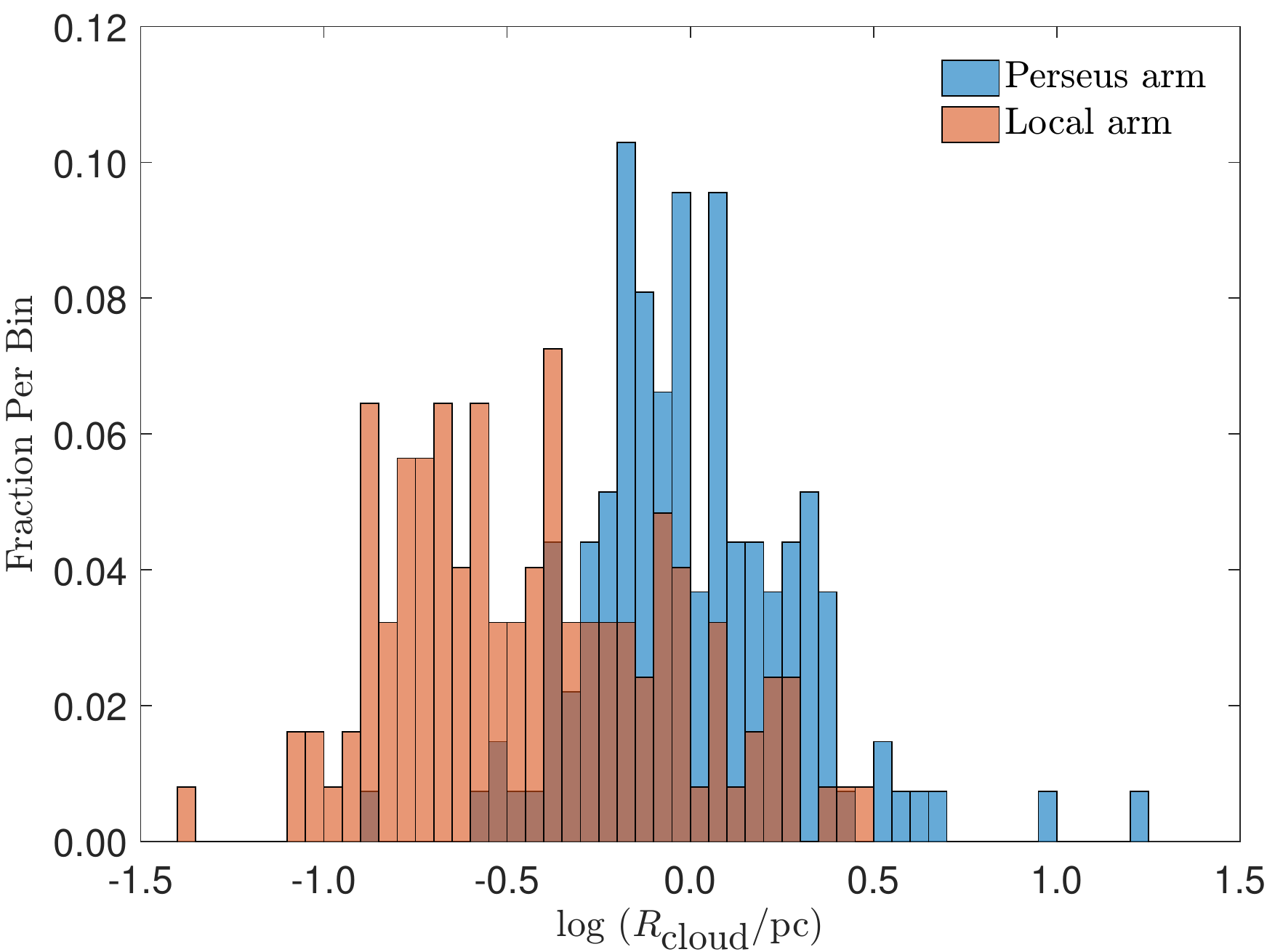}}
 \subfigure[$\Delta V_\mathrm{cloud}$]{\includegraphics[width=0.32\textwidth]{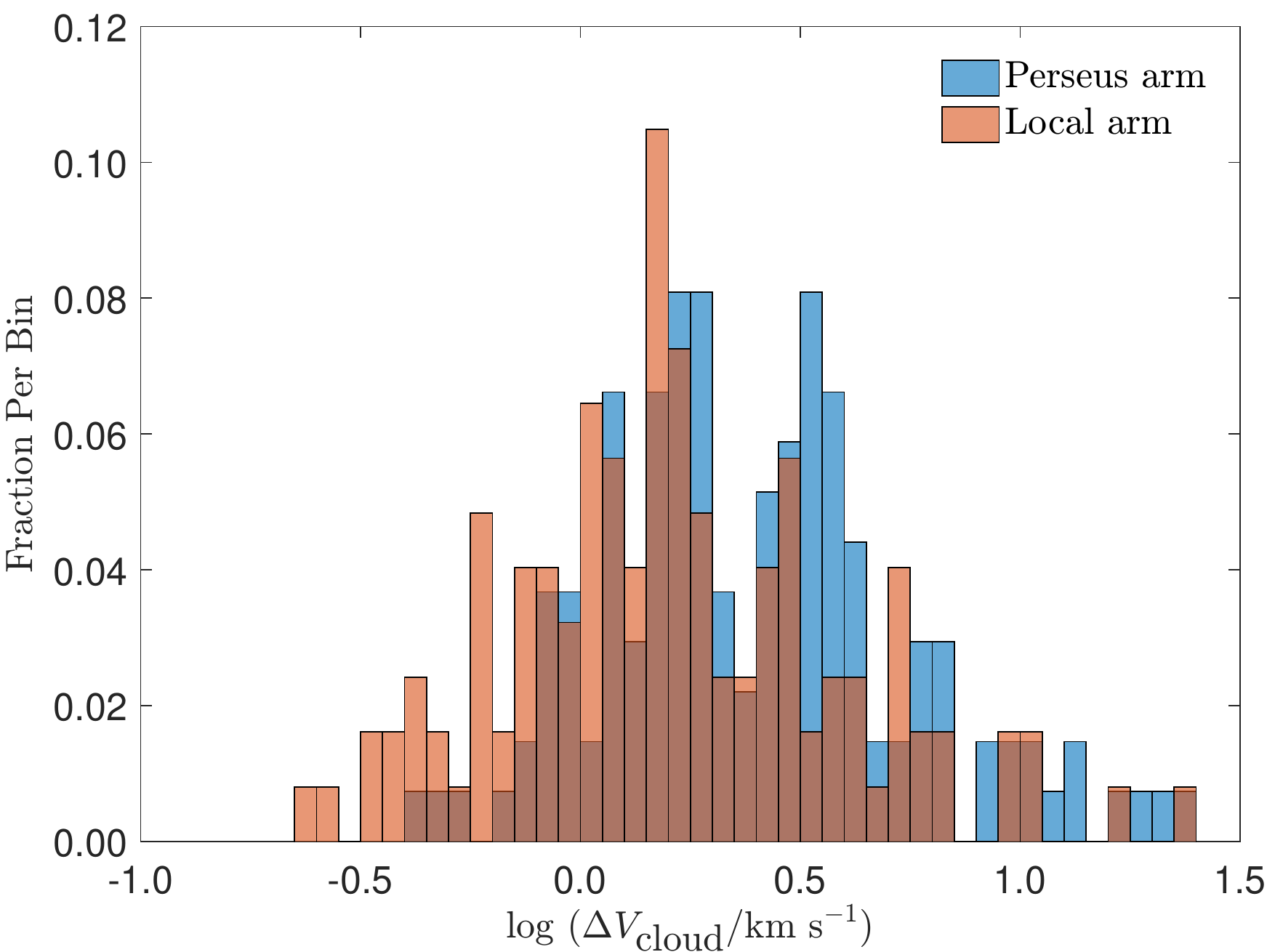}}
%\vspace{3.77cm}
\caption{The distributions of  (a) $M_\mathrm{cloud}$,  (b) $R_\mathrm{cloud}$ and  (c) $\Delta V_\mathrm{cloud}$ for clouds in the Perseus arm and the Local arm.}
\label{Fig:distr}
\end{figure}

Figure \ref{Fig:distr} shows that the two samples (one in the Perseus arm and the other one in the Local arm) are different from each other in properties of $M_\mathrm{cloud}$ and $R_\mathrm{cloud}$, but similar in properties of $\Delta V_\mathrm{cloud}$. The distances to the Perseus arm and the Local arm 
are significantly different, resulting in the different levels of beam dilution effect. A greater fraction of molecular gas is expected to be missed for the sample with a larger distance (such as the Perseus arm), and therefore bias the derived properties of the clouds (e.g., sizes and masses of the clouds), and the relationships between different properties. 
%This would result in the difference of the distribution of $M_\mathrm{cloud}$ and $R_\mathrm{cloud}$ between the two samples. 
To diminish the bias introduced by the different levels of beam dilution, 
we investigate the physical properties of clouds within individual spiral arm, e.g., we separately investigate turbulent support of outflow candidate of clouds in the Perseus arm and the Local arm.
%Although the entire sample can mitigate the effect of a different environment, they cannot alleviate the effect of dilution. Therefore, the related results for the entire sample below should be treated with caution.}

\begin{figure}[!ht]
%\figurenum{A. 5}
\vspace{0.2cm}
\centering
 \subfigure[$E_\mathrm{turb}$ vs. $R_\mathrm{cloud}$]{\includegraphics[width=0.49\textwidth]{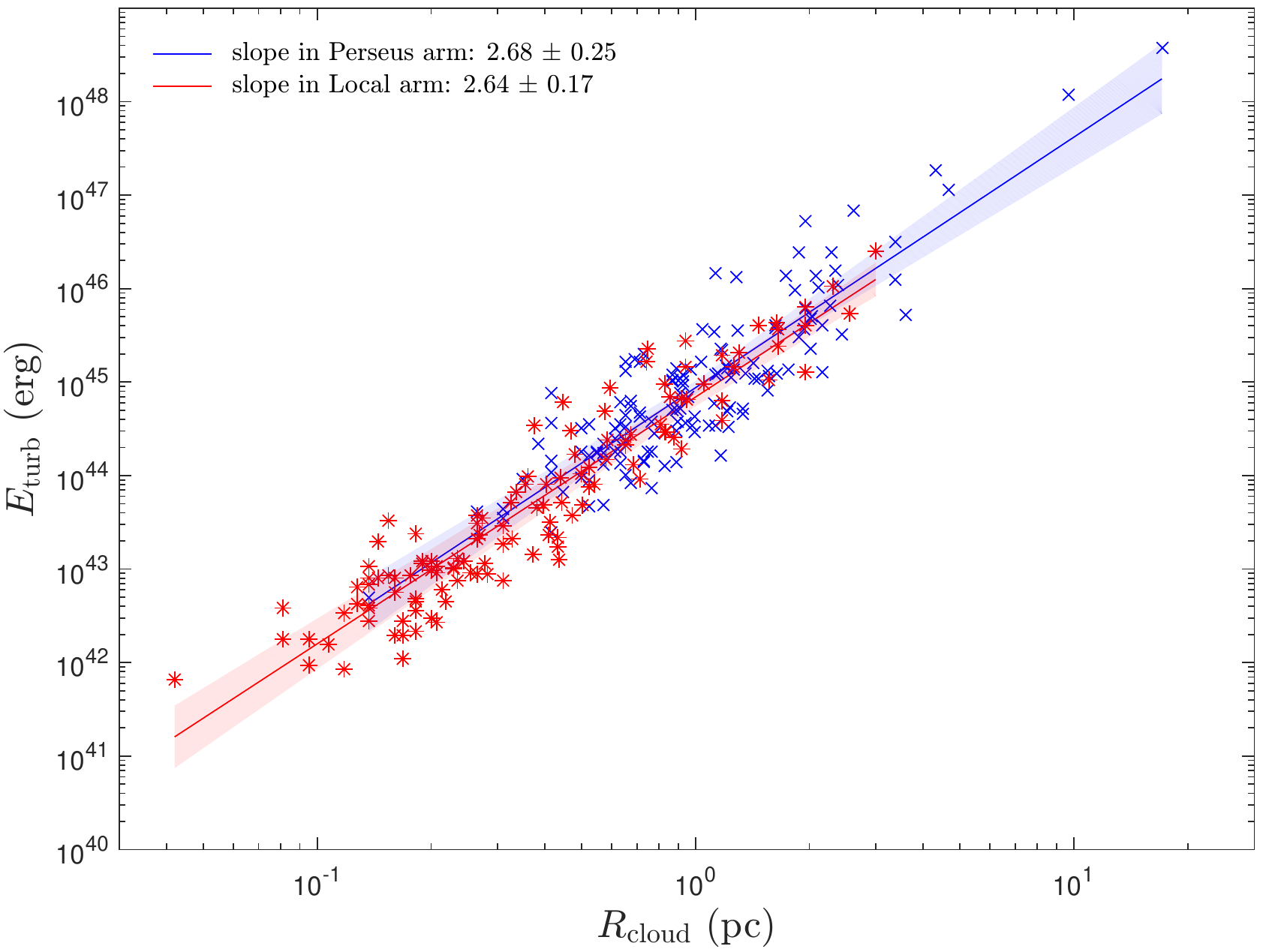}}
 \subfigure[$P_\mathrm{turb}$ vs. $R_\mathrm{cloud}$]{\includegraphics[width=0.49\textwidth]{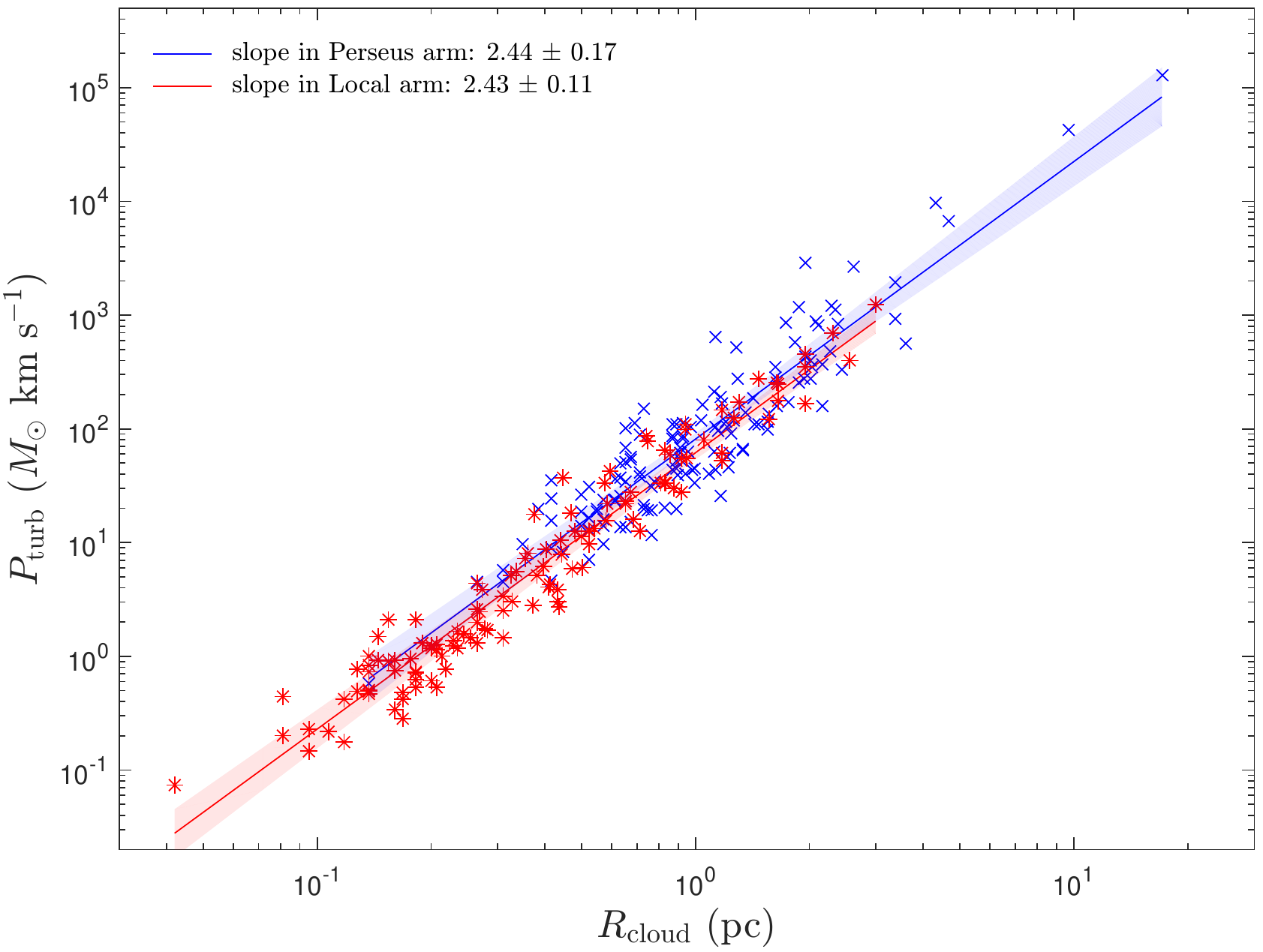}}
%\vspace{3.77cm}
\caption{(a) $E_\mathrm{turb}$ and (b) $P_\mathrm{turb}$ as functions of $R_\mathrm{cloud}$.}
\label{Fig:cep}
\end{figure}

Figure \ref{Fig:cep} shows $E_\mathrm{turb}$ and $P_\mathrm{turb}$ as functions of $R_\mathrm{cloud}$. The power-law indexes (PLIs) for the former one  are $2.68\pm0.25$ and $2.64\pm 0.17$ (with 95\% confidence using a least-square method, this method also applies to all fits below) for clouds in the Perseus arm and the Local arm, respectively, 
and the corresponding correlation coefficients (c.c.) are 0.89 and 0.94. For the latter one, the corresponding PLIs are $2.44\pm0.17$ and $2.43\pm 0.11$ with c.c. of 0.93 and 0.97. The expected PLIs for $E_\mathrm{turb}$-$R_\mathrm{cloud}$ from Larson's relations are 1.60 \citep[with assumption of $M_\mathrm{cloud}\propto R_\mathrm{cloud}^{1.1}$ and $\Delta V_\mathrm{cloud}\propto R_\mathrm{cloud}^{0.38}$,][]{L1981,TDE2018} or 1.84 \citep[with assumption of $\Delta V_\mathrm{cloud}\propto R_\mathrm{cloud}^{0.5}$ determined from large-scale observations,][see details about the linewidth-size relation in Section \ref{sec:LSRV}]{PN2002, MO2007, TDE2018}. The corresponding expected PLIs for $P_\mathrm{turb}$-$R_\mathrm{cloud}$ are 1.48 or 1.60. These facts indicate that, for clouds in the Local arm and the Perseus arm, $E_\mathrm{turb}$ and $P_\mathrm{turb}$ are correlated with $R_\mathrm{cloud}$ (with c.c. $\geq$ 0.89),  and the measured PLIs deviate from (larger than) those expected from Larson's relations. It is interesting to investigate whether outflow activity plays a role in impacting cloud properties.

\subsection{Turbulent Support}\label{sec:TSD}

\begin{figure}[!ht]
%\figurenum{A. 5}
\vspace{0.2cm}
\centering
 \subfigure[$E_{\mathrm{flow}}/E_{\mathrm{turb}}$ vs. $R_\mathrm{cloud}$]{\includegraphics[width=0.328\textwidth]{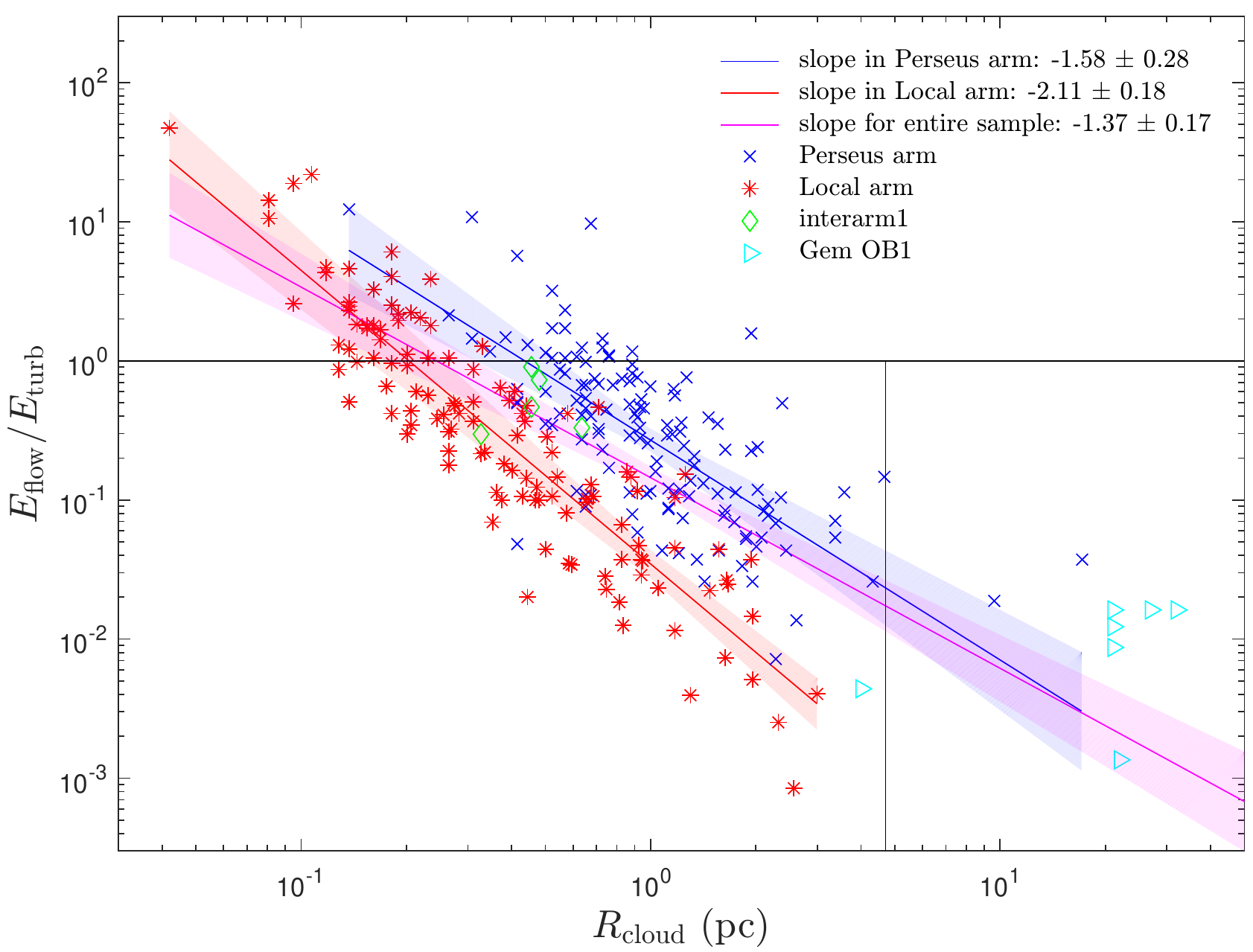}}
 \subfigure[$P_{\mathrm{flow}}/P_{\mathrm{turb}}$ vs. $R_\mathrm{cloud}$]{\includegraphics[width=0.328\textwidth]{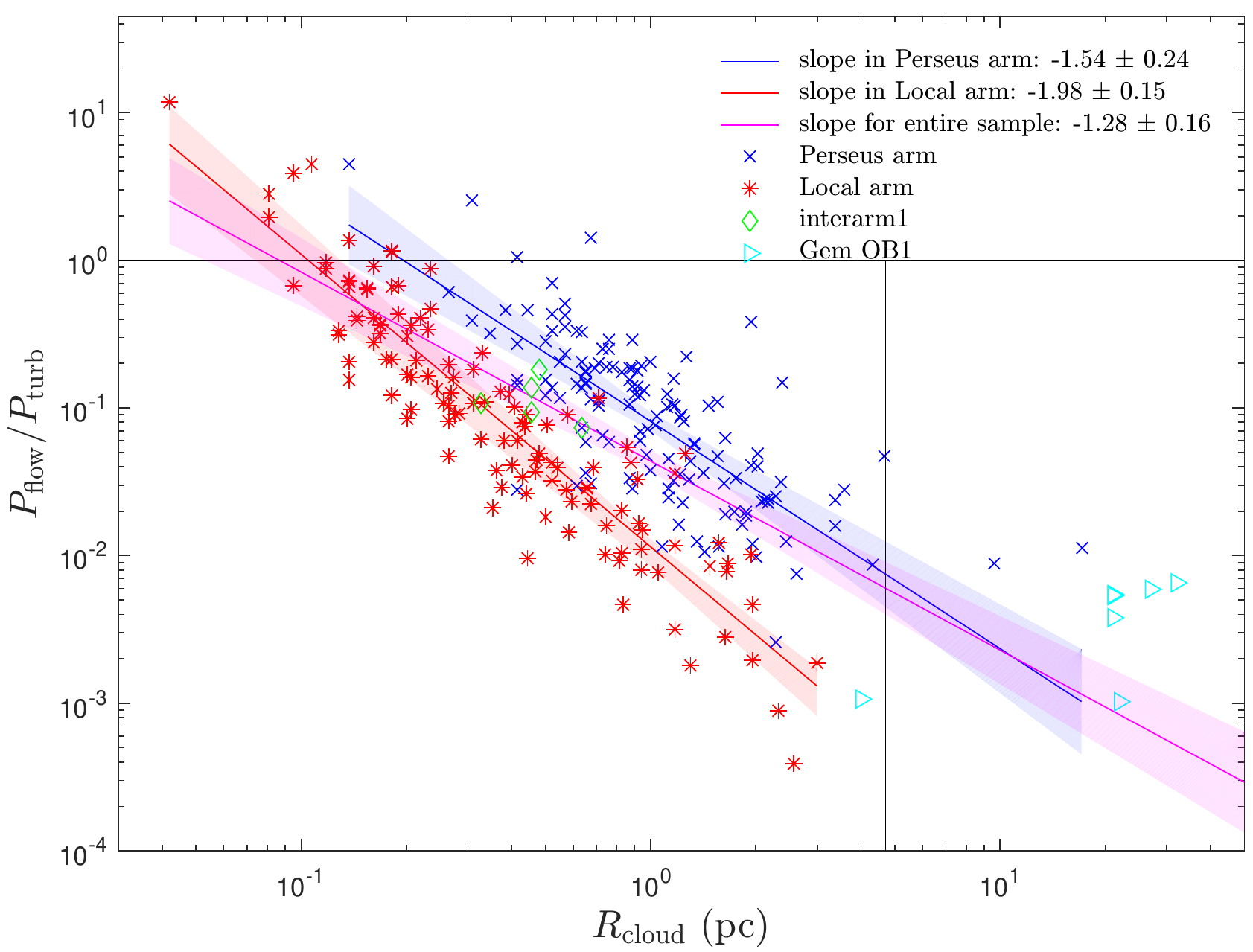}}
 \subfigure[$L_{\mathrm{flow}}/L_{\mathrm{turb}}$ vs. $R_\mathrm{cloud}$]{\includegraphics[width=0.328\textwidth]{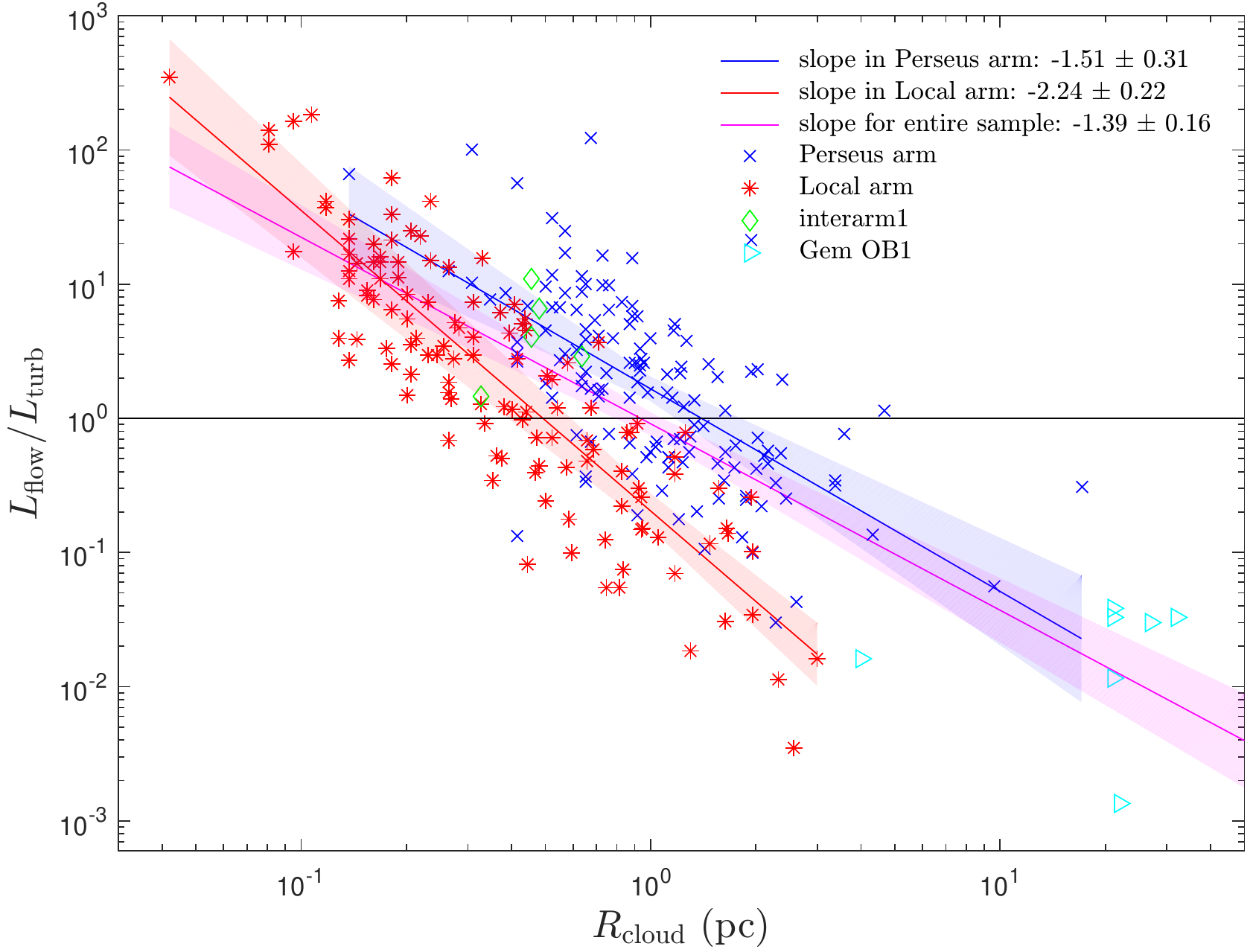}}
%\vspace{3.77cm}
\caption{(a) $E_{\mathrm{flow}}/E_{\mathrm{turb}}$, (b) $P_{\mathrm{flow}}/P_{\mathrm{turb}}$ and (c) $L_{\mathrm{flow}}/L_{\mathrm{turb}}$ as functions of $R_\mathrm{cloud}$. The data used to fit the relationships are located in the left side of the vertical line (totally 134 clouds for the sample of the Perseus arm) for panels (a) and (b).}
\label{Fig:ret}
\end{figure}

The ratios used to evaluate the turbulent support of the outflow candidates to their parent clouds are $E_{\mathrm{flow}}/E_{\mathrm{turb}}$, $P_{\mathrm{flow}}/P_{\mathrm{turb}}$ and $L_{\mathrm{flow}}/L_{\mathrm{turb}}$ (see Table \ref{Table:ratio}). These three ratios are plotted as functions of cloud radius, $R_\mathrm{cloud}$, in Figure \ref{Fig:ret}. The samples in \cite{LLX2018}, labeled as ``Gem OB1'', are also included to enlarge the sample to larger radii. Here, the ratio $P_{\mathrm{flow}}/P_{\mathrm{turb}}$ was calculated using equation (\ref{equ:2}) with data from \citet{LLX2018}.

Figure \ref{Fig:ret} shows that $L_{\mathrm{flow}}/L_{\mathrm{turb}}$ is much bigger than $E_{\mathrm{flow}}/E_{\mathrm{turb}}$; such result has been reported by many other researchers \citep[e.g.,][]{ABG2010, LLQ2015, LLX2018}. This result may arise from an overestimation of the driving wavelength (see Sections \ref{section:TSR}). Further discussion regarding three ratios as functions of $R_\mathrm{cloud}$ is presented in detail below, which is followed by a summary.

\subsubsection{Ratio of $E_{\mathrm{flow}}/E_{\mathrm{turb}}$ as a Function of $R_\mathrm{cloud}$}\label{subsection:RET}

The best-fitting power-law functions for the clouds in the Perseus arm, the Local arm and the entire sample are given by
\begin{subequations}
\begin{align}
  \log (E_{\mathrm{flow}}/E_{\mathrm{turb}}) = (-1.58 \pm 0.28) \log R_\mathrm{cloud}-0.57 \pm 0.08, c.c. = -0.70, \label{equ:retp}\\
  \log (E_{\mathrm{flow}}/E_{\mathrm{turb}}) = (-2.11 \pm 0.18) \log R_\mathrm{cloud}-1.46\pm 0.10, c.c. = -0.91, \label{equ:retl}\\
  \log (E_{\mathrm{flow}}/E_{\mathrm{turb}}) = (-1.37 \pm 0.17) \log R_\mathrm{cloud}-0.84 \pm 0.07, c.c. = -0.71, \label{equ:rett}
\end{align}
\end{subequations}
respectively. The PLIs and the scale factors (SFs) are catalogued in Table \ref{Table:CRR}.
%Taking the average radius of $\sim 1$ pc of the six sub-regions in the Perseus molecular cloud complex, the derived mean $E_{\mathrm{flow}}/E_{\mathrm{turb}}$ from equation (\ref{equ:retp})--(\ref{equ:rett}), i.e., $\sim 24\%$, is consisted with the average $E_{\mathrm{flow}}/E_{\mathrm{turb}}$ in \citet{ABG2010}.%We cataloged the fitted value of slope, intercept, $c.c.$ and CRR (see below) in table \ref{table:fits} to compare with other parameters used to estimate turbulent support (see Section \ref{section:TSR}) and disruptive effect (see Section \ref{section:DER}) of outflow candidates.

%\newgeometry{left=3cm,bottom=-1.5cm}
\rotate
%\rotatetable 
\begin{deluxetable*}{ccccccccccccccccccccccc}
%\renewcommand{\thetable}{\arabic{table}}
%\centering
\setlength\tabcolsep{1.3pt}
\tablecolumns{23}
\tabletypesize{\tiny}
%\movetabledown=10cm
%\tablewidth{2\linewidth}
%\adjustbox{center}{1.5\linewidth}
%\resizebox{\textwidth}{12mm}{}
\tablecaption{Effect of Beam Dilution, the Uncertainty of Distance and Multi-component on CRR\label{Table:CRR}}
%\begin{tabular}
\tablehead{
%\\
  \noalign{\smallskip} 
    \noalign{\smallskip} 
      \noalign{\smallskip} 
\colhead{Sample} & \multicolumn{3}{c}{$E_{\mathrm{flow}}/E_{\mathrm{turb}}$} & \colhead{} & \multicolumn{3}{c}{$P_{\mathrm{flow}}/P_{\mathrm{turb}}$} & \colhead{} & \multicolumn{3}{c}{$L_{\mathrm{flow}}/L_{\mathrm{turb}}$} & \colhead{} & \multicolumn{3}{c}{$E_{\mathrm{flow}}/E_{\mathrm{grav}}$} & \colhead{} & \multicolumn{3}{c}{$M_{\mathrm{esc}}/M_{\mathrm{cloud}}$} & \colhead{} & \multicolumn{2}{c}{$M_{\mathrm{esc}}/M_{\mathrm{flow}}$} \\
\cline{2-4} \cline{6-8}  \cline{10-12} \cline{14-16} \cline{18-20} \cline{22-23}
\colhead{ }  & \colhead{PLI} & \colhead{SF} & \colhead{CRR} & \colhead{} & \colhead{PLI} & \colhead{SF} & \colhead{CRR} & \colhead{} & \colhead{PLI} & \colhead{SF} & \colhead{CRR} & \colhead{} & \colhead{PLI} & \colhead{SF} & \colhead{CRR} & \colhead{} & \colhead{PLI} & \colhead{SF} & \colhead{CRR} & \colhead{} & \colhead{PLI} & \colhead{SF} \\
\colhead{ }  & \colhead{ } & \colhead{ } & \colhead{(pc)} & \colhead{} & \colhead{ } & \colhead{ } & \colhead{(pc)} & \colhead{} & \colhead{ } & \colhead{ } & \colhead{(pc)} & \colhead{} & \colhead{ } & \colhead{ } & \colhead{(pc)} & \colhead{} & \colhead{ } & \colhead{ } & \colhead{(pc)} & \colhead{} & \colhead{ } & \colhead{}
}
\startdata									
\multicolumn{23}{l}{Without regard to effects (all the physical values are with 95\% confidence)}\\
\hline
Perseus Arm      &	-1.58$\pm$0.28 	&	-0.57$\pm$0.08 	&	0.43$\pm$0.16 	& &	-1.54$\pm$0.24 	&	-1.09$\pm$0.06 	&	0.20$\pm$0.10 	& & -1.51$\pm$0.31 	&	 0.22$\pm$0.09 	&	1.45$\pm$0.58	& &	-2.37$\pm$0.28	&	-0.30$\pm$0.07	&	1.01$\pm$0.14	& &	-1.93$\pm$0.24	&	-0.96$\pm$0.07	&	0.32$\pm$0.10	& &	-0.45$\pm$0.06	&	0.68$\pm$0.02	\\
Local Arm        &	-2.11$\pm$0.18 	&	-1.46$\pm$0.10 	&	0.20$\pm$0.06 	& &	-1.98$\pm$0.15 	&	-1.94$\pm$0.09 	&	0.11$\pm$0.04 	& & -2.24$\pm$0.22 	&	-0.69$\pm$0.13 	&	0.49$\pm$0.16	& &	-2.91$\pm$0.18	&	-1.04$\pm$0.10	&	0.56$\pm$0.10	& &	-2.37$\pm$0.16	&	-1.73$\pm$0.09	&	0.19$\pm$0.06	& &	-0.52$\pm$0.06	&	0.73$\pm$0.03	\\
Entire sample&	-1.37$\pm$0.17 	&	-0.84$\pm$0.07 	&	0.24$\pm$0.10 	& &	-1.28$\pm$0.16 	&	-1.36$\pm$0.07 	&	0.09$\pm$0.06 	& & -1.39$\pm$0.16 	&	-0.04$\pm$0.08 	&	0.94$\pm$0.26	& &	-2.26$\pm$0.16	&	-0.52$\pm$0.07	&	0.80$\pm$0.12	& &	-1.73$\pm$0.15	&	-1.20$\pm$0.06	&	0.20$\pm$0.06	& &	-0.49$\pm$0.03	&	0.72$\pm$0.02	\\
Perseus+Local&	-1.36$\pm$0.17 	&	-0.84$\pm$0.08 	&	0.25$\pm$0.10 	& &	-1.28$\pm$0.16 	&	-1.36$\pm$0.07 	&	0.09$\pm$0.06 	& & -1.41$\pm$0.19 	&	-0.04$\pm$0.09 	&	0.95$\pm$0.28	& &	-2.25$\pm$0.16	&	-0.52$\pm$0.07	&	0.80$\pm$0.12	& &	-1.72$\pm$0.15	&	-1.19$\pm$0.07	&	0.20$\pm$0.06	& &	-0.53$\pm$0.04	&	0.70$\pm$0.02	\\
\hline                                                                
\multicolumn{23}{l}{Effect of beam dilution (set the disntance to the Perseus arm to be 600 pc,  all the physical values are with 95\% confidence)}\\                        
\hline                                                                
Perseus Arm      &	-1.58$\pm$0.28 	&	-0.57$\pm$0.08 	&	0.43$\pm$0.18 	& &	-1.54$\pm$0.24 	&	-1.88$\pm$0.14 	&	0.06$\pm$0.06 	& & -1.51$\pm$0.31 	&	-0.55$\pm$0.18 	&	0.44$\pm$0.30	& &	-2.37$\pm$0.28	&	-1.00$\pm$0.16	&	0.51$\pm$0.18	& &	-1.93$\pm$0.24	&	-1.69$\pm$0.14	&	0.14$\pm$0.08	& &	-0.45$\pm$0.06	&	0.71$\pm$0.03	\\
Local Arm      &	-2.11$\pm$0.18 	&	-1.46$\pm$0.10 	&	0.20$\pm$0.06 	& &	-1.98$\pm$0.15 	&	-1.94$\pm$0.09 	&	0.11$\pm$0.04 	& & -2.24$\pm$0.22 	&	-0.69$\pm$0.13 	&	0.49$\pm$0.16	& &	-2.91$\pm$0.18	&	-1.04$\pm$0.10	&	0.56$\pm$0.10	& &	-2.37$\pm$0.16	&	-1.73$\pm$0.09	&	0.19$\pm$0.06	& &	-0.52$\pm$0.06	&	0.73$\pm$0.03	\\
Perseus+Local	&	-1.89$\pm$0.16 	&	-1.45$\pm$0.09 	&	0.17$\pm$0.06 	& &	-1.79$\pm$0.14 	&	-1.94$\pm$0.08 	&	0.08$\pm$0.04 	& & -1.92$\pm$0.19 	&	-0.66$\pm$0.11 	&	0.45$\pm$0.14	& &	-2.68$\pm$0.16	&	-1.05$\pm$0.09	&	0.53$\pm$0.10	& &	-2.19$\pm$0.14	&	-1.74$\pm$0.08	&	0.16$\pm$0.04	& &	-0.48$\pm$0.04	&	0.72$\pm$0.03	\\
\hline                                                               
\multicolumn{23}{l}{Effect of the uncertainty of distance (``average value'' $\pm$ ``standard deviation'' of one thousand tests)}\\                            
\hline                                                               
Perseus Arm   &	-1.53$\pm$0.04 	&	-0.57$\pm$0.01 	&	0.42$\pm$0.02 	& &	-1.49$\pm$0.03 	&	-1.09$\pm$0.01 	&	0.19$\pm$0.02 	& & -1.61$\pm$0.04 	&	-0.21$\pm$0.01 	&	0.74$\pm$0.02	& &	-2.32$\pm$0.03	&	-0.30$\pm$0.01	&	1.00$\pm$0.02	& &	-1.88$\pm$0.03	&	-0.96$\pm$0.01	&	0.31$\pm$0.02	& &	-0.46$\pm$0.01	&	0.68$\pm$0.00	\\
Local Arm   &	-1.85$\pm$0.09 	&	-1.38$\pm$0.04 	&	0.18$\pm$0.04 	& &	-1.73$\pm$0.08 	&	-1.87$\pm$0.04 	&	0.08$\pm$0.02 	& & -1.96$\pm$0.10 	&	-0.61$\pm$0.04 	&	0.49$\pm$0.06	& &	-2.67$\pm$0.08	&	-0.97$\pm$0.04	&	0.56$\pm$0.04	& &	-2.14$\pm$0.09	&	-1.66$\pm$0.04	&	0.17$\pm$0.02	& &	-0.52$\pm$0.01	&	0.73$\pm$0.00	\\
\hline                                                            
\multicolumn{23}{l}{Effect of multi-component (``average value'' $\pm$ ``standard deviation'' of three thousand tests)}\\                  
\hline                                                            
Perseus Arm     &	-1.50$\pm$0.05 	&	-0.55$\pm$0.01 	&	0.43$\pm$0.02 	& &	-1.46$\pm$0.04 	&	-1.07$\pm$0.01 	&	0.19$\pm$0.01 	& & -1.39$\pm$0.05 	&	-0.25$\pm$0.01 	&	0.66$\pm$0.02	& &	-2.24$\pm$0.06	&	-0.27$\pm$0.01	&	1.03$\pm$0.02	& &	-1.84$\pm$0.05	&	-0.94$\pm$0.01	&	0.31$\pm$0.02	& &	-0.44$\pm$0.01	&	0.67$\pm$0.00	\\
Local Arm    &	-2.07$\pm$0.03 	&	-1.41$\pm$0.01 	&	0.21$\pm$0.02 	& &	-1.93$\pm$0.02 	&	-1.90$\pm$0.01 	&	0.10$\pm$0.01 	& & -2.19$\pm$0.03 	&	-0.63$\pm$0.02 	&	0.52$\pm$0.02	& &	-2.85$\pm$0.04	&	-0.97$\pm$0.02	&	0.58$\pm$0.02	& &	-2.33$\pm$0.03	&	-1.69$\pm$0.01	&	0.19$\pm$0.01	& &	-0.52$\pm$0.01	&	0.73$\pm$0.00	
\enddata
\tablecomments{0.00 represents $<0.005$.}
\end{deluxetable*}

The critical radius (CRR hereafter) at which the fitted line of the ratio equals unity was used to describe $E_{\mathrm{flow}}/E_{\mathrm{turb}}$ as a function of $R_\mathrm{cloud}$.\footnote{The characteristic radius CRR of the ratio $E_{\mathrm{flow}}/E_{\mathrm{turb}}$ (abbreviated as TE) is denoted as CRR$_\mathrm{TE}$. This naming convention also applies to the other ratios, such as $P_{\mathrm{flow}}/P_{\mathrm{turb}}$, $L_{\mathrm{flow}}/L_{\mathrm{turb}}$, $E_{\mathrm{flow}}/E_{\mathrm{grav}}$ (see Section \ref{sec:dege}) and $M_{\mathrm{esc}}/M_{\mathrm{cloud}}$ (see Section \ref{sec:demc}), which are abbreviated as TM, TL, GE and MC, respectively.} From our results it can be concluded that the turbulence driven by the ejecta of the outflow activity are enough to maintain the turbulence at the scale below the CRR if the outflows can couple to the dense gas where stars are forming, and therefore limit the star formation rate in their host clouds.
The CRR$_\mathrm{TE}$ (catalogued in Table \ref{Table:CRR}) is $0.43\pm0.16$, $0.20\pm0.06$ and $0.24\pm0.10$ pc with 95\% confidence for the clouds in the Perseus arm, the Local arm and for the entire sample, respectively. These values are consistent with the results of \citet{BHM2009} and \citet{ABG2010}, which collectively suggest a turbulence driving scale of $\lesssim 0.4$ pc.

\subsubsection{Ratio of $P_{\mathrm{flow}}/P_{\mathrm{turb}}$ as a Function of $R_\mathrm{cloud}$}\label{sec:tsp}

The best-fitting power law functions for the clouds in the Perseus arm, the Local arm and the entire sample are
\begin{subequations}
\begin{align}
  \log (P_{\mathrm{flow}}/P_{\mathrm{turb}}) = (-1.54 \pm 0.24) \log R_\mathrm{cloud} - 1.09 \pm 0.06, c.c. = -0.74, \label{equ:rptp}\\
  \log (P_{\mathrm{flow}}/P_{\mathrm{turb}}) = (-1.98 \pm 0.15) \log R_\mathrm{cloud} - 1.94 \pm 0.09, c.c. = -0.92, \label{equ:rptl}\\
  \log (P_{\mathrm{flow}}/P_{\mathrm{turb}}) = (-1.28 \pm 0.16) \log R_\mathrm{cloud} - 1.36 \pm 0.07, c.c. = -0.72, \label{equ:rptt}
\end{align}
\end{subequations}
respectively. CRR$_\mathrm{TM}$ is smaller than CRR$_\mathrm{TE}$, where the former is $0.20\pm0.10$, $0.11\pm0.04$ and $0.09\pm0.06$ pc for the clouds in the Perseus arm, the Local arm and for the entire sample, respectively.
%That could result from that a fraction of outflow momentum is hidden in the form of atomic gas and dissociated in shocks \citep{DC2007}. A&A 471,873 Considering that we may underestimate the total $P_{\mathrm{flow}}$ by a factor of $\sim$ 2 \citep[combine tables1 -- 3 in][]{DC2007},\footnote{The inclination angle was fixed to 60$\degr$ which was close to the inclination angle used for inclination correction in Paper II.  The total $P_{\mathrm{flow}}$ was underestimed by a factor of 1.7 for $\eta$ = 0.1 (an underdensity-matched case) and $\eta$ = 1 (a dense jet case) with typical cut-off velocities of 2 km s$^{-1}$, .} CRR$_\mathrm{TM}$ becomes $\sim 0.4$, $\sim0.3$ and $\sim 0.3$ pc for clouds in the Perseus arm, the Local arm and for the entire sample, respectively. These three values are consisting with  CRR$_\mathrm{TE}$.

\subsubsection{Ratio of $L_{\mathrm{flow}}/L_{\mathrm{turb}}$ as a Function of $R_\mathrm{cloud}$}\label{sec:rlr}

The best-fitting power-law functions for the clouds in the Perseus arm, the Local arm and the entire sample are
\begin{subequations}
\begin{align}
  \log (L_{\mathrm{flow}}/L_{\mathrm{turb}}) = (-1.51 \pm 0.31) \log R_\mathrm{cloud} + 0.22 \pm 0.09, c.c. = -0.65, \label{equ:rltp}\\
  \log (L_{\mathrm{flow}}/L_{\mathrm{turb}}) = (-2.24 \pm 0.22) \log R_\mathrm{cloud} - 0.69 \pm 0.13, c.c. = -0.88, \label{equ:rltl}\\
  \log (L_{\mathrm{flow}}/L_{\mathrm{turb}}) = (-1.39 \pm 0.16) \log R_\mathrm{cloud} - 0.04 \pm 0.08, c.c. = -0.72, \label{equ:rltt}
\end{align}
\end{subequations}
respectively. CRR$_\mathrm{TL}$ is much higher than CRR$_\mathrm{TE}$ and CRR$_\mathrm{TM}$, which is $1.45\pm0.58$, $0.49\pm0.16$ and $0.94\pm0.26$ pc for the clouds in the Perseus arm, the Local arm and the entire sample, respectively. However, we note that CRR$_\mathrm{TL}$ is likely overestimated and therefore these values are highly uncertain (see Sections \ref{section:TSR}).

\subsubsection{Summaries of Turbulent Support}\label{section:summary for TS}

The three ratios defining turbulent support (i.e., $E_{\mathrm{flow}}/E_{\mathrm{turb}}$, $P_{\mathrm{flow}}/P_{\mathrm{turb}}$ and $L_{\mathrm{flow}}/L_{\mathrm{turb}}$) as functions of $R_\mathrm{cloud}$ showed negative PLIs (see Table \ref{Table:CRR}). The Local arm presented the steepest slope. One reason for this might be that the amount of outflow activities in the Local arm is less than those in other regions. For instance, the number of outflow candidates which were associated with each cloud in the Local arm was less than that in the Perseus arm (see Table \ref{Table:relation} for details).\footnote{On average, $\sim 2.1$ outflow candidates were associated with a single cloud in the Perseus arm, and $\sim1.3$ in the Local arm.} Therefore, when $R_\mathrm{cloud}$ increases, the quantities related to outflow activities (such as $E_\mathrm{flow}$, $P_\mathrm{flow}$, $L_\mathrm{flow}$ and $M_\mathrm{flow}$) increase slower in the Local arm than those in the Perseus arm, but the cloud's properties (such as $E_{\mathrm{turb}}$ and $P_{\mathrm{turb}}$) increase at a similar rate in both arms (e.g., see Figure \ref{Fig:cep}), resulting in a steeper slope in the Local arm. 

CRR$_\mathrm{TE}$, CRR$_\mathrm{TM}$ and CRR$_\mathrm{TL}$ showed different values (see Table \ref{Table:CRR}).
CRR$_\mathrm{TE}$ and CRR$_\mathrm{TM}$ were respectively $\sim 0.2 $ and $\sim 0.4$ pc for clouds in the Perseus arm, and $\sim 0.1$ and $\sim 0.2$ pc for those in the Local arm. The strength of outflow activity might influence the value of the critical radius, because the critical radius in the Perseus arm with stronger outflow activities was higher than that in the Local arm. CRR$_\mathrm{TL}$ provided less useful physical information because $L_{\mathrm{flow}}/L_{\mathrm{turb}}$ is highly uncertain.

The spatial resolution of the telescope may have little effect on creating the differences in the CRR$_\mathrm{TE}$ and the CRR$_\mathrm{TM}$, but the conclusion for CRR$_\mathrm{TM}$ should be viewed with caution. Two possible reasons are presented as follows. First, the physical scale, whether the minimum deconvolution radius \footnote{The minimum deconvolution radius is equal to the minimum cloud radius (see Section \ref{section:pclouds}) with the area of a cloud corresponding to three pixels (i.e., the minimum number of pixels is two both in the direction of $l$ and $b$).}(respectively $\sim 0.14$ and $\sim 0.04$ pc for clouds in the Perseus arm and the Local arm) or the half-width of the telescope resolution ($\sim 0.24$ and $\sim 0.07$ pc), is less than the CRR$_\mathrm{TE}$ ($\sim 0.43$ and $\sim 0.20$ pc). Similarly, these two physical scales are less than the CRR$_\mathrm{TM}$ ($\sim 0.11$ pc) for clouds in the Local arm. For clouds in the Perseus arm, CRR$_\mathrm{TM}$ ($\sim 0.20$ pc) is larger than the minimum deconvolution radius, but is slightly less than the half-width of the telescope resolution, and only one cloud has radius below CRR$_\mathrm{TM}$. 

Second, lower spatial resolution would be more likely to result in multi-component clouds (i.e., large clouds that contain some small ones). The effect of multi-component is discussed in detail in Section \ref{sec:emc}. The results show that the PLIs and the CRRs for ratios of $E_{\mathrm{flow}}/E_{\mathrm{turb}}$ and $P_{\mathrm{flow}}/P_{\mathrm{turb}}$ roughly remain unchanged (see Table \ref{Table:CRR}), implying that resolution may also have little effect.

%Therefore, the effect of spatial resolution on CRR$_\mathrm{TM}$ for clouds in the Perseus arm should be further investigated with improved spatial resolution. may not make the CRR$_\mathrm{TE}$ and the CRR$_\mathrm{TM}$ being changed

The measured ratios, $E_{\mathrm{flow}}/E_{\mathrm{turb}}$ and $P_{\mathrm{flow}}/P_{\mathrm{turb}}$ exceeded the corresponding specific values of the fitted lines (at the same radius) for the vast majority of clouds with $R_{\mathrm{cloud}}\gtrsim 4.7$ pc in the Perseus arm and the entire sample (see clouds on the right side of the vertical line in panels (a) and (b) of Figure \ref{Fig:ret}). And they were approximately along a horizontal straight line, likely indicating that the break of the scale of turbulence driven by outflow feedback is $\gtrsim 4.7$ pc.

\subsection{Potential Disruptive Effect}\label{sec:DED}

\begin{figure}[!ht]
%\figurenum{A. 5}
\vspace{0.2cm}
\centering
\subfigure[$E_{\mathrm{flow}}/E_{\mathrm{grav}}$ vs. $R_\mathrm{cloud}$]{\includegraphics[width=0.328\textwidth]{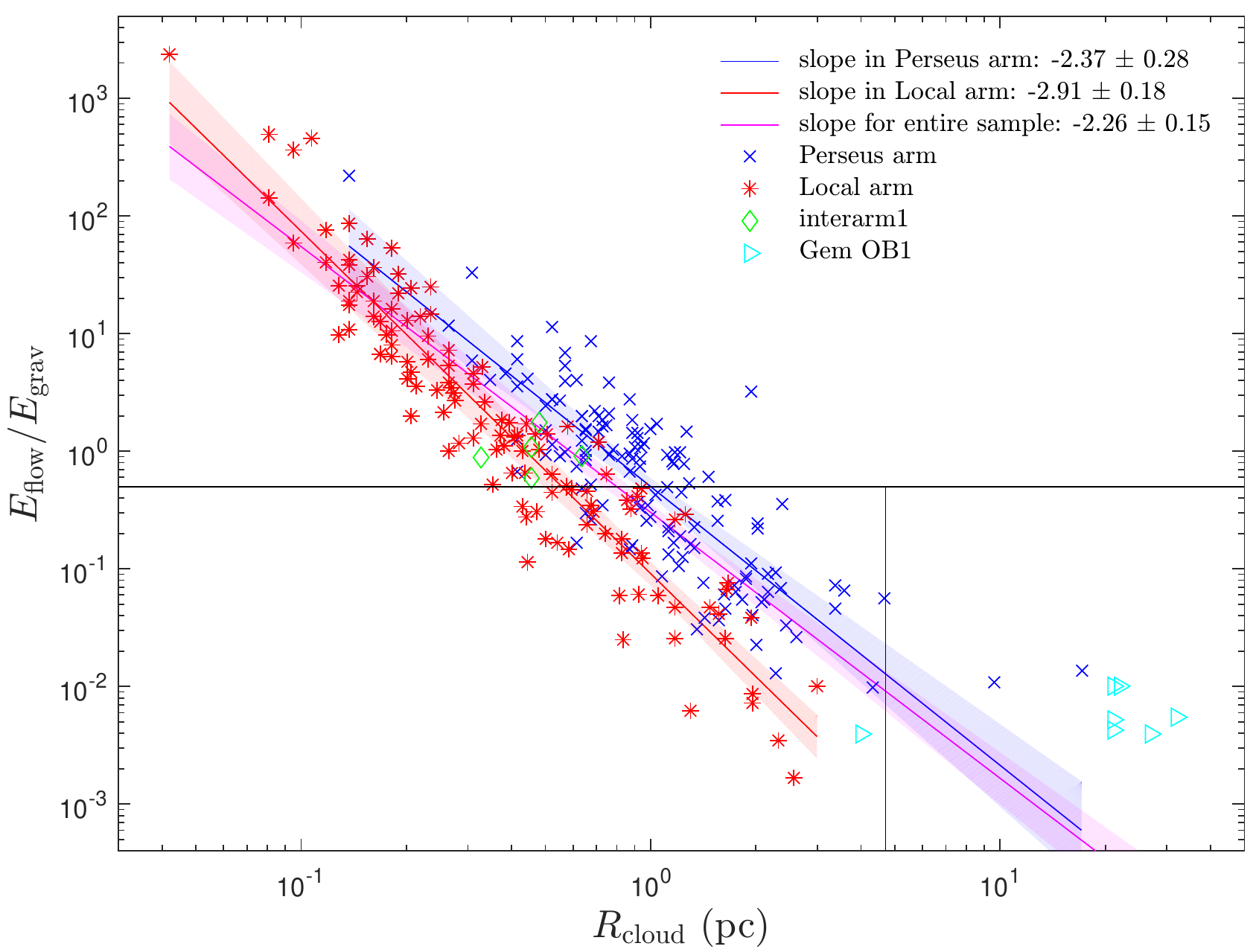}}
\subfigure[$M_{\mathrm{esc}}/M_{\mathrm{cloud}}$ vs. $R_\mathrm{cloud}$]{\includegraphics[width=0.328\textwidth]{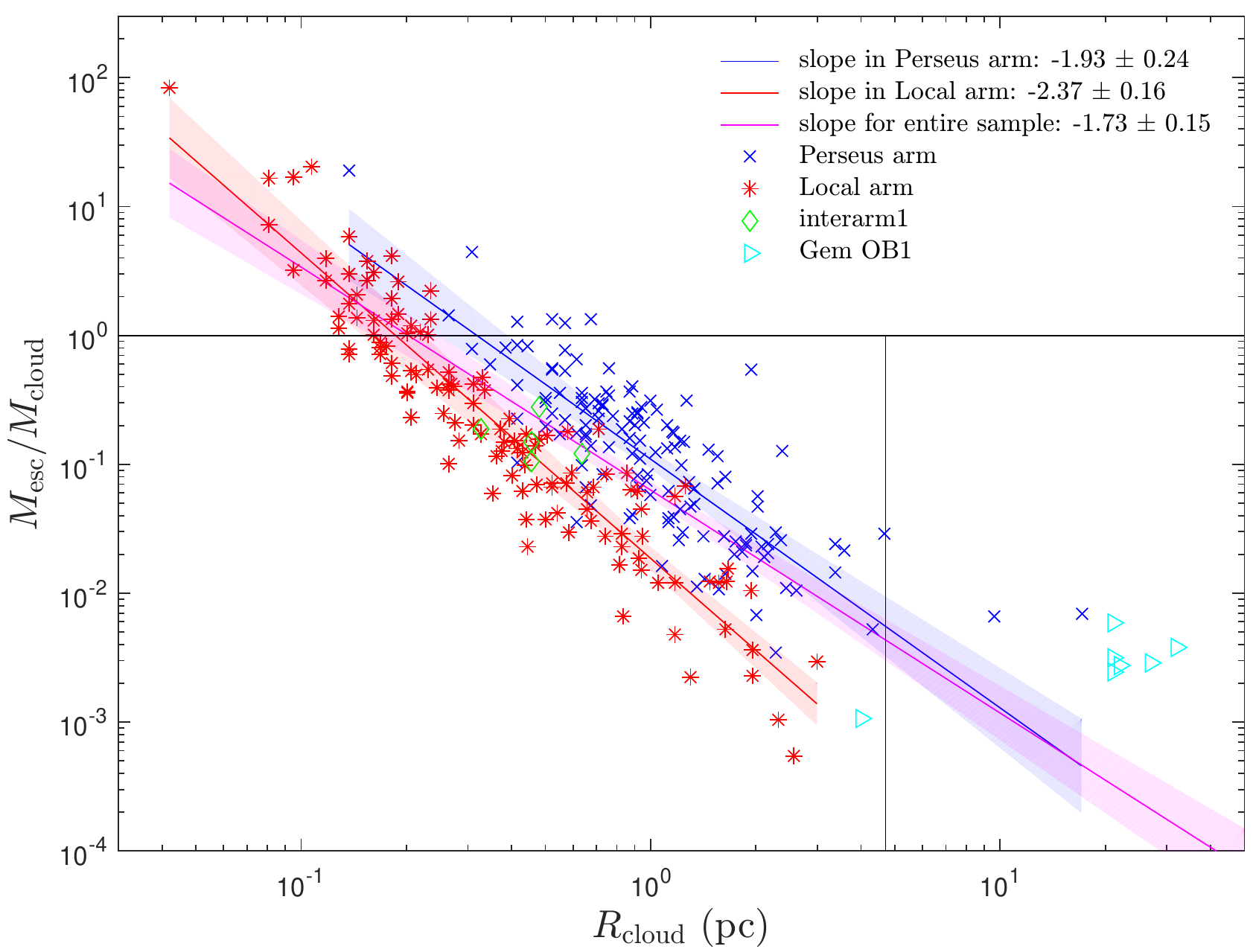}}
\subfigure[$M_{\mathrm{esc}}/M_{\mathrm{flow}}$ vs. $R_\mathrm{cloud}$]{\includegraphics[width=0.328\textwidth]{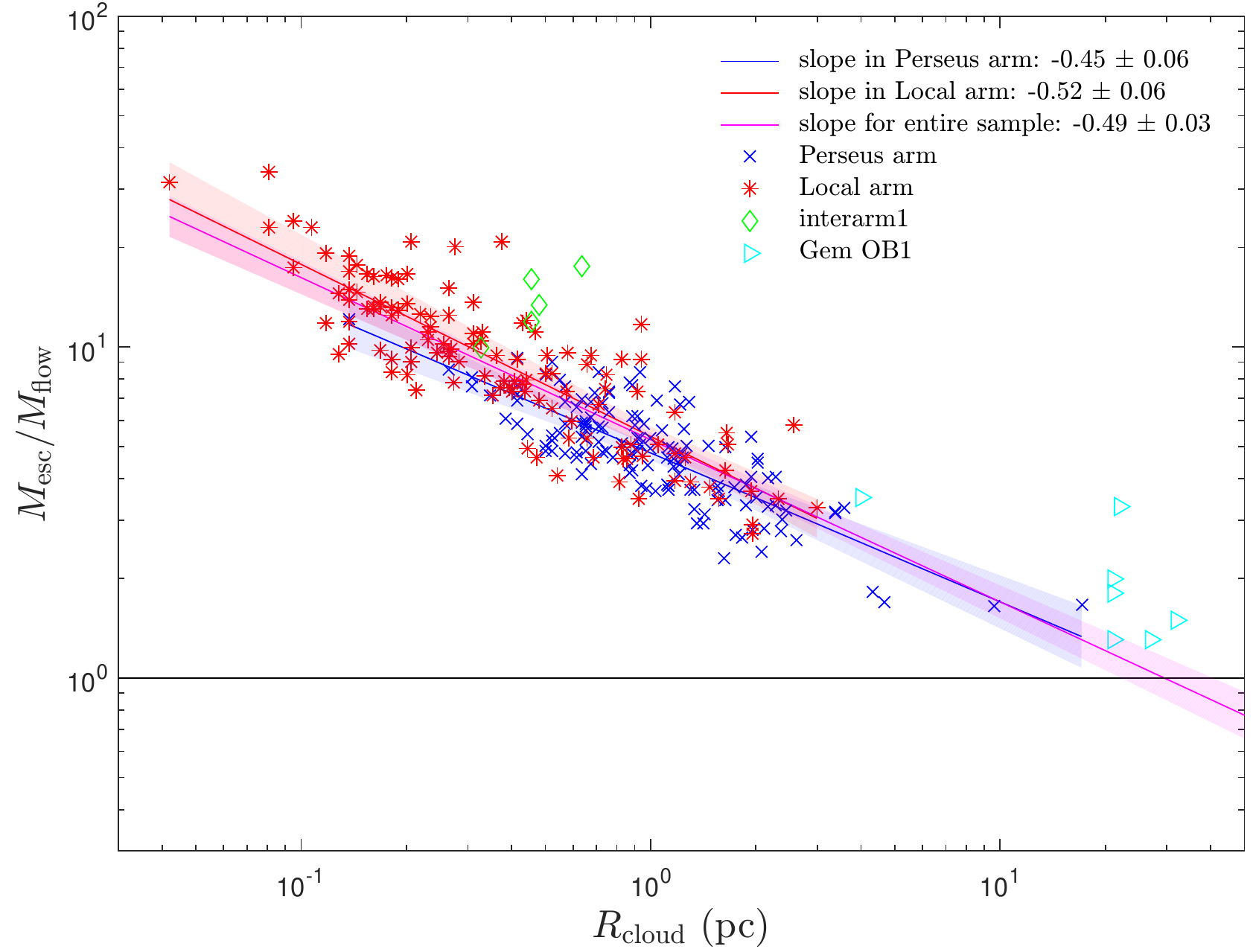}}
%\vspace{3.77cm}
\caption{(a) $E_{\mathrm{flow}}/E_{\mathrm{grav}}$, (b) $M_{\mathrm{esc}}/M_{\mathrm{cloud}}$ and (c) $M_{\mathrm{esc}}/M_{\mathrm{flow}}$ as functions of $R_\mathrm{cloud}$. The data used to fit the relationships are located in the left side of the vertical line (totally 134 clouds for the sample of the Perseus arm) for panels (a) and (b).}
\label{Fig:reg}
\end{figure}

As stated in Section \ref{section:DER}, the quantities used to estimate the potential disruptive effect of the outflow candidates include $E_{\mathrm{flow}}/E_{\mathrm{grav}}$, $M_\mathrm{esc}/M_{\mathrm{cloud}}$ and $M_{\mathrm{esc}}/M_{\mathrm{flow}}$. Figure \ref{Fig:reg} shows these three ratios as functions of $R_\mathrm{cloud}$, where the ratios in the study of \citet{LLX2018}, labeled as ``Gem OB1'', are also included. We comment on each ratio in detail and made a summary in the following subsection.

\subsubsection{Ratio of $E_{\mathrm{flow}}/E_{\mathrm{grav}}$ as a Function of $R_\mathrm{cloud}$}\label{sec:dege}

The best-fitting power-law functions for the clouds in the Perseus arm, the Local arm and the entire sample are
\begin{subequations}
\begin{align}
  \log (E_{\mathrm{flow}}/E_{\mathrm{grav}}) = (-2.37 \pm 0.28) \log R_\mathrm{cloud} -0.30 \pm 0.07, c.c. = -0.83, \label{equ:rgtp}\\
  \log (E_{\mathrm{flow}}/E_{\mathrm{grav}}) = (-2.91 \pm 0.18) \log R_\mathrm{cloud} - 1.04 \pm 0.10, c.c. = -0.95, \label{equ:rgtl}\\
  \log (E_{\mathrm{flow}}/E_{\mathrm{grav}}) = (-2.26 \pm 0.16) \log R_\mathrm{cloud} - 0.52 \pm 0.07, c.c. = -0.87, \label{equ:rgtt}
\end{align}
\end{subequations}
respectively.

Similar to $E_{\mathrm{flow}}/E_{\mathrm{turb}}$, we also defined the CRR (i.e., the cloud radius where the ratio of the fitted line equaled 0.5, i.e., $\eta_\mathrm{out}$ = 1) for $E_{\mathrm{flow}}/E_{\mathrm{grav}}$ (denoted as CRR$_\mathrm{GE}$). \footnote{If $\eta_\mathrm{out} \gtrsim$ 1, the outflow candidates will potentially disperse the clump/cloud material away from the parent clump/cloud, and hence suppress any subsequent star formation \cite[e.g.,][]{NL2014}.}CRR$_\mathrm{GE}$ (see the expression in Section \ref{subsection:RET}) is $1.01\pm 0.14$, $0.56\pm0.10$ and $0.80\pm0.12$ pc for the clouds in the Perseus arm, the Local arm and the entire sample, respectively.
%Because few cloud/clump radii are exceeded $\sim$ 0.9 pc in previous observations \citep[e.g.,][]{ABG2010, NL2014, DHB2016}, the upper limits of $\sim$ 1.3 and $\sim$ 4.9 pc for CRR$_\mathrm{GE}$ and MIR$_\mathrm{GE}$ are consist with those three studies.

\subsubsection{Ratio of $M_{\mathrm{esc}}/M_{\mathrm{cloud}}$ as a Function of $R_\mathrm{cloud}$}\label{sec:demc}

The best-fitting power-law functions for the clouds in the Perseus arm, the Local arm and the entire sample are
\begin{subequations}
\begin{align}
  \log (M_{\mathrm{esc}}/M_{\mathrm{cloud}}) = (-1.93 \pm 0.24) \log R_\mathrm{cloud} - 0.96 \pm 0.07, c.c. = -0.81, \label{equ:rmtp}\\
  \log (M_{\mathrm{esc}}/M_{\mathrm{cloud}}) = (-2.37 \pm 0.16) \log R_\mathrm{cloud} - 1.73 \pm 0.09, c.c. = -0.94, \label{equ:rmtl}\\
  \log (M_{\mathrm{esc}}/M_{\mathrm{cloud}}) = (-1.73 \pm 0.15) \log R_\mathrm{cloud} - 1.20 \pm 0.06, c.c. = -0.82, \label{equ:rmtt}
\end{align}
\end{subequations}
respectively.

CRR$_\mathrm{MC}$  (the cloud radius where the $M_{\mathrm{esc}}/M_{\mathrm{cloud}}$ of the fitted line equals unity, see the expression in Section \ref{subsection:RET}) is $0.32\pm0.10$, $0.19\pm0.06$ and $0.20\pm0.06$ pc for the clouds in the Perseus arm, the Local arm and the entire sample, respectively. They are close to the values in the case of $E_{\mathrm{flow}}/E_{\mathrm{turb}}$ (i.e., CRR$_\mathrm{TE}$).

The ratio $M_{\mathrm{esc}}/M_{\mathrm{cloud}}$ exceeds unity in environments with specific mass-to-flux ratios \citep[e.g., see figure 8 in][]{OC2017}. In addition, magnetic fields might provide more effective coupling between outflows and molecular clouds \citep[][and references therein]{DR2005, FRC2014}. Constraining the properties of the magnetic fields in these regions is therefore of great interest in the future.

\subsubsection{Ratio of $M_{\mathrm{esc}}/M_{\mathrm{flow}}$ as a Function of $R_\mathrm{cloud}$}\label{section:mmf}

The best-fitting results for the clouds in the Perseus arm, the Local arm and for the entire sample are
\begin{subequations}\label{equ:resf}
\begin{align}
  \log (M_{\mathrm{esc}}/M_{\mathrm{flow}}) = (-0.45 \pm 0.06) \log R_\mathrm{cloud} + 0.68 \pm 0.02, c.c. = -0.82, \label{equ:rftp}\\
  \log (M_{\mathrm{esc}}/M_{\mathrm{flow}}) = (-0.52 \pm 0.06) \log R_\mathrm{cloud} + 0.73 \pm 0.03, c.c. = -0.86, \label{equ:rftl}\\
  \log (M_{\mathrm{esc}}/M_{\mathrm{flow}}) = (-0.49 \pm 0.03) \log R_\mathrm{cloud} + 0.72 \pm 0.02, c.c. = -0.88, \label{equ:rftt}
\end{align}
\end{subequations}
respectively.

$M_{\mathrm{esc}}/M_{\mathrm{flow}}$ ranges from $\sim 1.6$ to $\sim34$, indicating that the outflow activities have the potential to entrain and accelerate ambient gas up to ten times their own mass. Such powerful processes could destroy the gas in the immediate vicinity of the outflow candidates.

The negative slopes in Equation (\ref{equ:resf}) indicate that outflows in smaller clouds were more effective in dispersing gas than those in larger clouds. This suggests that some physical mechanisms probably cancel out the action of entraining or accelerating the ambient gas surrounding the protostars via outflows in larger clouds.

\subsubsection{Summaries of Potential Disruptive Effect}\label{sec:sde}

Three ratios related to potential disruptive effect (i.e., $E_{\mathrm{flow}}/E_{\mathrm{grav}}$, $M_{\mathrm{esc}}/M_{\mathrm{cloud}}$, and $M_{\mathrm{esc}}/M_{\mathrm{flow}}$) as functions of $R_{\mathrm{cloud}}$ showed negative PLIs (see Table \ref{Table:CRR}). The Local arm showed the steepest slopes for the first two ratios. Similar to the explanation for the cases of turbulent support (see Section \ref{section:summary for TS}), one reason is probably that the amount of outflow activities in the Local arm is less than those in other regions. $M_{\mathrm{esc}}/M_{\mathrm{flow}}$ against $R_{\mathrm{cloud}}$  showed a similar slope for the Perseus arm, the Local arm and the entire sample. The negative value of this slope implies that some physical mechanisms may enable the ambient gas surrounding the protostars to resist dispersal by outflows in larger clouds.

CRR$_\mathrm{GE}$ and CRR$_\mathrm{MC}$ showed different values, where CRR$_\mathrm{GE}$ is $\sim 1.0$ and $\sim 0.6$ pc respectively for clouds in the Perseus arm and the Local arm, and the corresponding values of CRR$_\mathrm{MC}$ are $\sim 0.3$ and $\sim 0.2$ pc. CRR$_\mathrm{MC}$ is similar to the CRRs found for turbulent support in Section \ref{section:summary for TS} (i.e., $\sim 0.1$ -- 0.4 pc). In addition, similar to the discussion of turbulent support (see Section \ref{section:summary for TS}), the strength of the outflow activity might also influence the value of these characteristic radii above, and the spatial resolution of the telescope may have little effect on creating the differences in the CRR$_\mathrm{GE}$ and the CRR$_\mathrm{MC}$.

There were signs that $E_{\mathrm{flow}}/E_{\mathrm{grav}}$ and $M_{\mathrm{esc}}/M_{\mathrm{cloud}}$ as functions of $R_{\mathrm{cloud}}$ were approximately along a horizontal straight line for clouds with $R_{\mathrm{cloud}}\gtrsim 4.7 $ pc in the Perseus arm and the entire sample (see clouds on the right side of the vertical line in panels (a) and (b) of Figure \ref{Fig:reg}). This likely indicates that the possible scale break of potential disruptive effect of outflow feedback is $\gtrsim 4.7$ pc.

\subsection{Uncertainties of Turbulent Support and Potential Disruptive Effect}\label{sec:effect}

\subsubsection{Effects of Beam Dilution and the Uncertainty of Distance}\label{sec:effect-dis}

\begin{figure}[!htb]
%\figurenum{A. 5}
\vspace{0.2cm}
\centering
 \subfigure[$E_{\mathrm{flow}}/E_{\mathrm{turb}}$ vs. $R_\mathrm{cloud}$]{\includegraphics[width=0.328\textwidth]{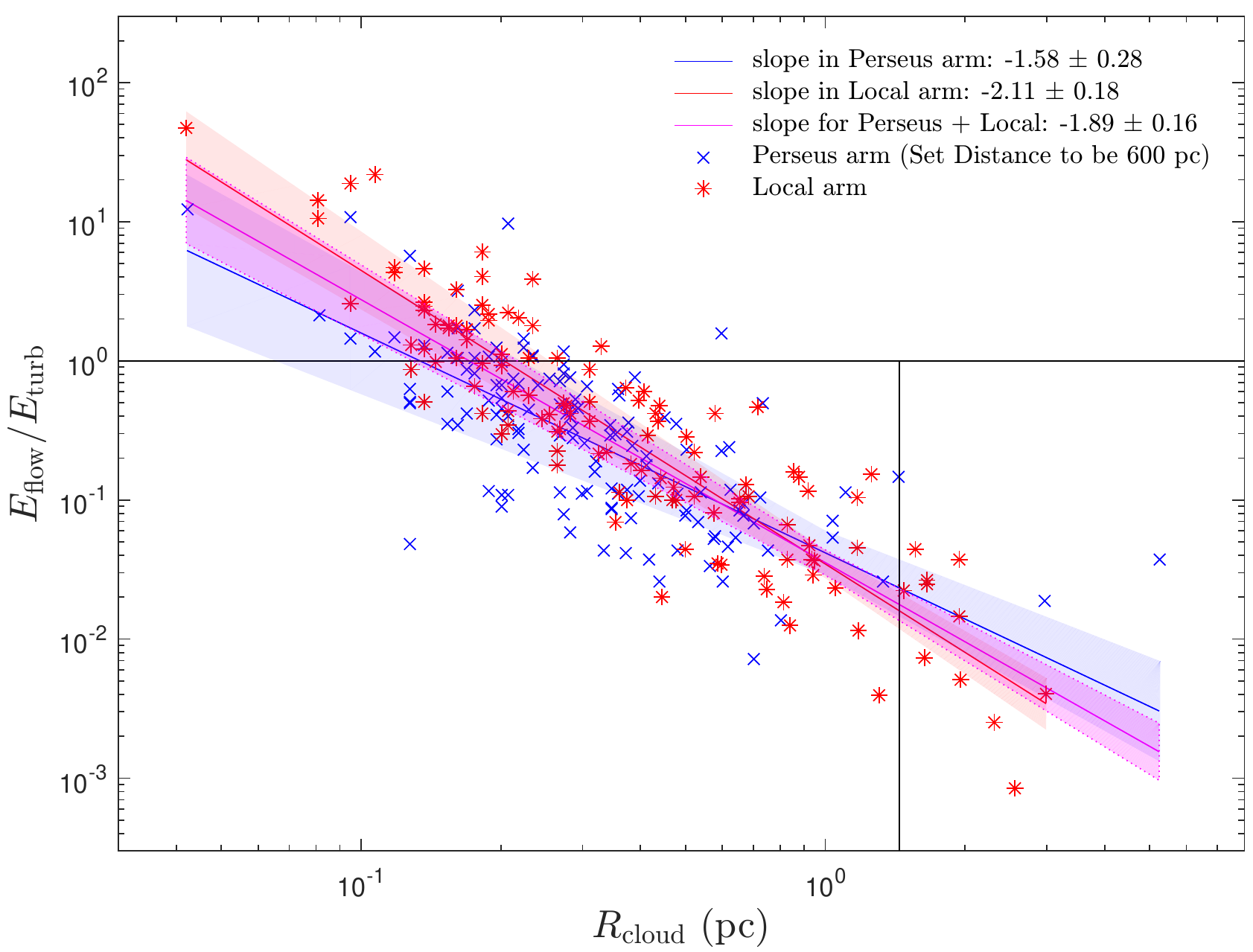}}
 \subfigure[$P_{\mathrm{flow}}/P_{\mathrm{turb}}$ vs. $R_\mathrm{cloud}$]{\includegraphics[width=0.328\textwidth]{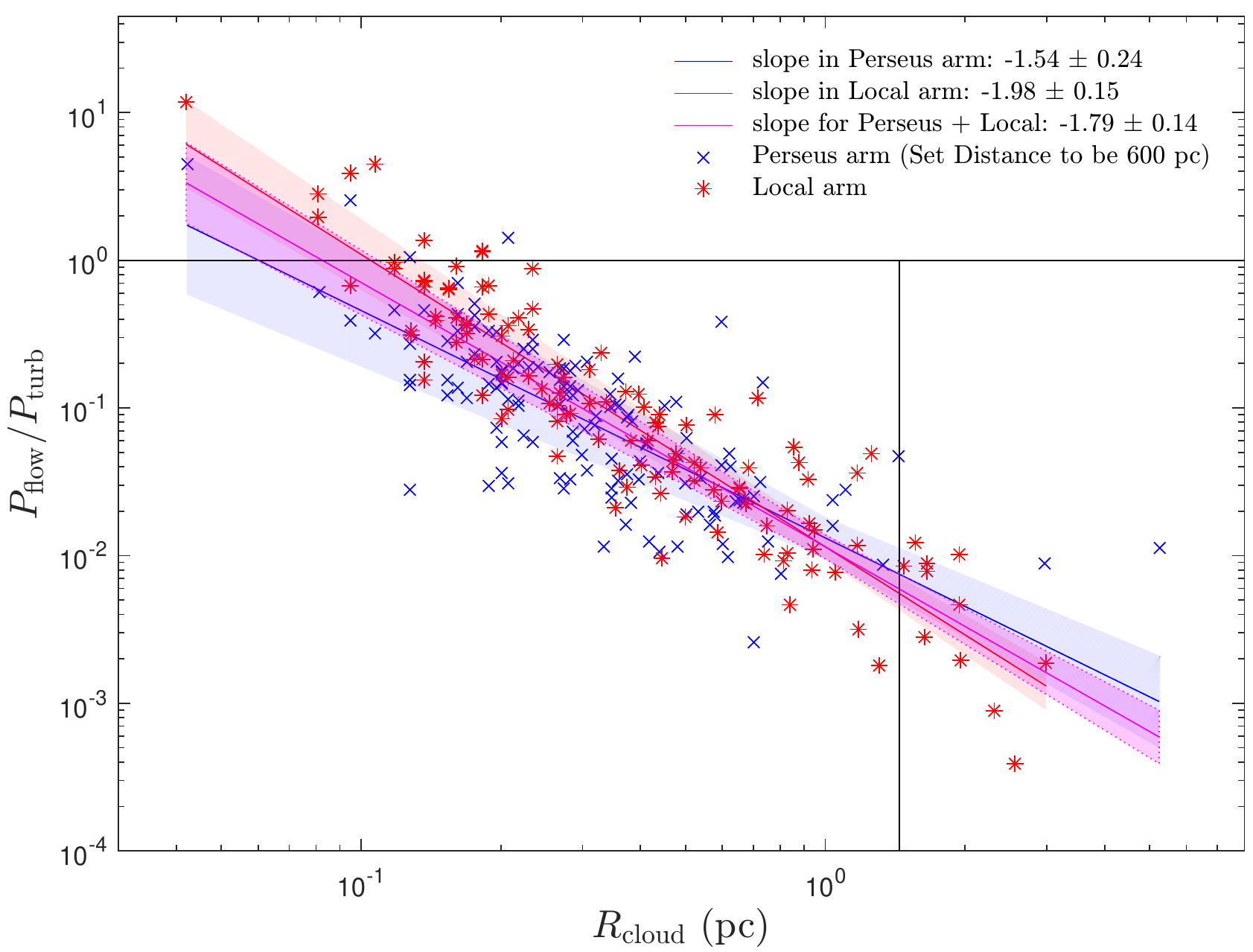}}
 \subfigure[$L_{\mathrm{flow}}/L_{\mathrm{turb}}$ vs. $R_\mathrm{cloud}$]{\includegraphics[width=0.328\textwidth]{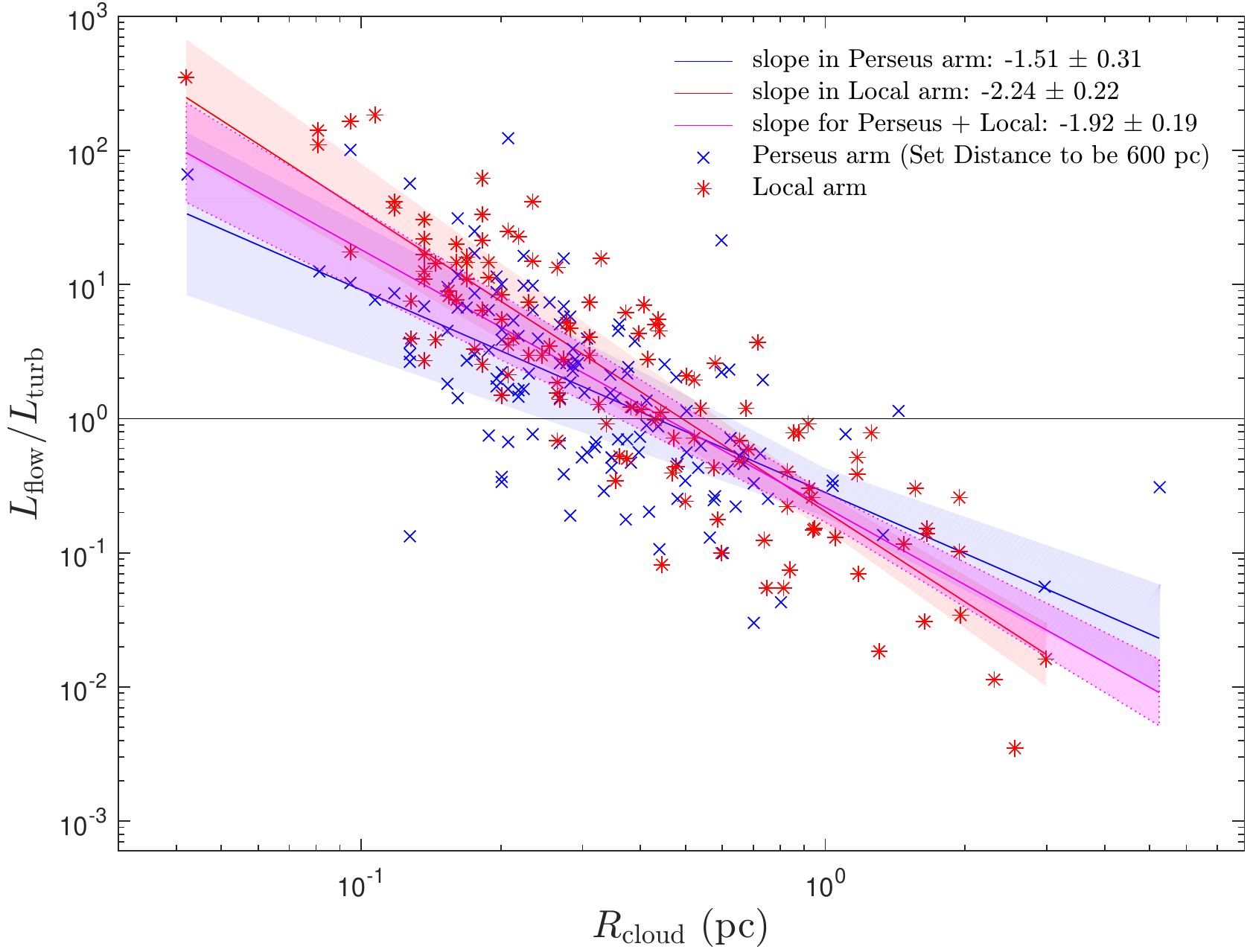}}
\subfigure[$E_{\mathrm{flow}}/E_{\mathrm{grav}}$ vs. $R_\mathrm{cloud}$]{\includegraphics[width=0.328\textwidth]{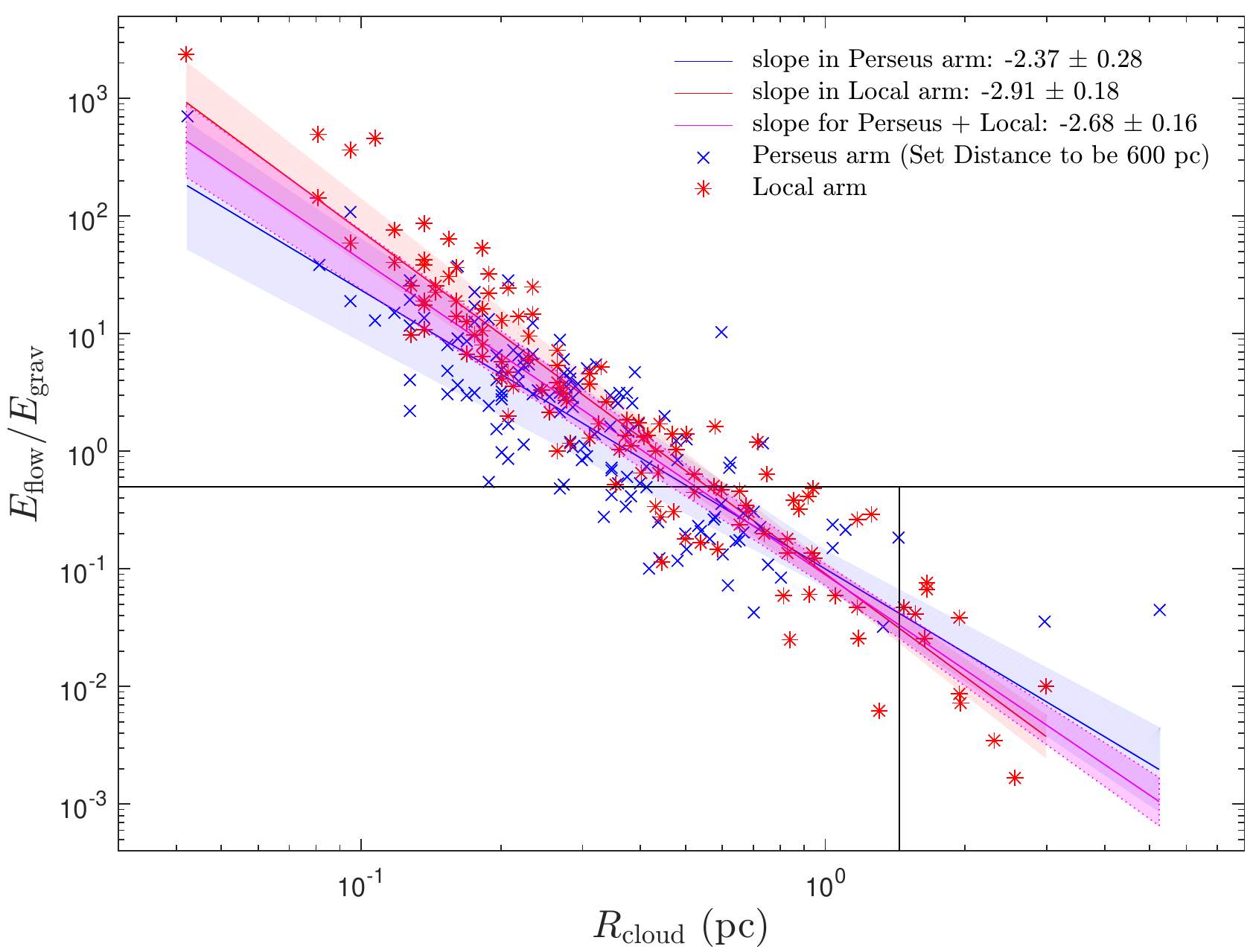}}
\subfigure[$M_{\mathrm{esc}}/M_{\mathrm{cloud}}$ vs. $R_\mathrm{cloud}$]{\includegraphics[width=0.328\textwidth]{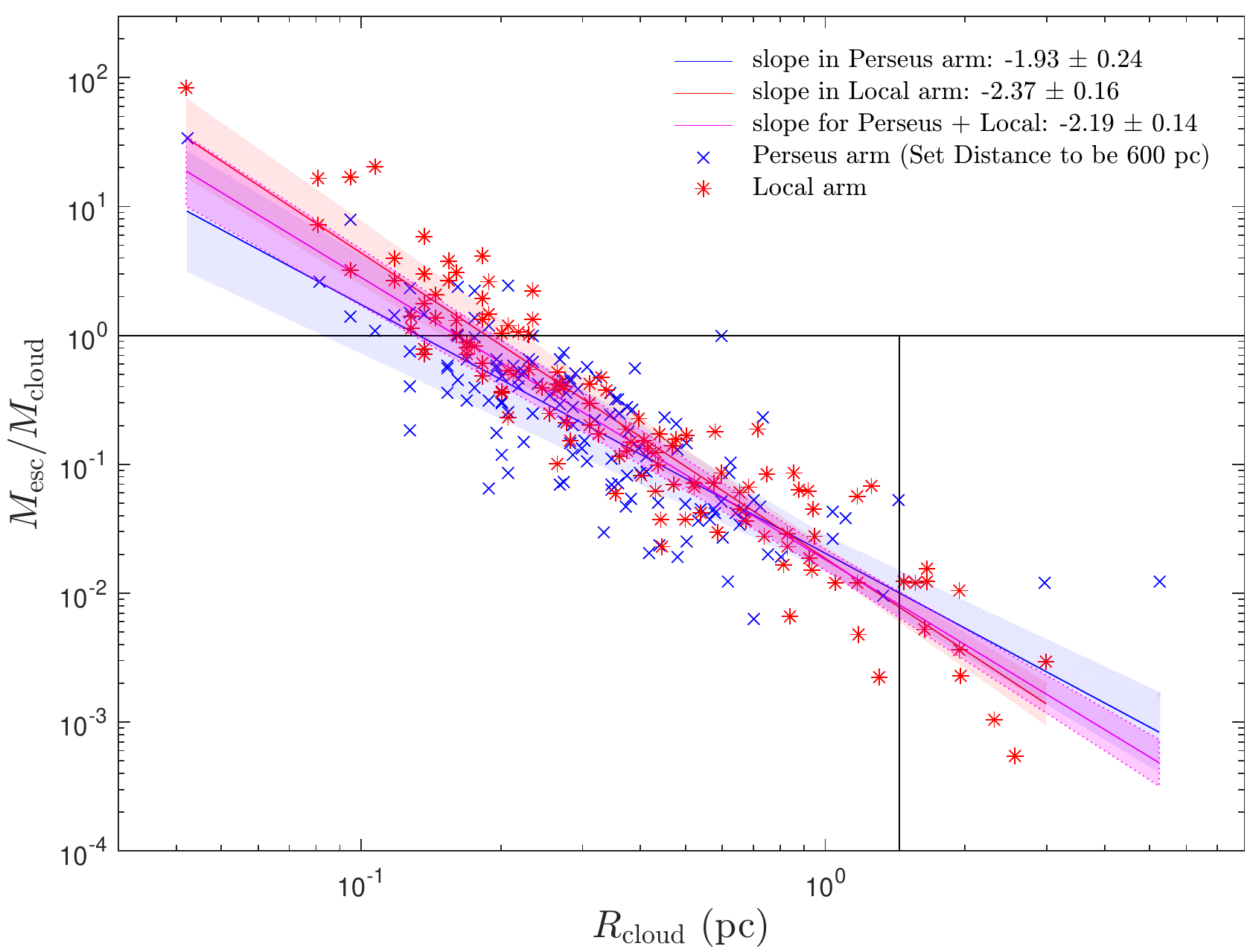}}
\subfigure[$M_{\mathrm{esc}}/M_{\mathrm{flow}}$ vs. $R_\mathrm{cloud}$]{\includegraphics[width=0.328\textwidth]{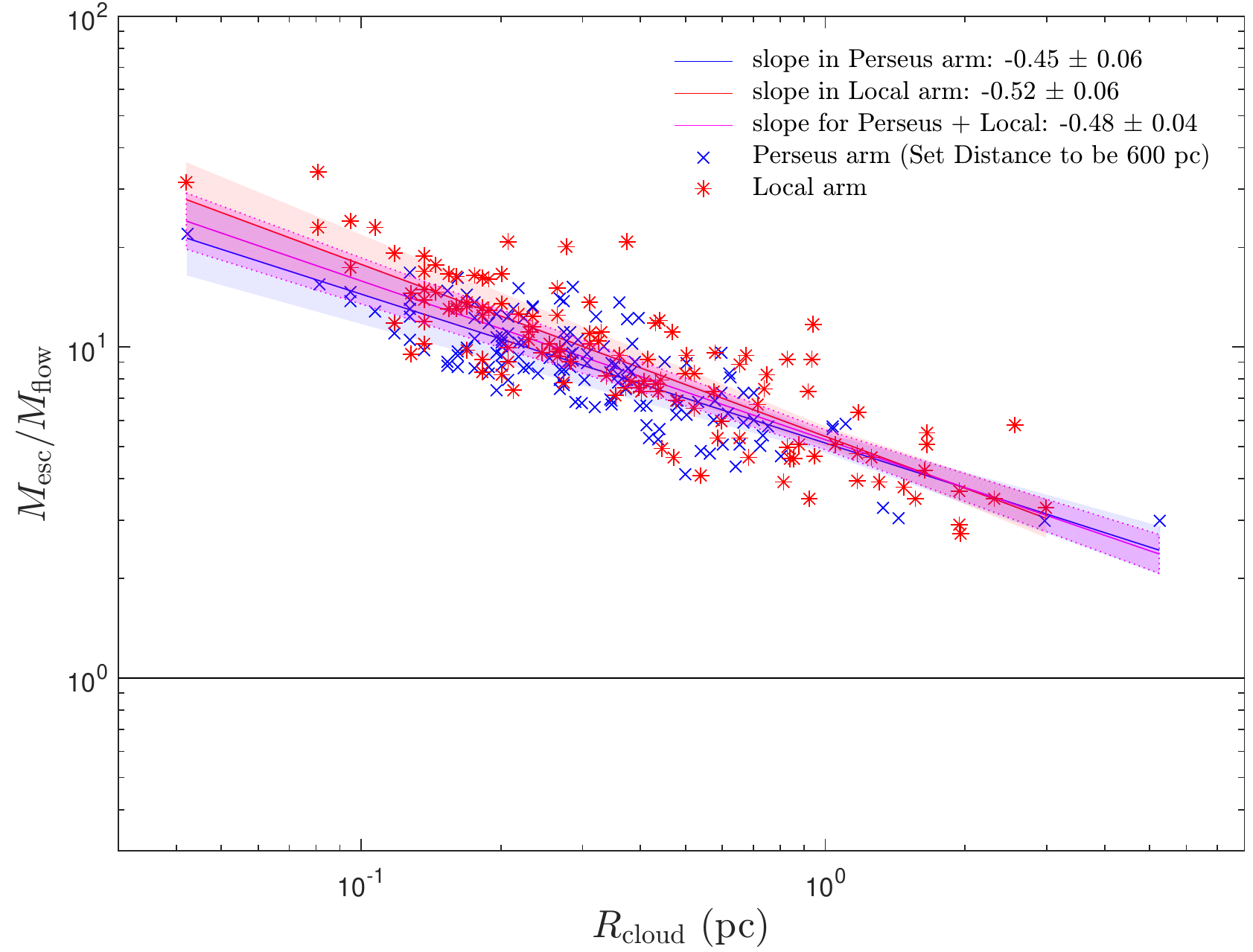}}
%\vspace{3.77cm}
\caption{(a) $E_{\mathrm{flow}}/E_{\mathrm{turb}}$, (b) $P_{\mathrm{flow}}/P_{\mathrm{turb}}$,  (c)$L_{\mathrm{flow}}/L_{\mathrm{turb}}$, (d) $E_{\mathrm{flow}}/E_{\mathrm{grav}}$, (e) $M_{\mathrm{esc}}/M_{\mathrm{cloud}}$ and (f) $M_{\mathrm{esc}}/M_{\mathrm{flow}}$ as functions of $R_\mathrm{cloud}$. To keep in line with the treatments in Sections \ref{sec:TSD} and \ref{sec:DED}, two clouds (see the blue crosses on the right side of the vertical line) are excluded when fit the relationships for panels (a), (b), (d) and (e).}
\label{Fig:dilu}
\end{figure}

To estimate the effect of beam dilution, we set the distance to the Perseus arm to be the same as that to the Local arm (i.e., 600 pc). The corresponding results are plotted in Figure \ref{Fig:dilu} and cataloged in Table \ref{Table:CRR}. The result shows that: both PLI and CRR for the sample of \textbf{Perseus + Local} after changing distance are consistent with those for the clouds in the Local arm, and are different from the sample of Perseus + Local before changing distance. This indicates that the effect of beam dilution is probably coupled with the effect of distance. 
%In addition, the CRR for the clouds in the Perseus arm is different before and after changing the distance of the Perseus arm.

\begin{figure}[!ht]
 %\figurenum{1-cloud 1}
 \centering
 \vspace{-0.55cm}
 \subfigure[$E_{\mathrm{flow}}/E_{\mathrm{turb}}$ vs. $R_\mathrm{cloud}$]{\includegraphics[width=0.3\textwidth,trim={50 10 180 78},clip]{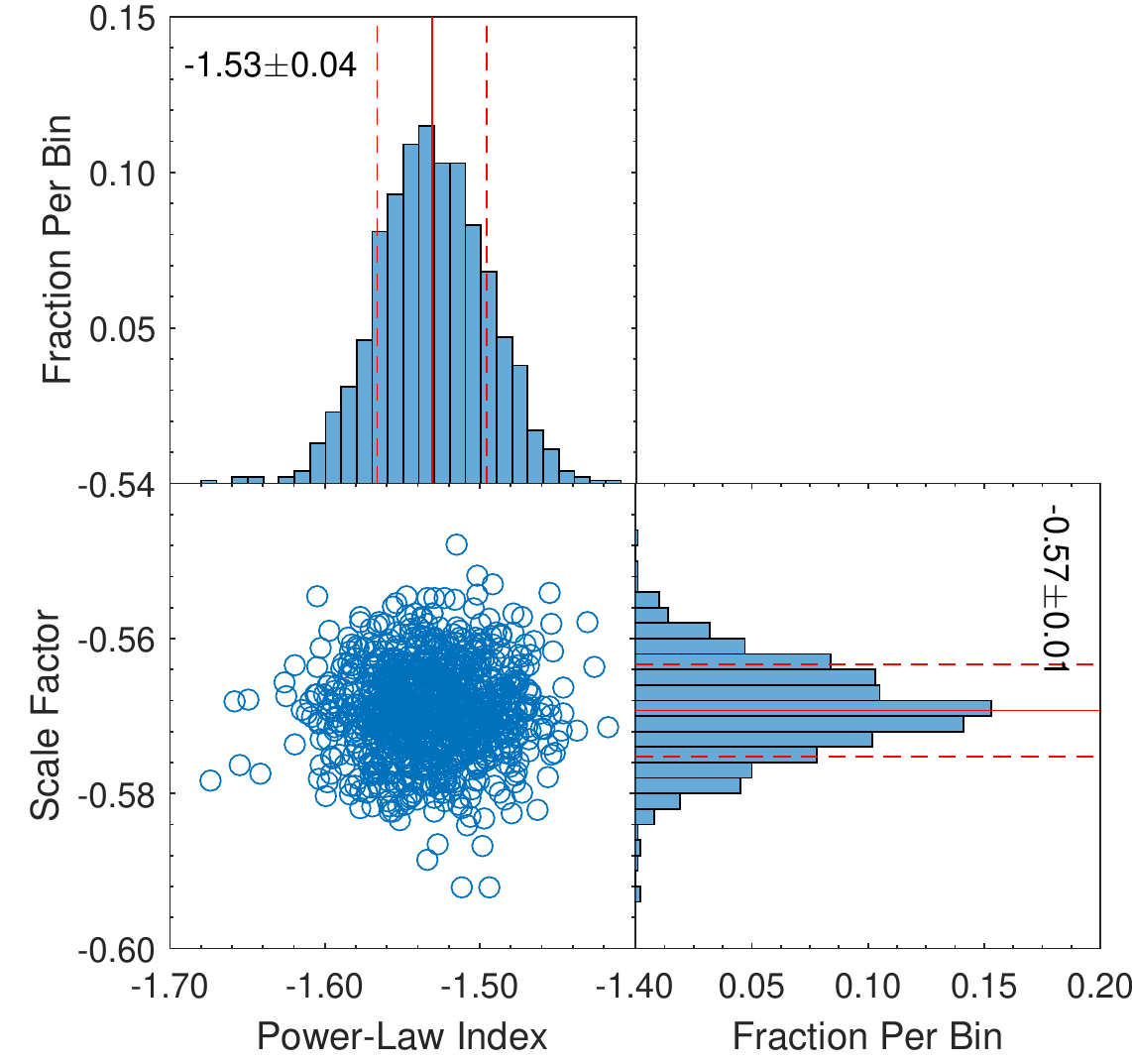}}
 \subfigure[$P_{\mathrm{flow}}/P_{\mathrm{turb}}$ vs. $R_\mathrm{cloud}$]{\includegraphics[width=0.3\textwidth,trim={50 10 180 78},clip]{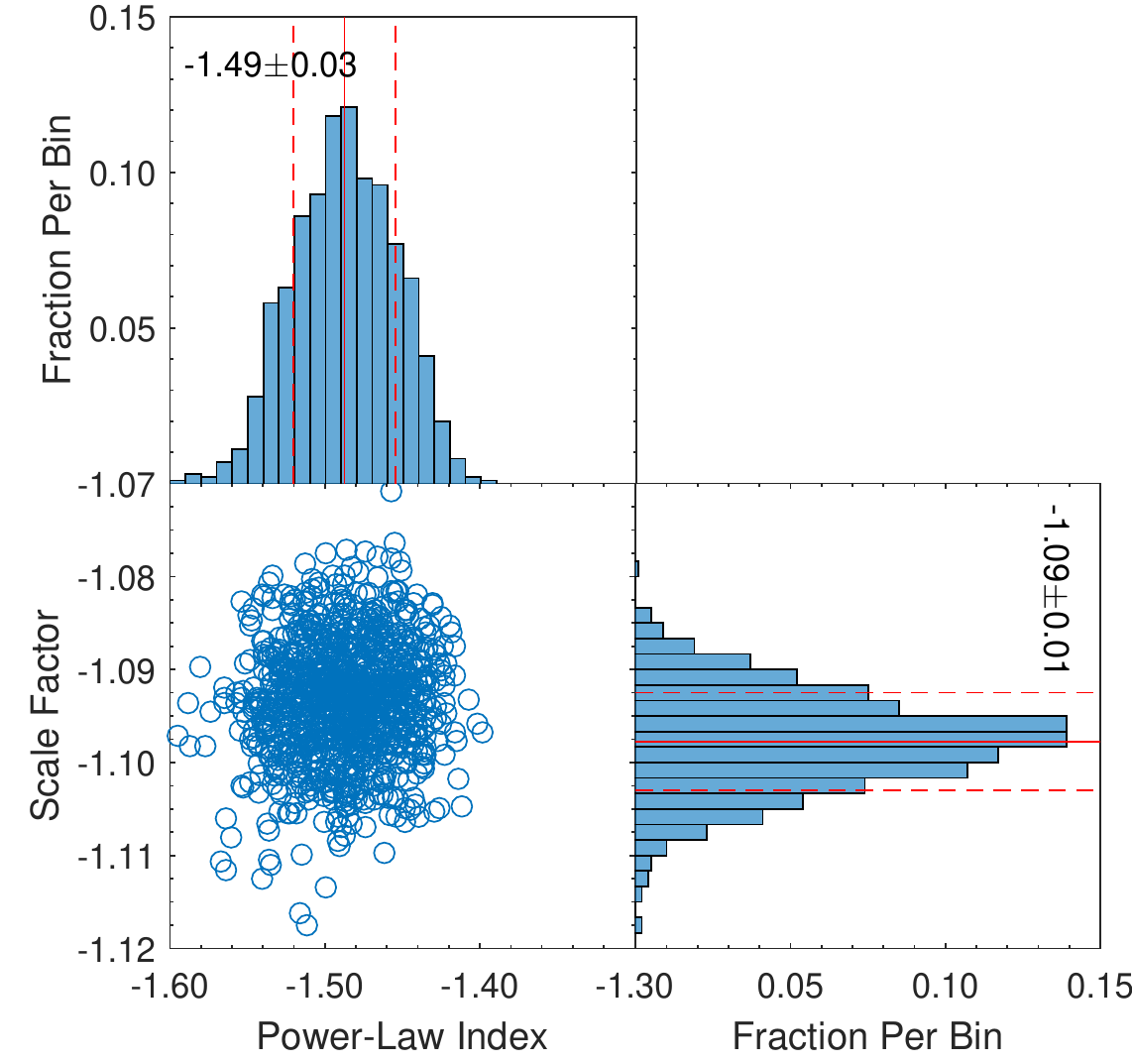}}
  \subfigure[$L_{\mathrm{flow}}/L_{\mathrm{turb}}$ vs. $R_\mathrm{cloud}$]{\includegraphics[width=0.3\textwidth,trim={50 10 180 78},clip]{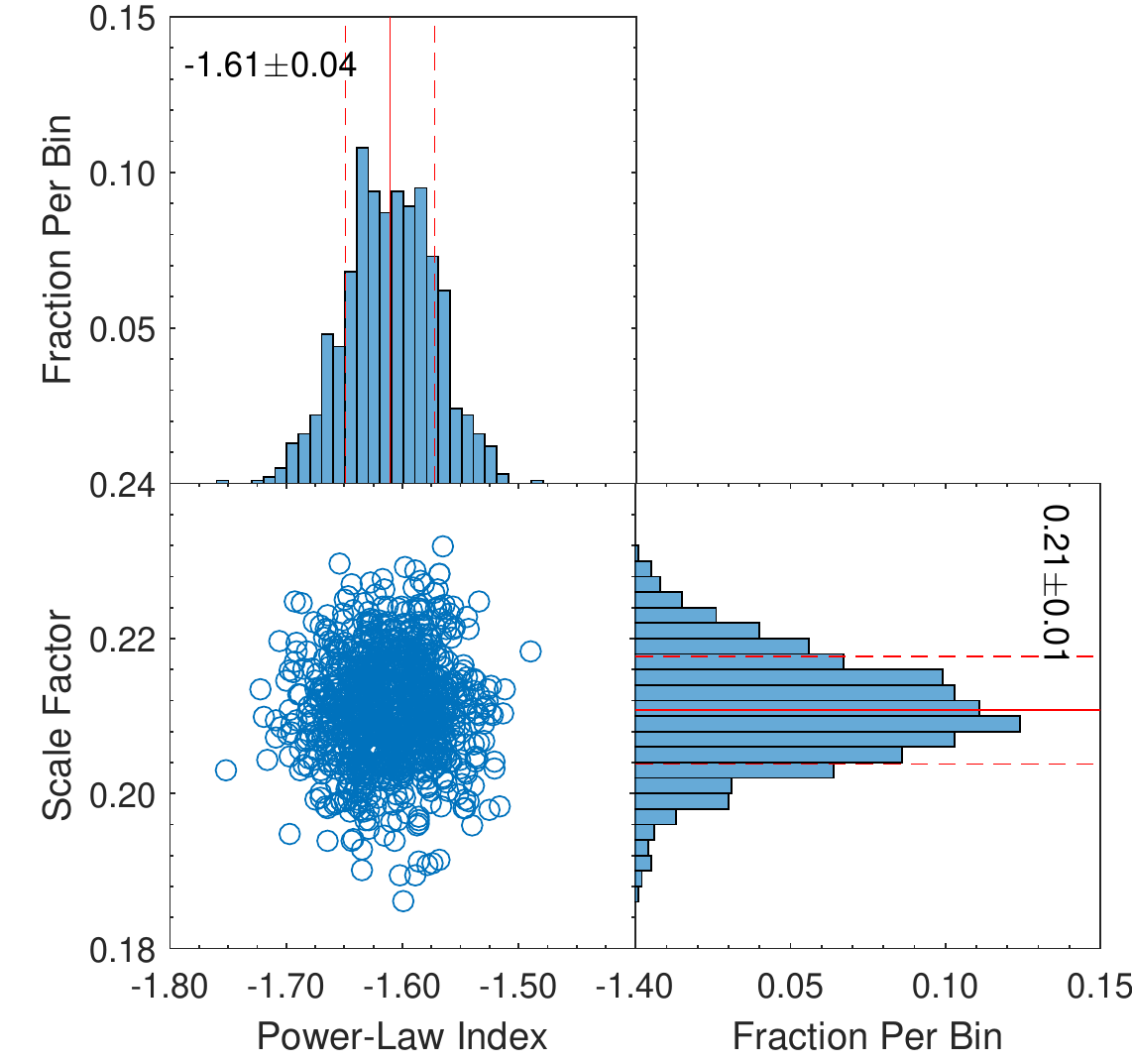}}
   \subfigure[$E_{\mathrm{flow}}/E_{\mathrm{grav}}$ vs. $R_\mathrm{cloud}$]{\includegraphics[width=0.3\textwidth,trim={50 10 180 78},clip]{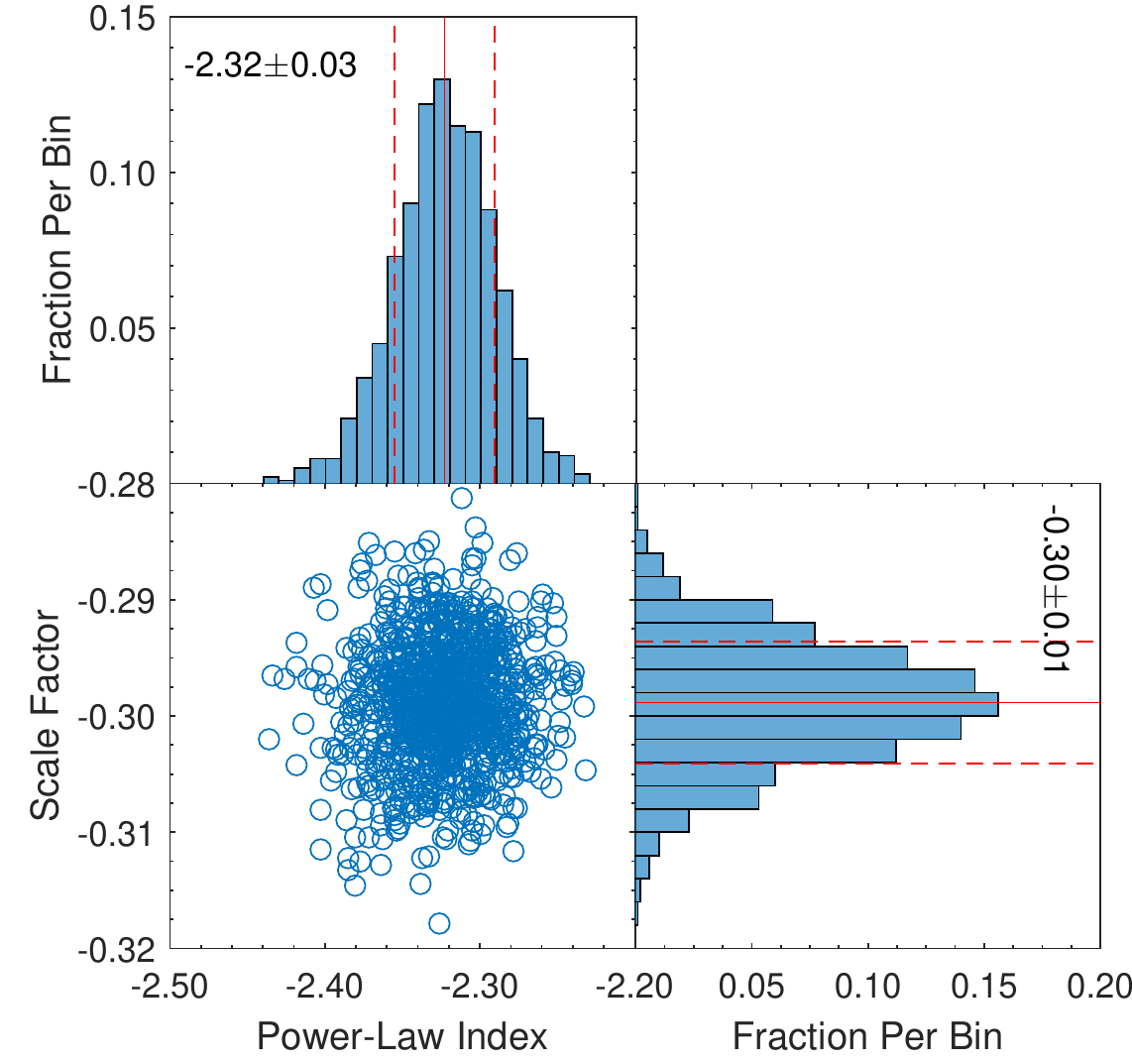}}
 \subfigure[$M_{\mathrm{esc}}/M_{\mathrm{cloud}}$ vs. $R_\mathrm{cloud}$]{\includegraphics[width=0.3\textwidth,trim={50 10 180 78},clip]{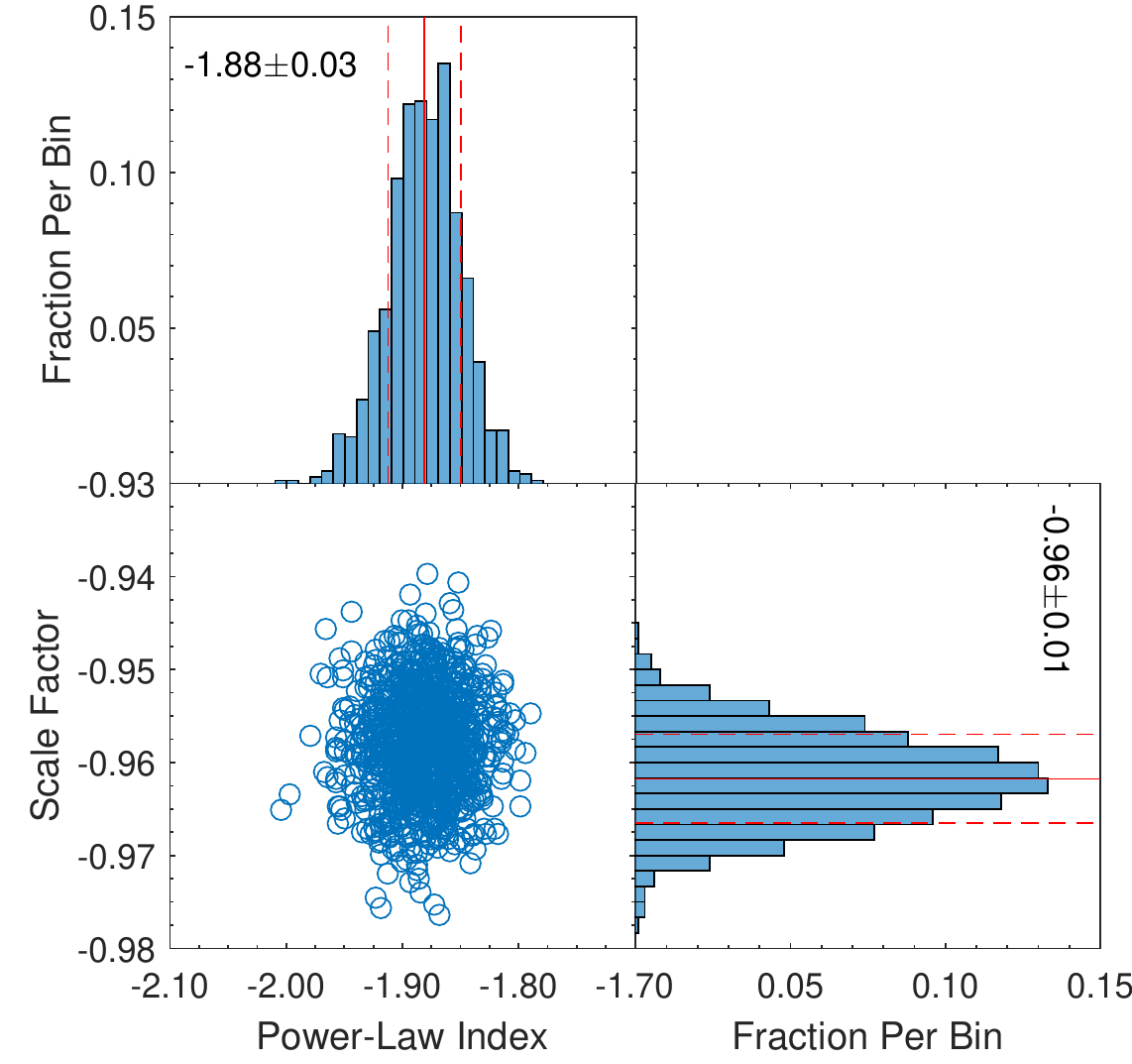}}
  \subfigure[$M_{\mathrm{esc}}/M_{\mathrm{flow}}$ vs. $R_\mathrm{cloud}$]{\includegraphics[width=0.3\textwidth,trim={50 10 180 78},clip]{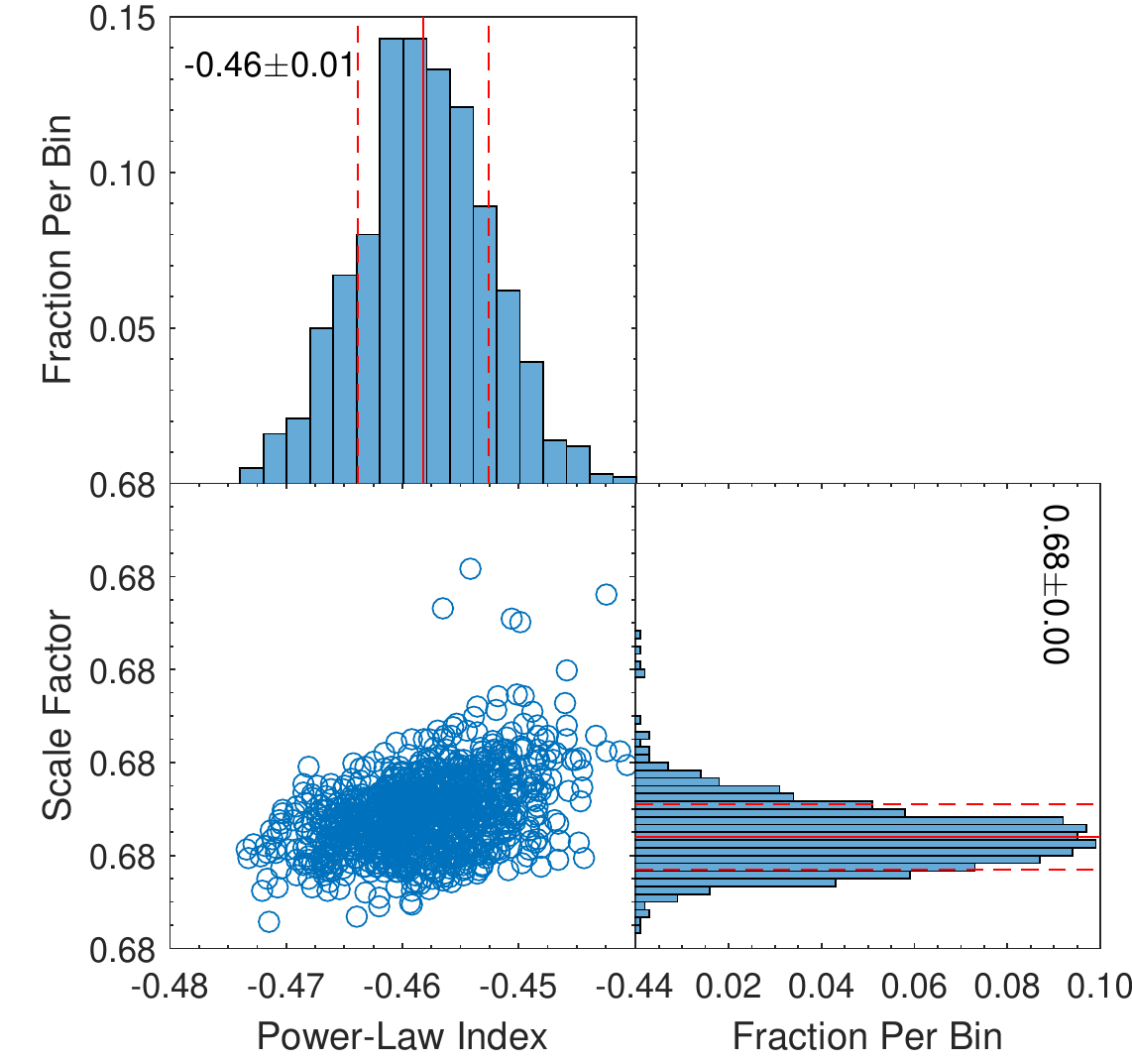}}
   \subfigure[$E_{\mathrm{flow}}/E_{\mathrm{turb}}$ vs. $R_\mathrm{cloud}$]{\includegraphics[width=0.3\textwidth,trim={50 10 180 78},clip]{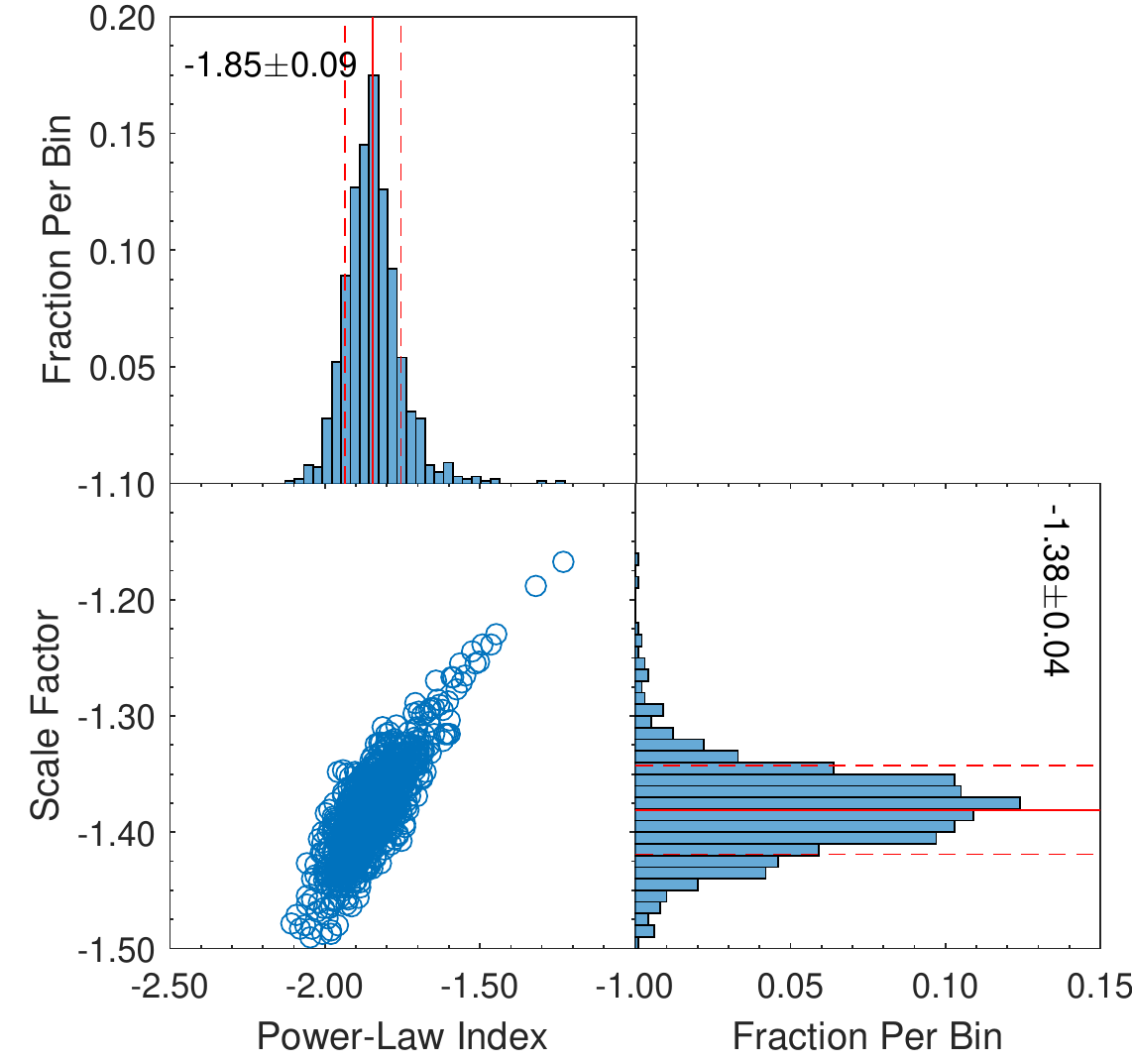}}
 \subfigure[$P_{\mathrm{flow}}/P_{\mathrm{turb}}$ vs. $R_\mathrm{cloud}$]{\includegraphics[width=0.3\textwidth,trim={50 10 180 78},clip]{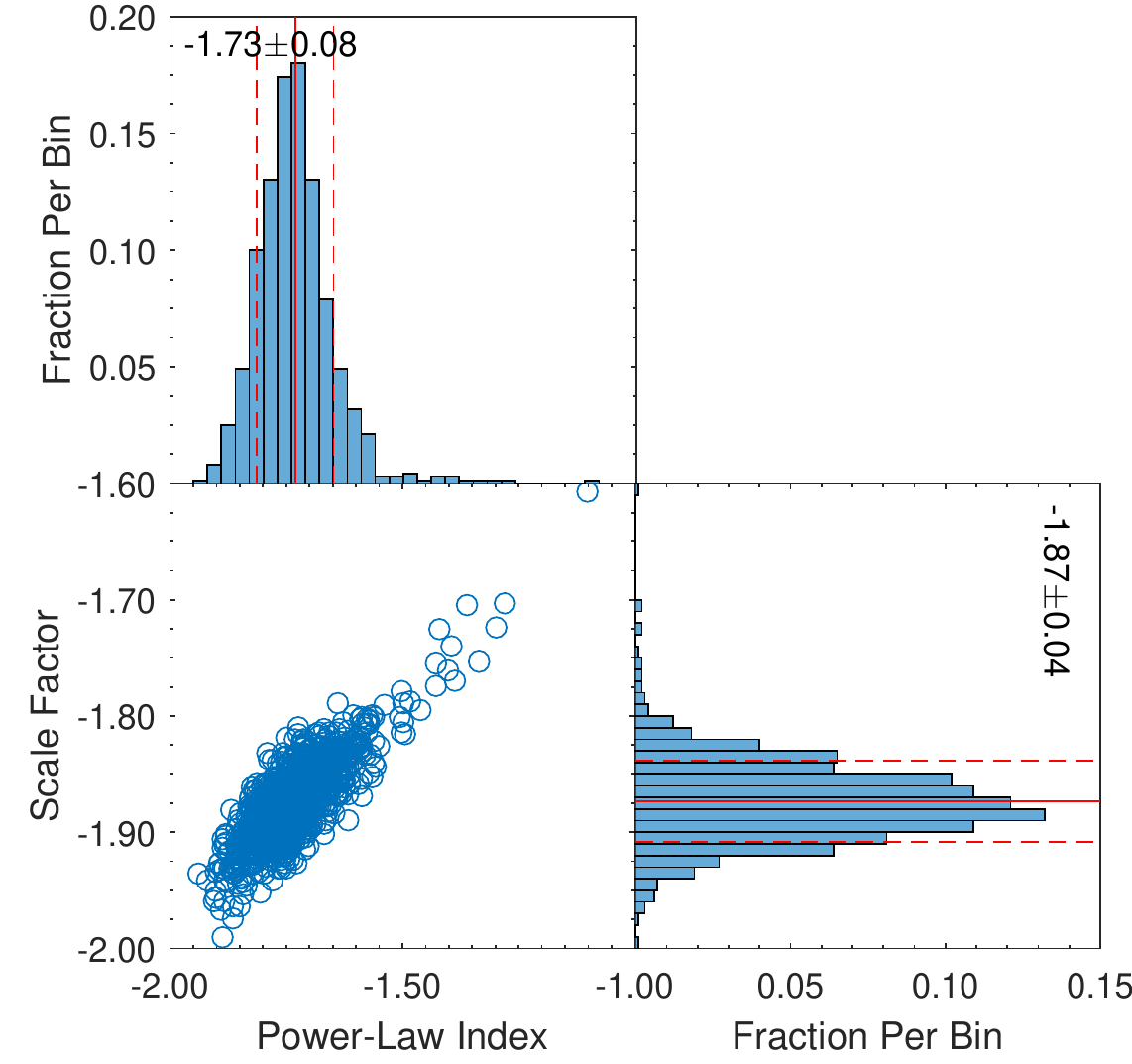}}
  \subfigure[$L_{\mathrm{flow}}/L_{\mathrm{turb}}$ vs. $R_\mathrm{cloud}$]{\includegraphics[width=0.3\textwidth,trim={50 10 180 78},clip]{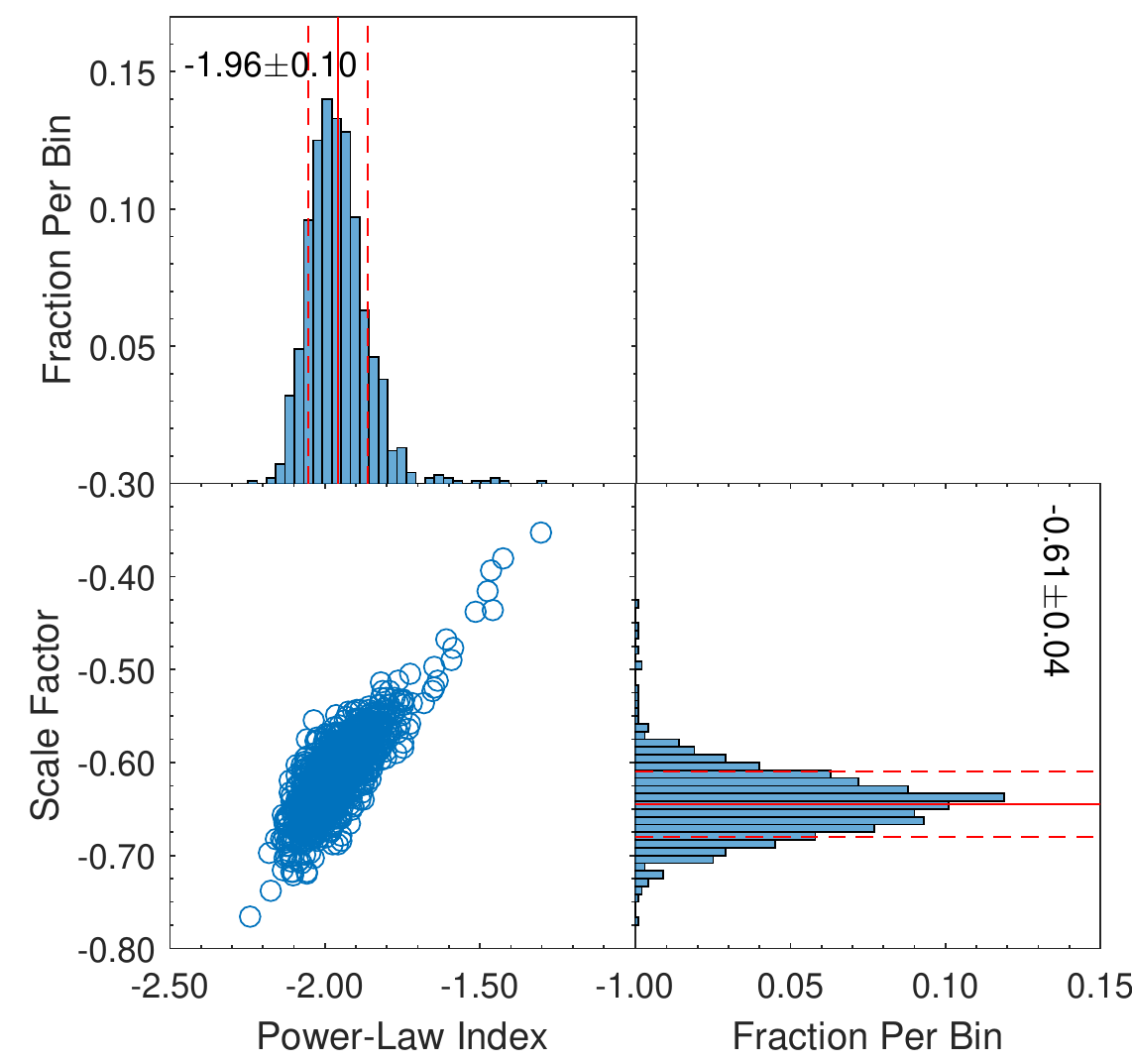}}
   \subfigure[$E_{\mathrm{flow}}/E_{\mathrm{grav}}$ vs. $R_\mathrm{cloud}$]{\includegraphics[width=0.3\textwidth,trim={50 10 180 78},clip]{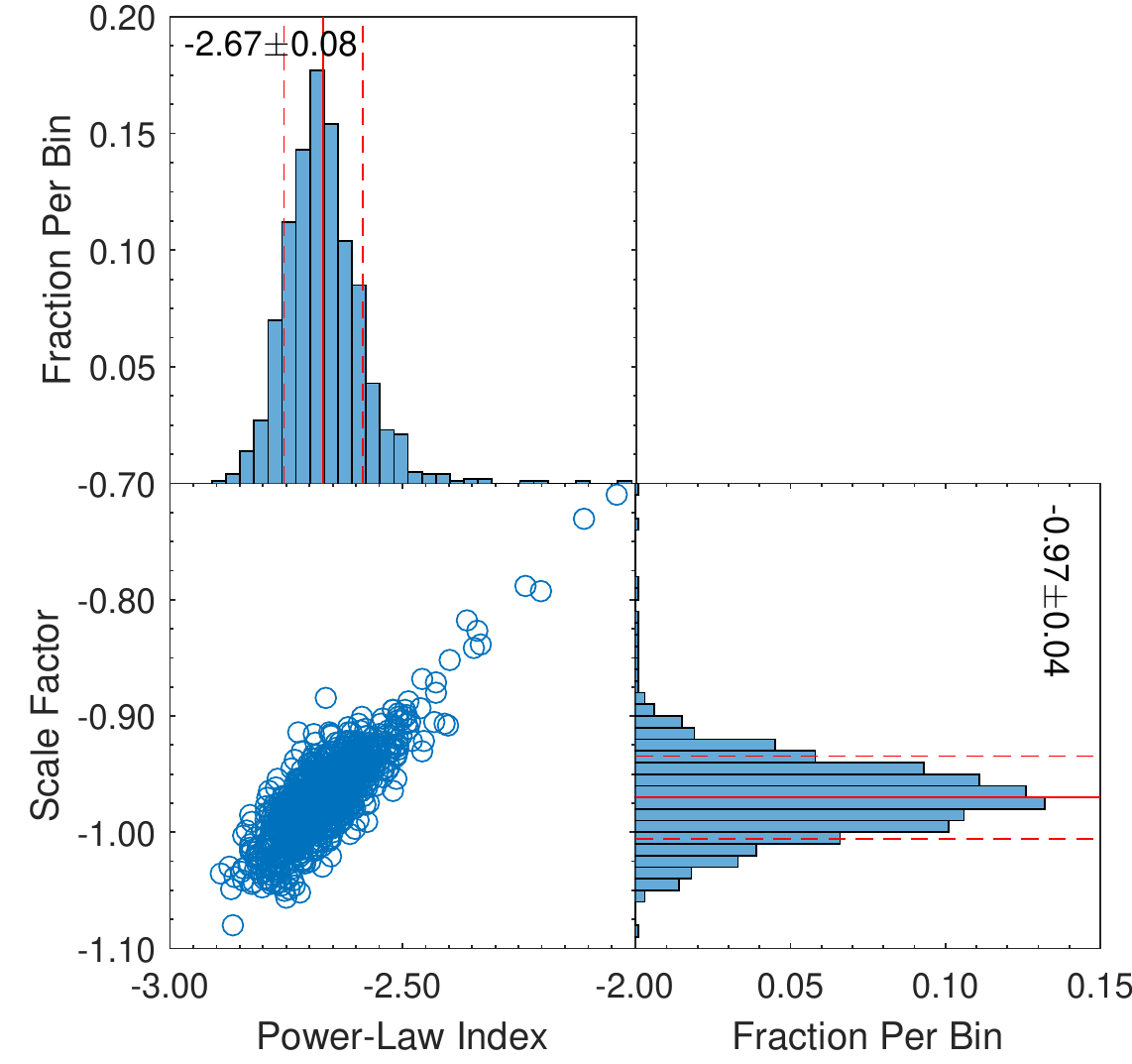}}
 \subfigure[$M_{\mathrm{esc}}/M_{\mathrm{cloud}}$ vs. $R_\mathrm{cloud}$]{\includegraphics[width=0.3\textwidth,trim={50 10 180 78},clip]{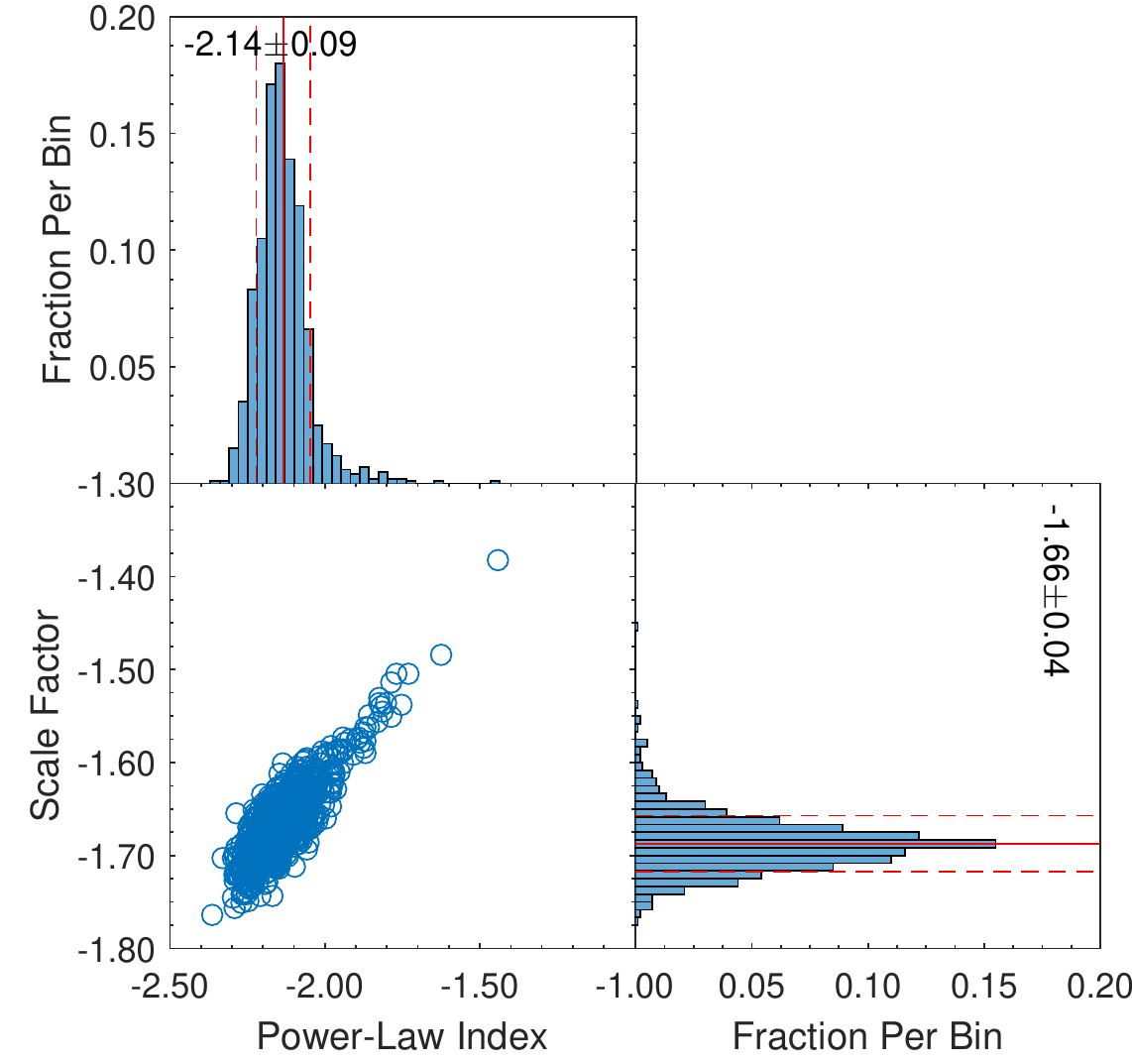}}
  \subfigure[$M_{\mathrm{esc}}/M_{\mathrm{flow}}$ vs. $R_\mathrm{cloud}$]{\includegraphics[width=0.3\textwidth,trim={50 10 180 78},clip]{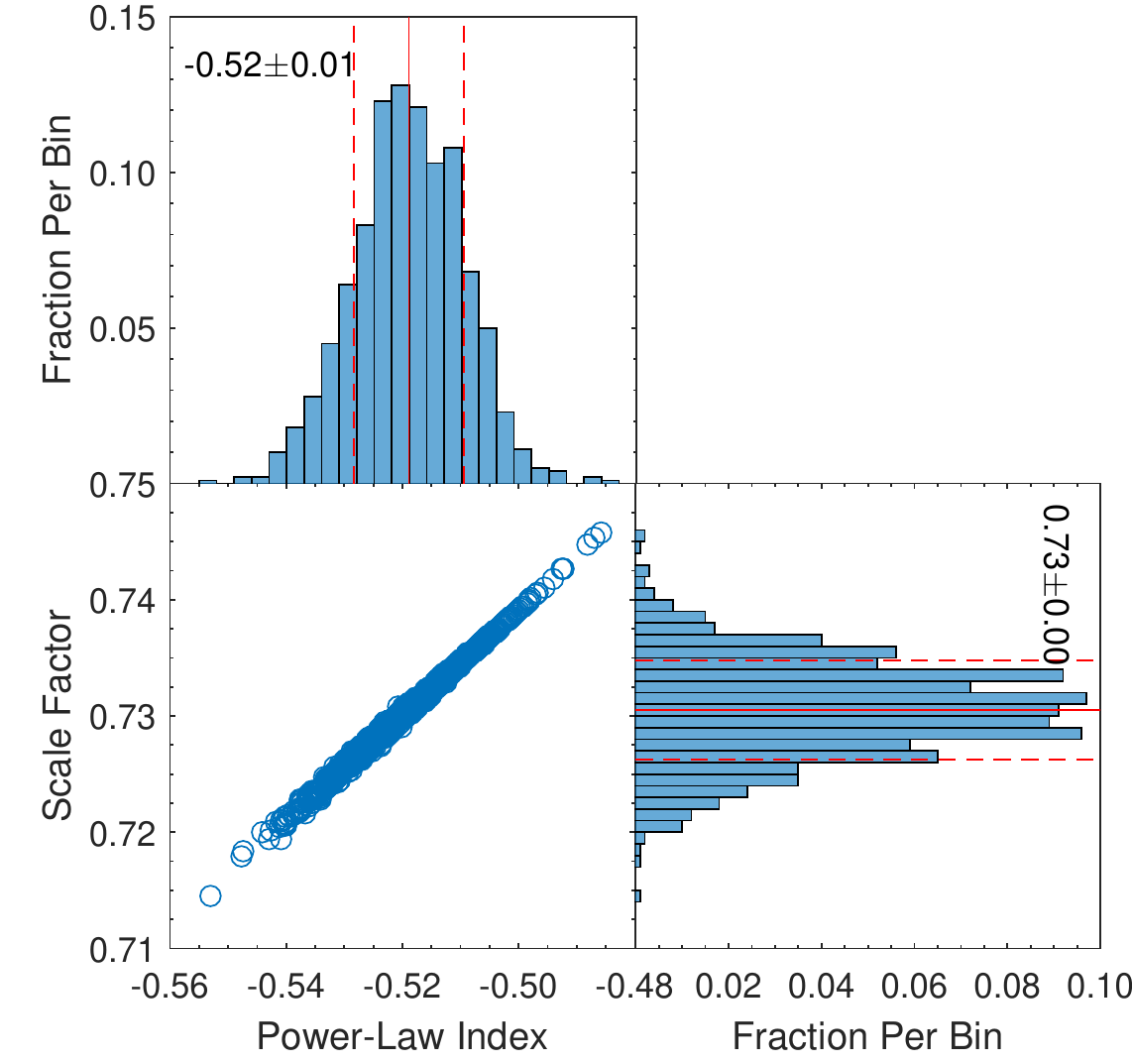}}
  \vspace{-0.1cm}
\caption{The effect of the uncertainty of distance: the Perseus arm (a -- f); the Local arm (g -- l).}
\label{fig:distance}
\end{figure}

To test the effect of the uncertainty of distance, we have constructed simulated cloud samples, \footnote{Similar to the treatment showed in Figure \ref{Fig:dilu}, the number of clouds is 134 for the cases in $E_{\mathrm{flow}}/E_{\mathrm{turb}}$, $P_{\mathrm{flow}}/P_{\mathrm{turb}}$, $E_{\mathrm{flow}}/E_{\mathrm{grav}}$ and $M_{\mathrm{esc}}/M_{\mathrm{cloud}}$, and 136 for the cases in $L_{\mathrm{flow}}/L_{\mathrm{turb}}$ and $M_{\mathrm{esc}}/M_{\mathrm{flow}}$.\label{footnote}}where the distance is randomly generated according to the probability density function that obeys a normal distribution. The parameters of the generated distance are: 1) for the Perseus arm, the expected value is 1950 pc, the standard deviation is $\sim$ 213 pc, and the range of values is (1450, 2450) pc; 2) for the Local arm, the corresponding values are 600 pc, $\sim$ 170 pc, and (200, 1000) pc, respectively \citep[see][]{SYX2020}. The results after one thousand tests are plotted in Figure \ref{fig:distance} and catalogued in Table \ref{Table:CRR}. The PLIs and CRRs are roughly consistent with the results of Sections \ref{sec:TSD} and \ref{sec:DED}, except in the case of $L_{\mathrm{flow}}/L_{\mathrm{turb}}$ where $L_{\mathrm{flow}}/L_{\mathrm{turb}}$ and CRR$_\mathrm{TL}$ are highly uncertain as stated in Sections \ref{section:TSR} and \ref{sec:rlr}. This indicates that the uncertainties of distance and beam dilution which are probably coupled together (see discussion above), have little effect on the conclusions of Sections \ref{section:summary for TS} and \ref{sec:sde}.

\subsubsection{Effect of Multi-component} \label{sec:emc}

To test the effect of multi-component, we have artificially merged some clouds to be one cloud. Similar to the proportion of the clouds with multi-velocity components, the proportion of cloud with multi-components (i.e., the cloud which merges two or more old clouds) are set to be $\sim 19\%$ and $\sim 17\%$, resulting in the total numbers of clouds in the new sample being 119 and 111 respectively for clouds in the Perseus arm and the Local arm. For clouds in the Perseus arm, a set of random numbers with the sum of 134 or 136 (see Figure \ref{Fig:dilu} or Footnote \ref{footnote}) and the number of elements of 119 are created to determine which clouds should be merged to be one. For clouds in the Local arm, the corresponding value of the sum and the number of elements are 124 and 111, respectively. The radius of the new cloud is set to be the maximum value of the following terms: the sum of the radius of old clouds multiply a random number ranging from zero to one (determine the overlaps of the old clouds); and the maximum radius of old clouds. Other physical parameters are revised accordingly. 

\begin{figure}[!ht]
 %\figurenum{1-cloud 1}
 \centering
 \vspace{-0.5cm}
 \subfigure[$E_{\mathrm{flow}}/E_{\mathrm{turb}}$ vs. $R_\mathrm{cloud}$]{\includegraphics[width=0.3\textwidth,trim={50 10 180 78},clip]{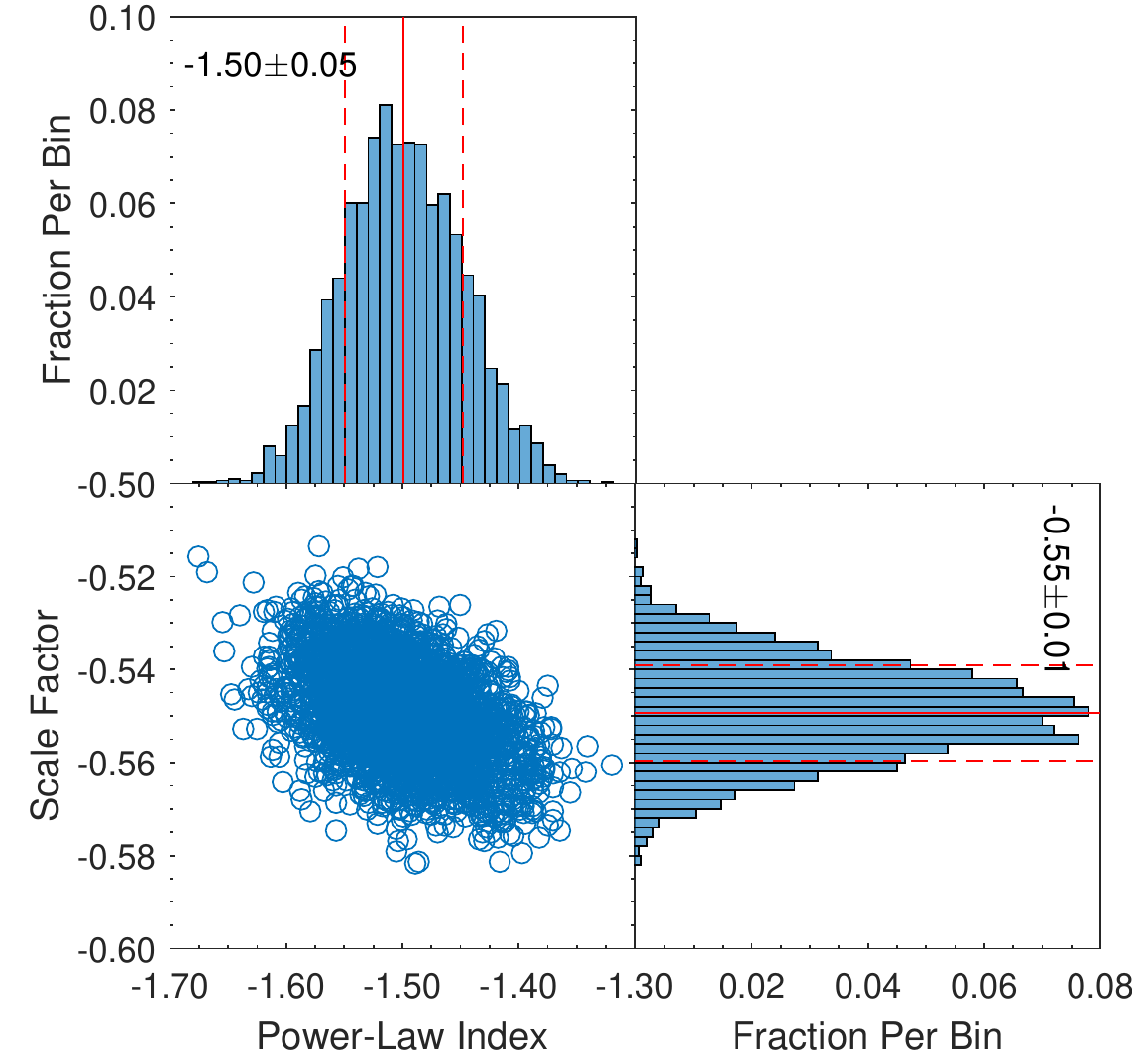}}
 \subfigure[$P_{\mathrm{flow}}/P_{\mathrm{turb}}$ vs. $R_\mathrm{cloud}$]{\includegraphics[width=0.3\textwidth,trim={50 10 180 78},clip]{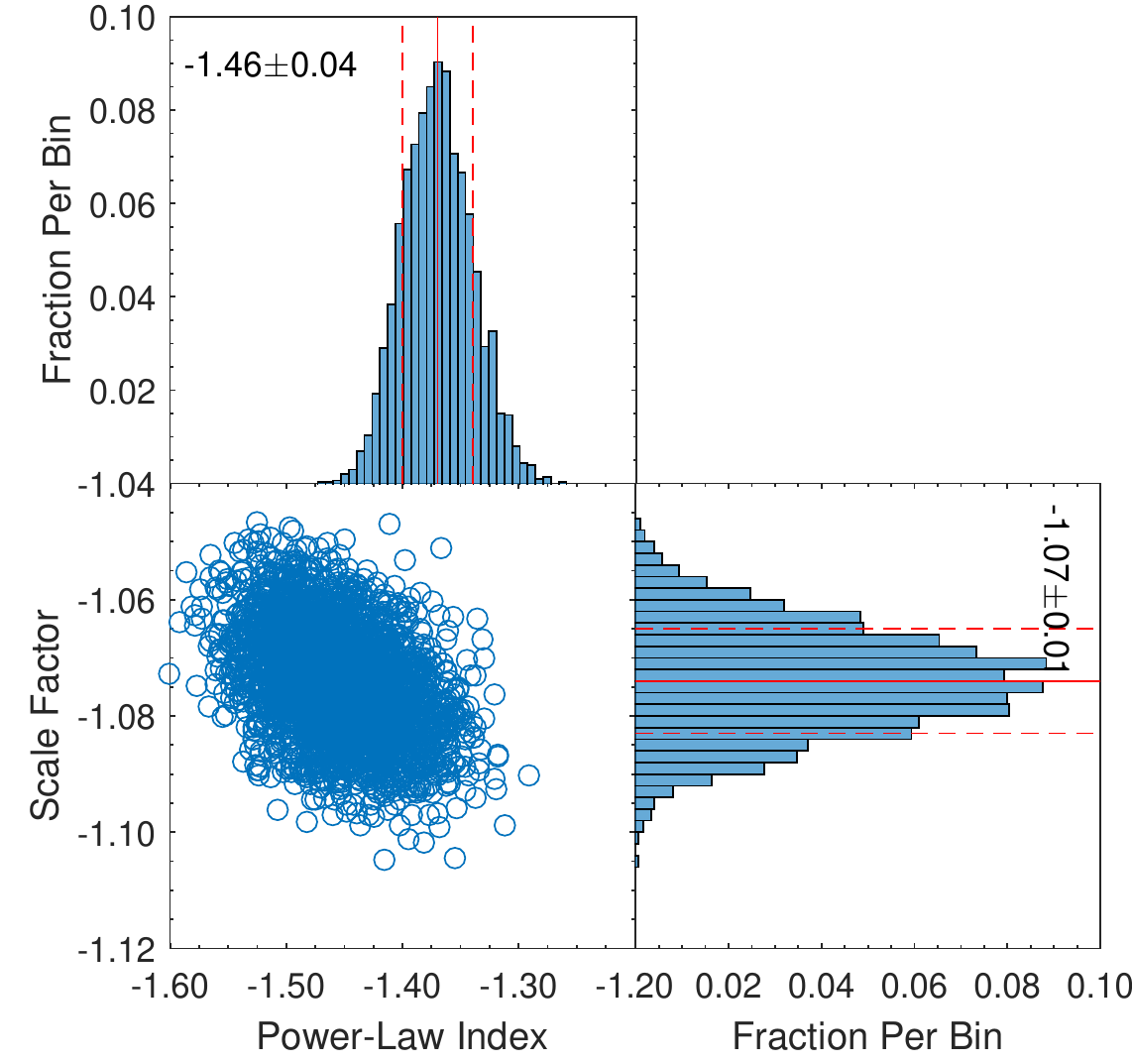}}
  \subfigure[$L_{\mathrm{flow}}/L_{\mathrm{turb}}$ vs. $R_\mathrm{cloud}$]{\includegraphics[width=0.3\textwidth,trim={50 10 180 78},clip]{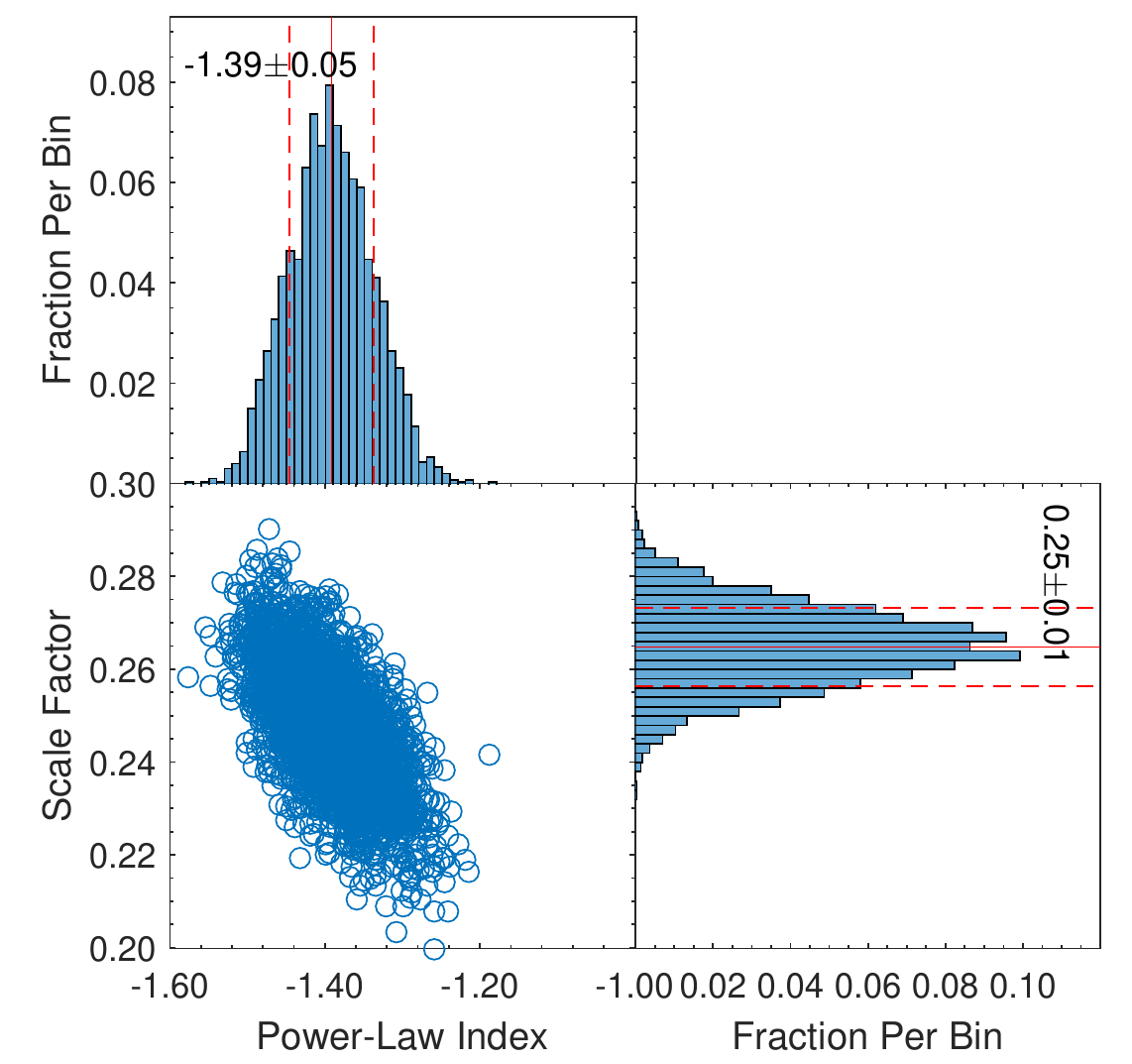}}
   \subfigure[$E_{\mathrm{flow}}/E_{\mathrm{grav}}$ vs. $R_\mathrm{cloud}$]{\includegraphics[width=0.3\textwidth,trim={50 10 180 78},clip]{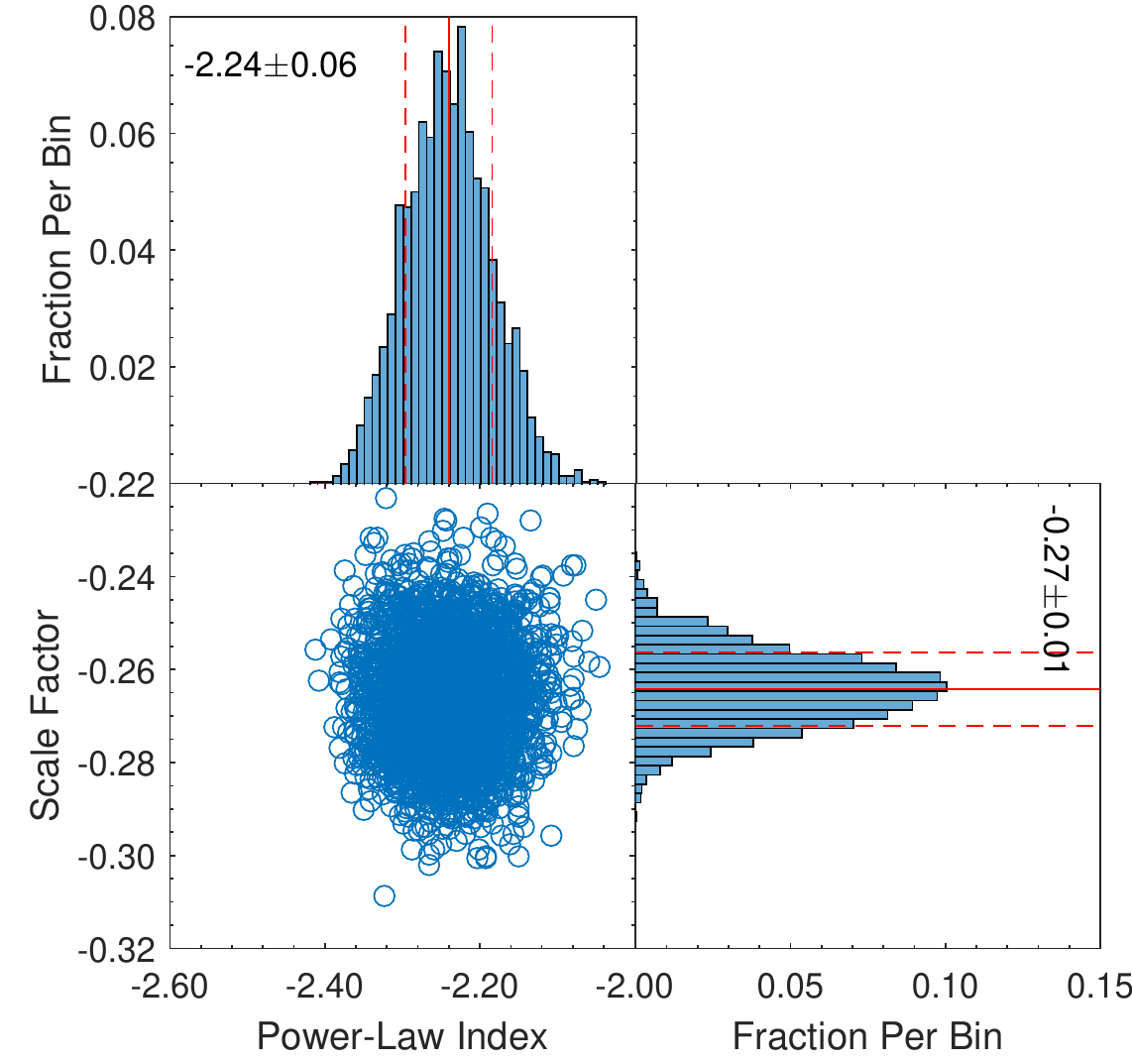}}
 \subfigure[$M_{\mathrm{esc}}/M_{\mathrm{cloud}}$ vs. $R_\mathrm{cloud}$]{\includegraphics[width=0.3\textwidth,trim={50 10 180 78},clip]{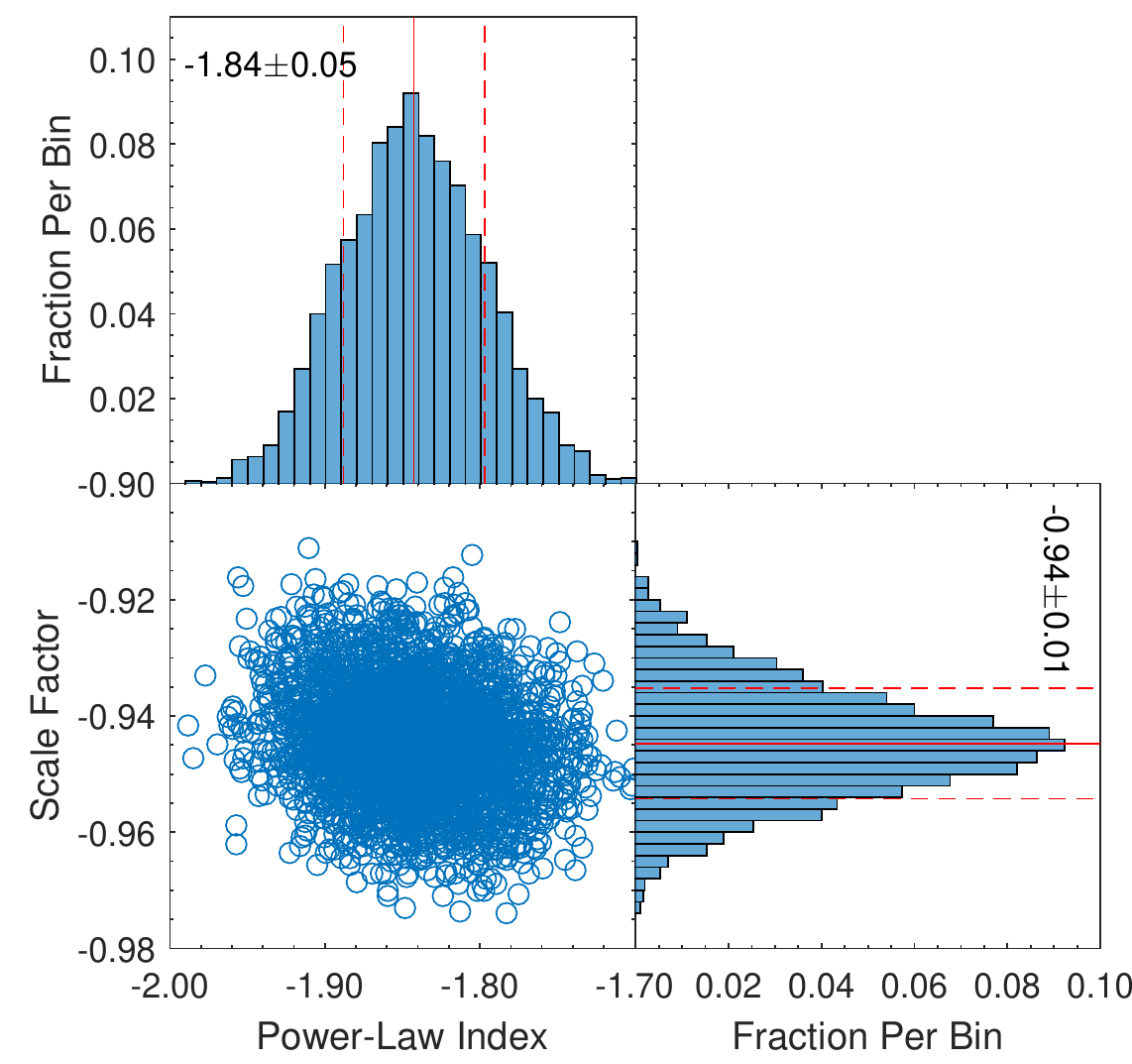}}
  \subfigure[$M_{\mathrm{esc}}/M_{\mathrm{flow}}$ vs. $R_\mathrm{cloud}$]{\includegraphics[width=0.3\textwidth,trim={50 10 180 78},clip]{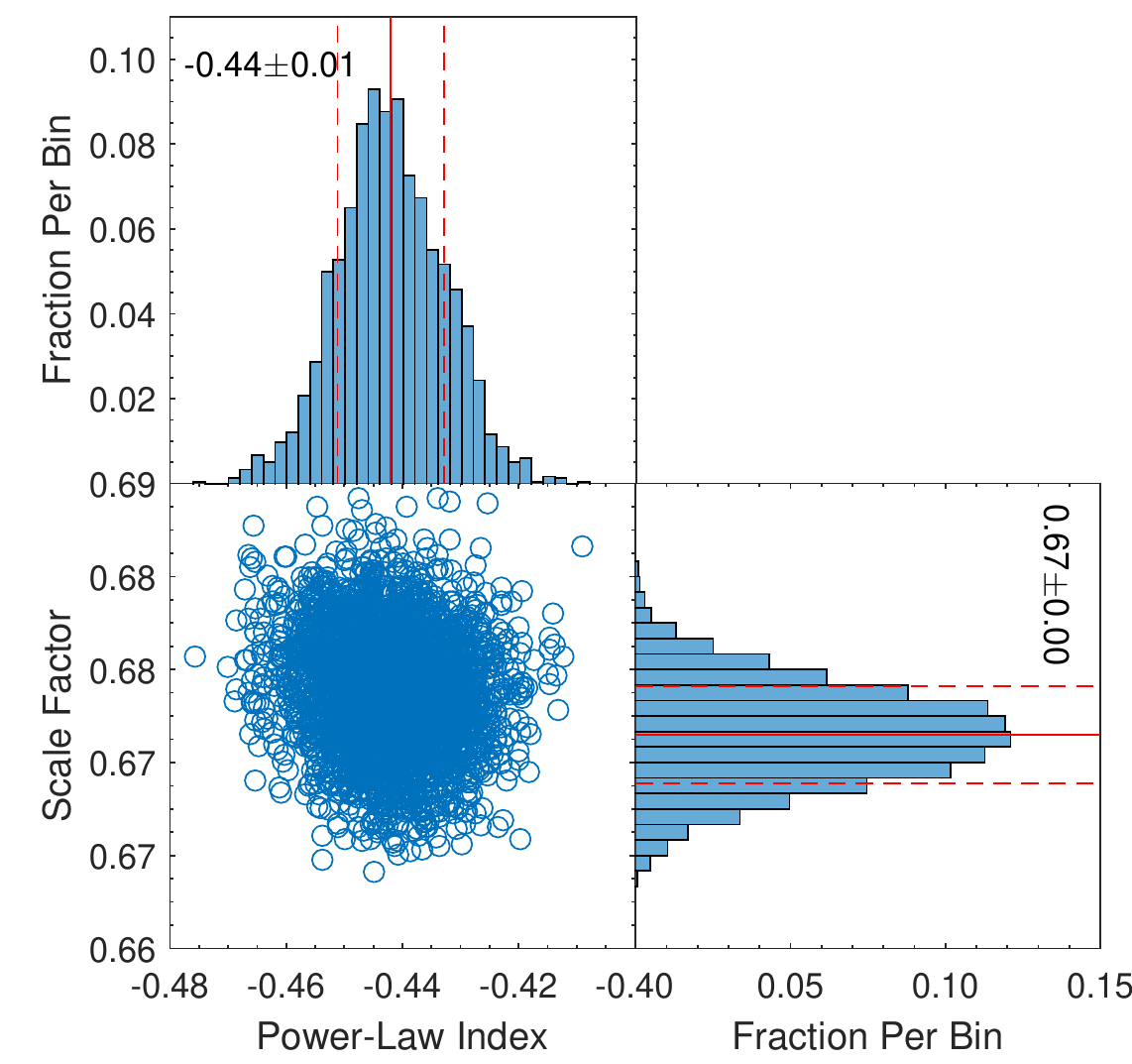}}
   \subfigure[$E_{\mathrm{flow}}/E_{\mathrm{turb}}$ vs. $R_\mathrm{cloud}$]{\includegraphics[width=0.3\textwidth,trim={50 10 180 78},clip]{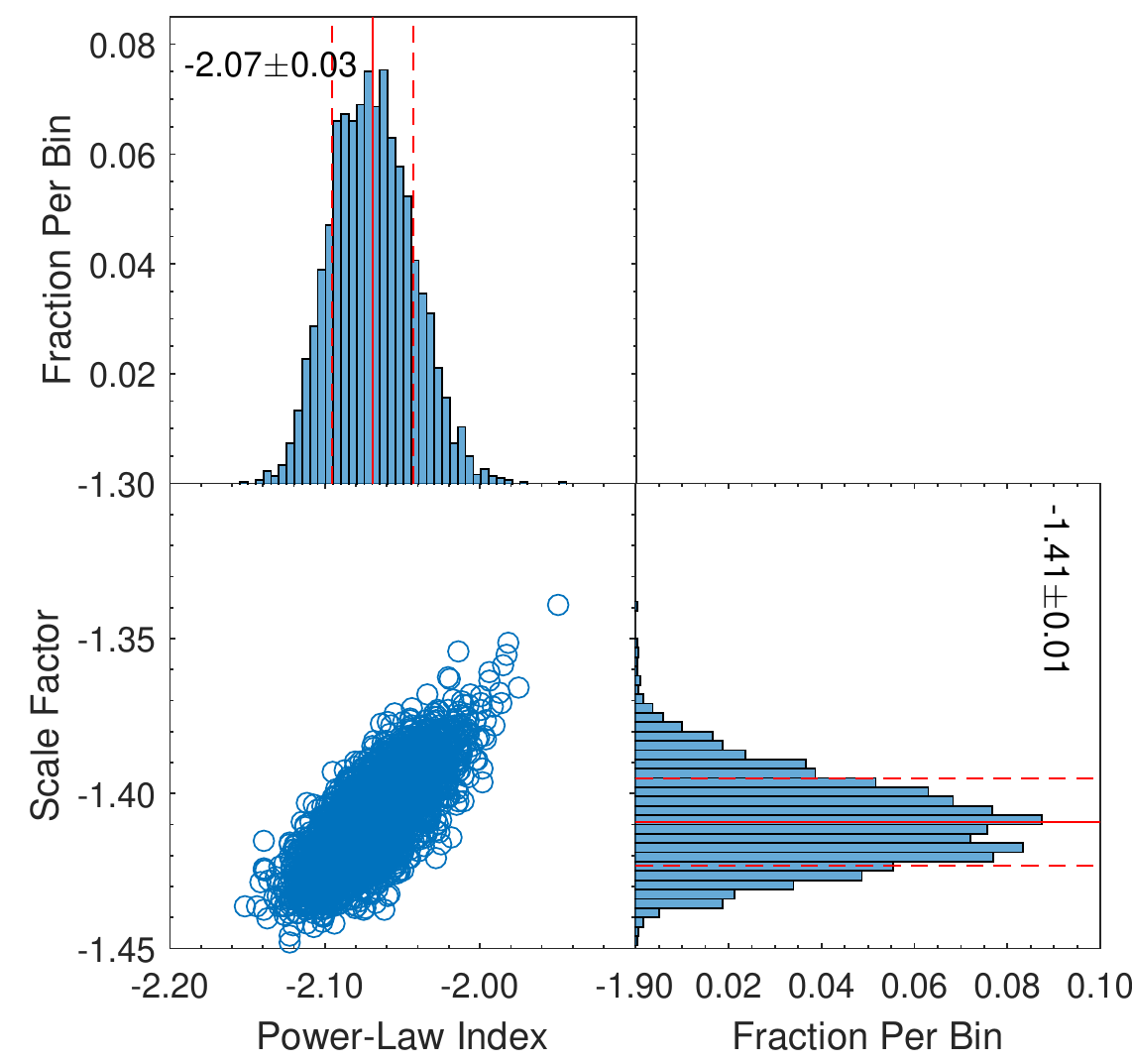}}
 \subfigure[$P_{\mathrm{flow}}/P_{\mathrm{turb}}$ vs. $R_\mathrm{cloud}$]{\includegraphics[width=0.3\textwidth,trim={50 10 180 78},clip]{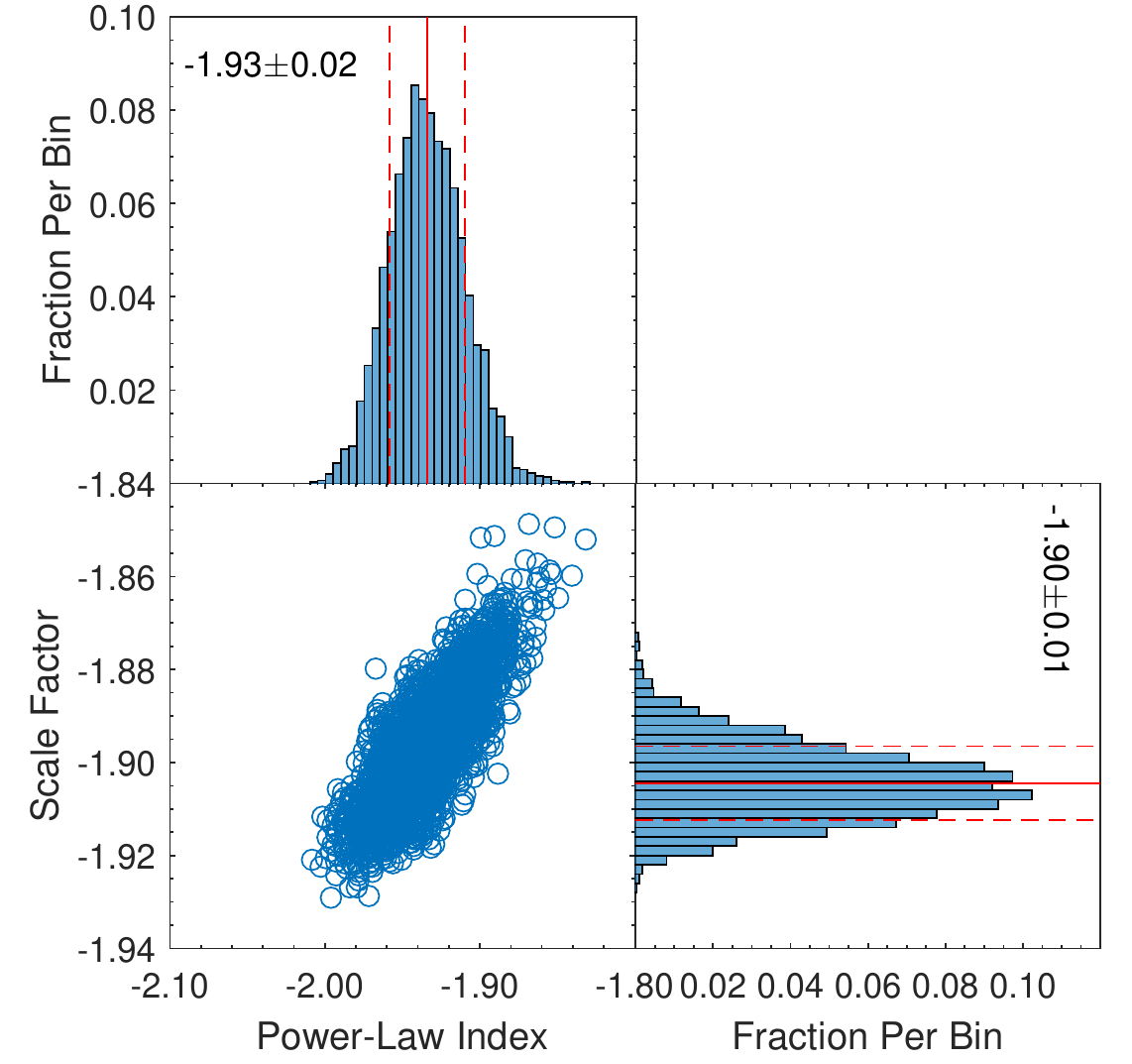}}
  \subfigure[$L_{\mathrm{flow}}/L_{\mathrm{turb}}$ vs. $R_\mathrm{cloud}$]{\includegraphics[width=0.3\textwidth,trim={50 10 180 78},clip]{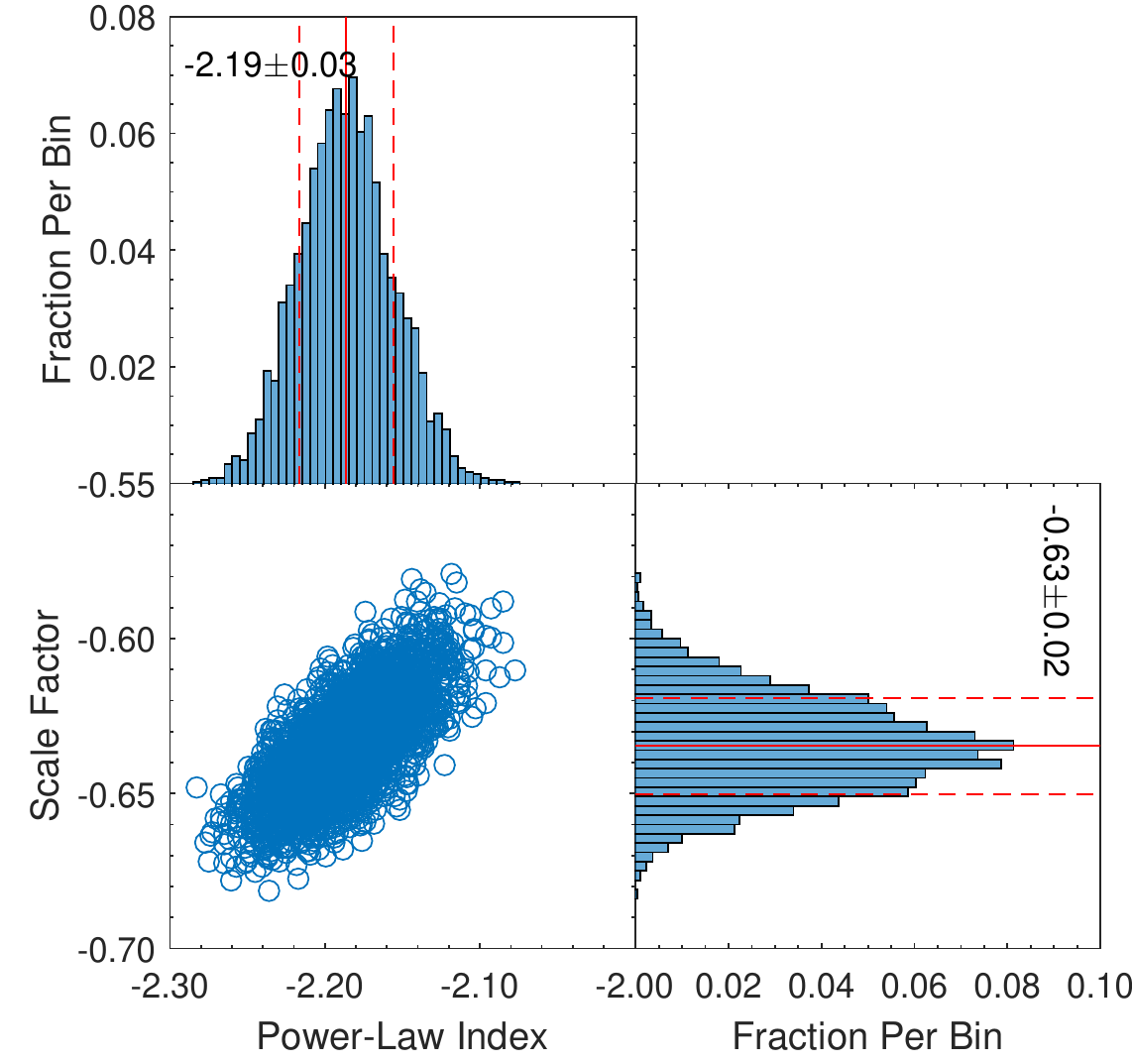}}
   \subfigure[$E_{\mathrm{flow}}/E_{\mathrm{grav}}$ vs. $R_\mathrm{cloud}$]{\includegraphics[width=0.3\textwidth,trim={50 10 180 78},clip]{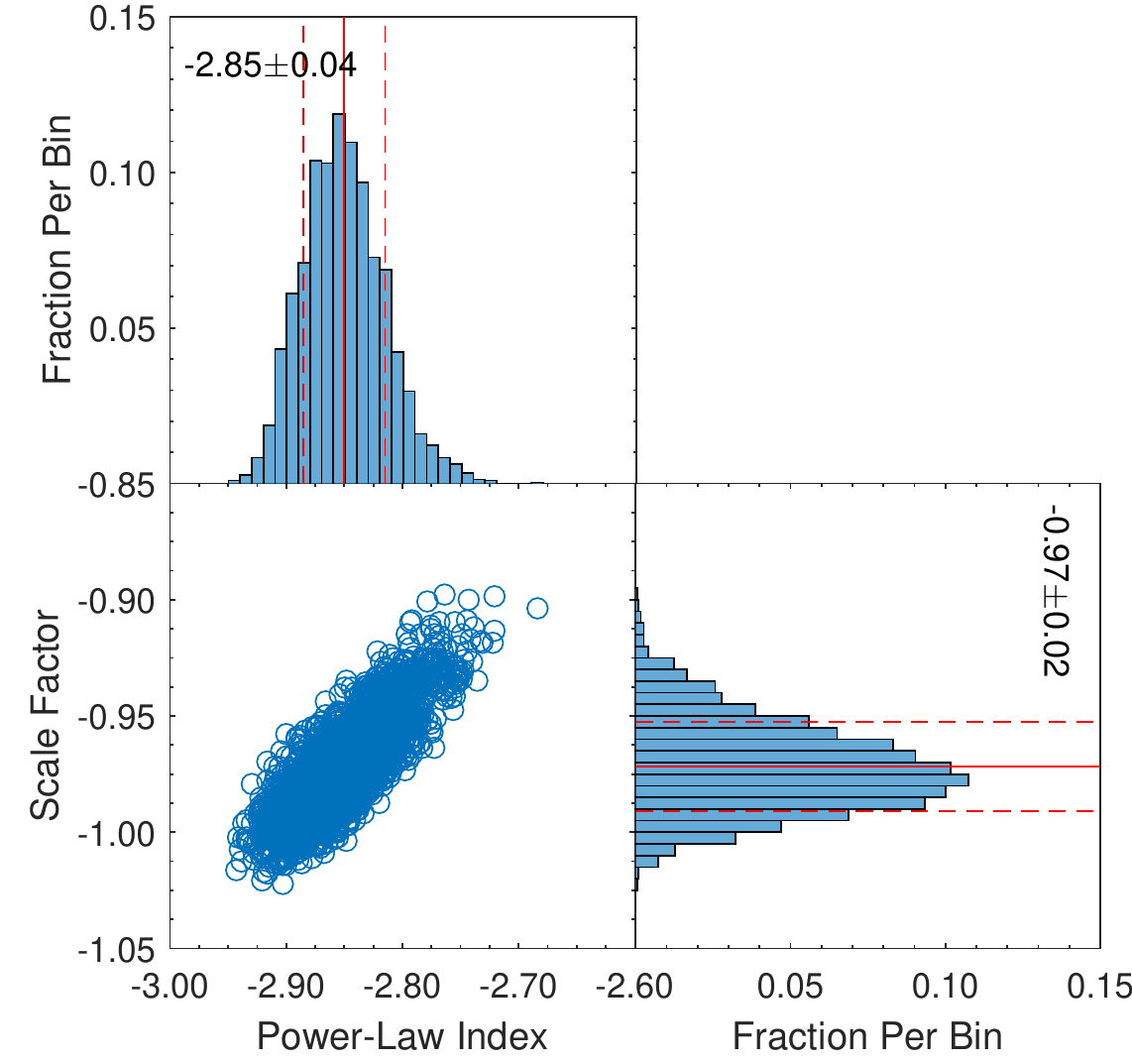}}
 \subfigure[$M_{\mathrm{esc}}/M_{\mathrm{cloud}}$ vs. $R_\mathrm{cloud}$]{\includegraphics[width=0.3\textwidth,trim={50 10 180 78},clip]{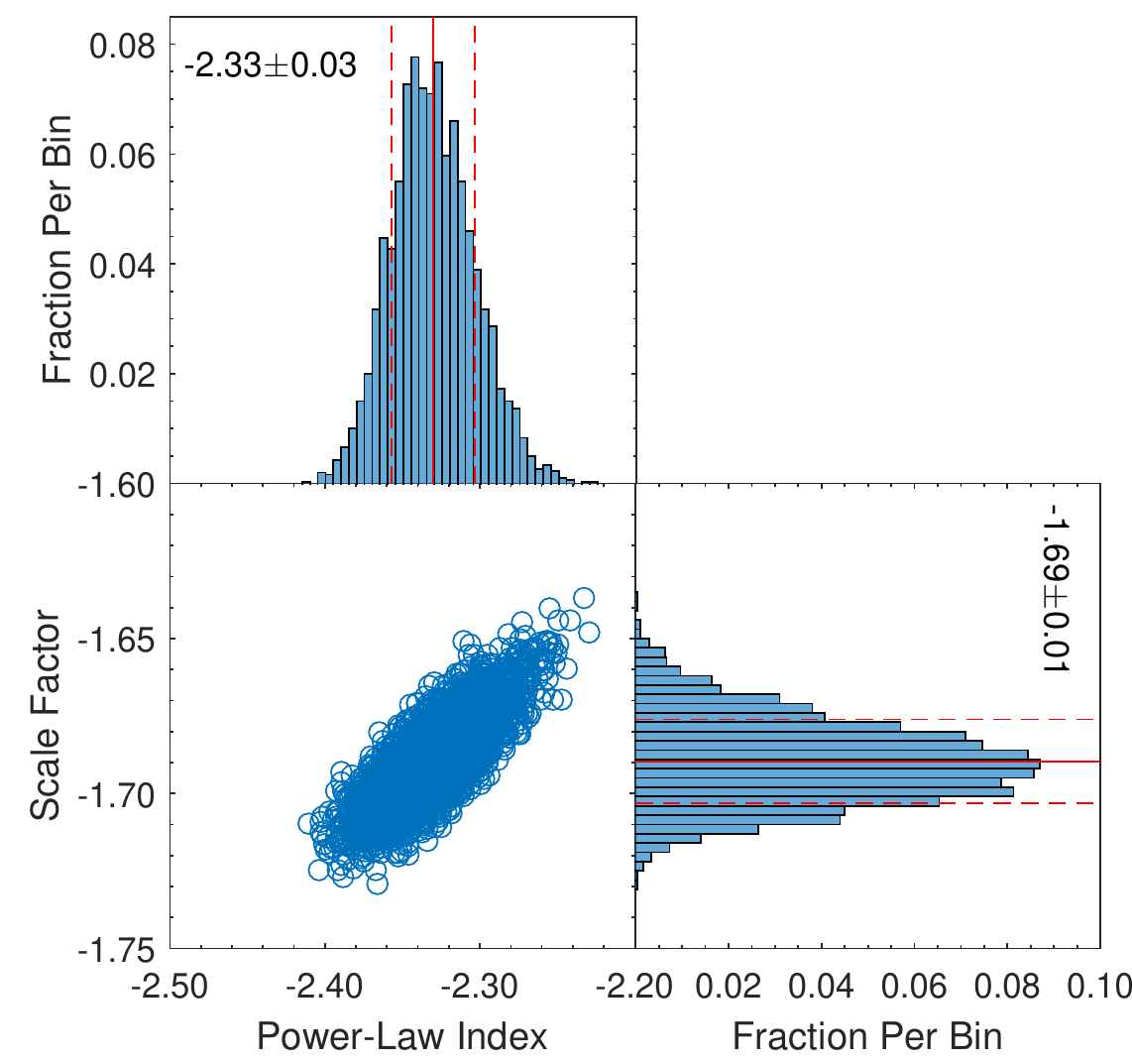}}
  \subfigure[$M_{\mathrm{esc}}/M_{\mathrm{flow}}$ vs. $R_\mathrm{cloud}$]{\includegraphics[width=0.3\textwidth,trim={50 10 180 78},clip]{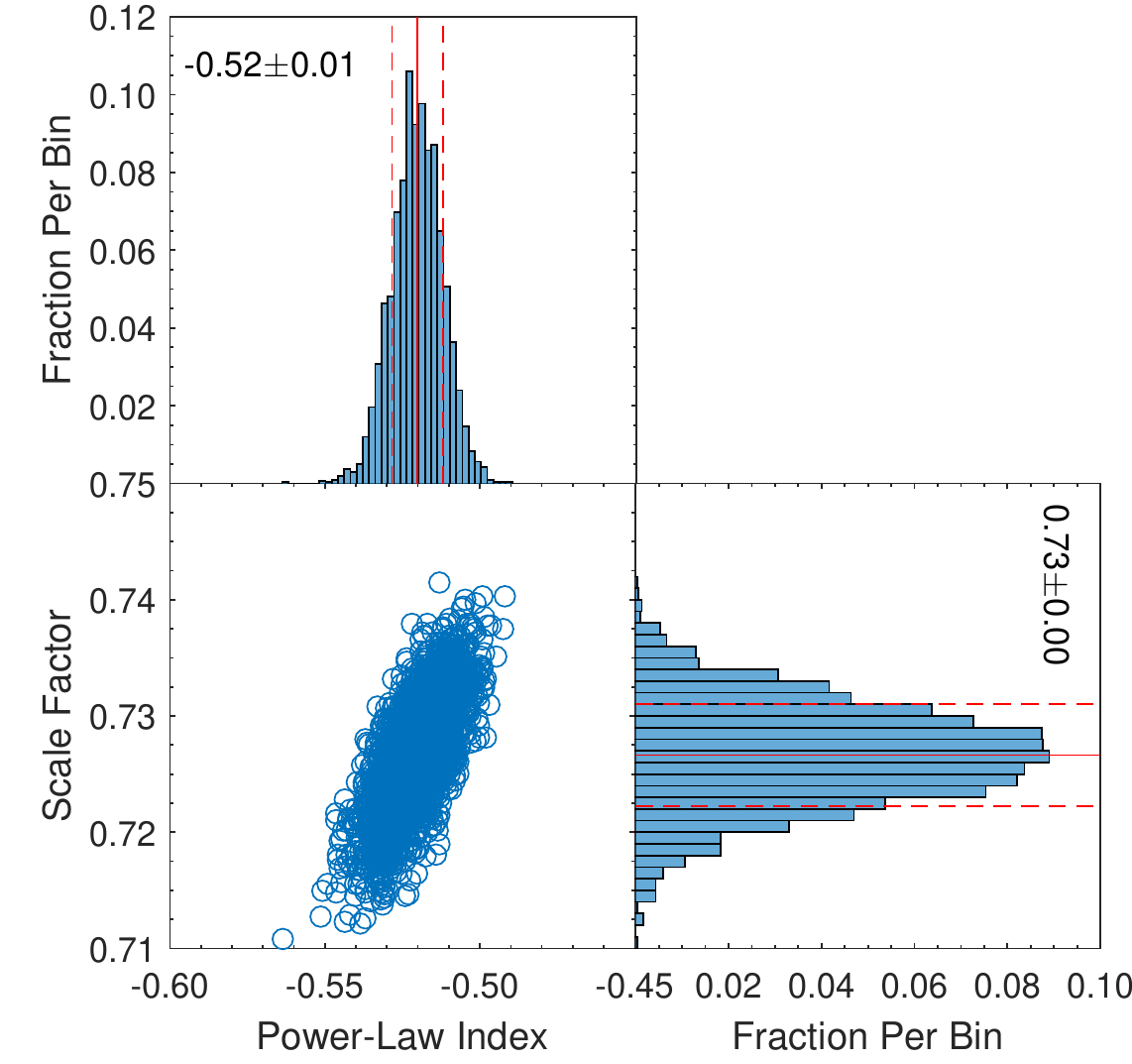}}
\caption{The effect of multi-component: the Perseus arm (a -- f); the Local arm (g -- l).}
\label{fig:mul}
\end{figure}

Three thousand tests are needed to stabilize the result (see Figure \ref{fig:mul} and Table \ref{Table:CRR}). Similar to the effect of beam dilution and the uncertainty of distance, the results show that multi-component has little effect on the conclusions of Sections \ref{section:summary for TS} and \ref{sec:sde}.

Throughout what has been done in Sections \ref{sec:TSD} -- \ref{sec:effect}, differences of both PLIs and CRRs between the samples in the Perseus arm and the Local arm roughly remain unchanged for four ratios (i.e.,  $E_{\mathrm{flow}}/E_{\mathrm{turb}}$, $P_{\mathrm{flow}}/P_{\mathrm{turb}}$, $E_{\mathrm{flow}}/E_{\mathrm{grav}}$ and $M_{\mathrm{esc}}/M_{\mathrm{cloud}}$, see Table \ref{Table:CRR}). This indicates that these differences are probably resulted from the environment of the cloud such as outflow activity discussed in Sections \ref{section:summary for TS} and \ref{sec:sde} or other star-forming activities.

\subsection{Linewidth-Size Relation}\label{sec:LSRV}

\subsubsection{Feedback of Outflow Activity}

\begin{figure}[!ht]
\vspace{0.2cm}
%\figurenum{A. 5}
\centering
 \subfigure[The Perseus Arm]{\includegraphics[width=0.49\textwidth]{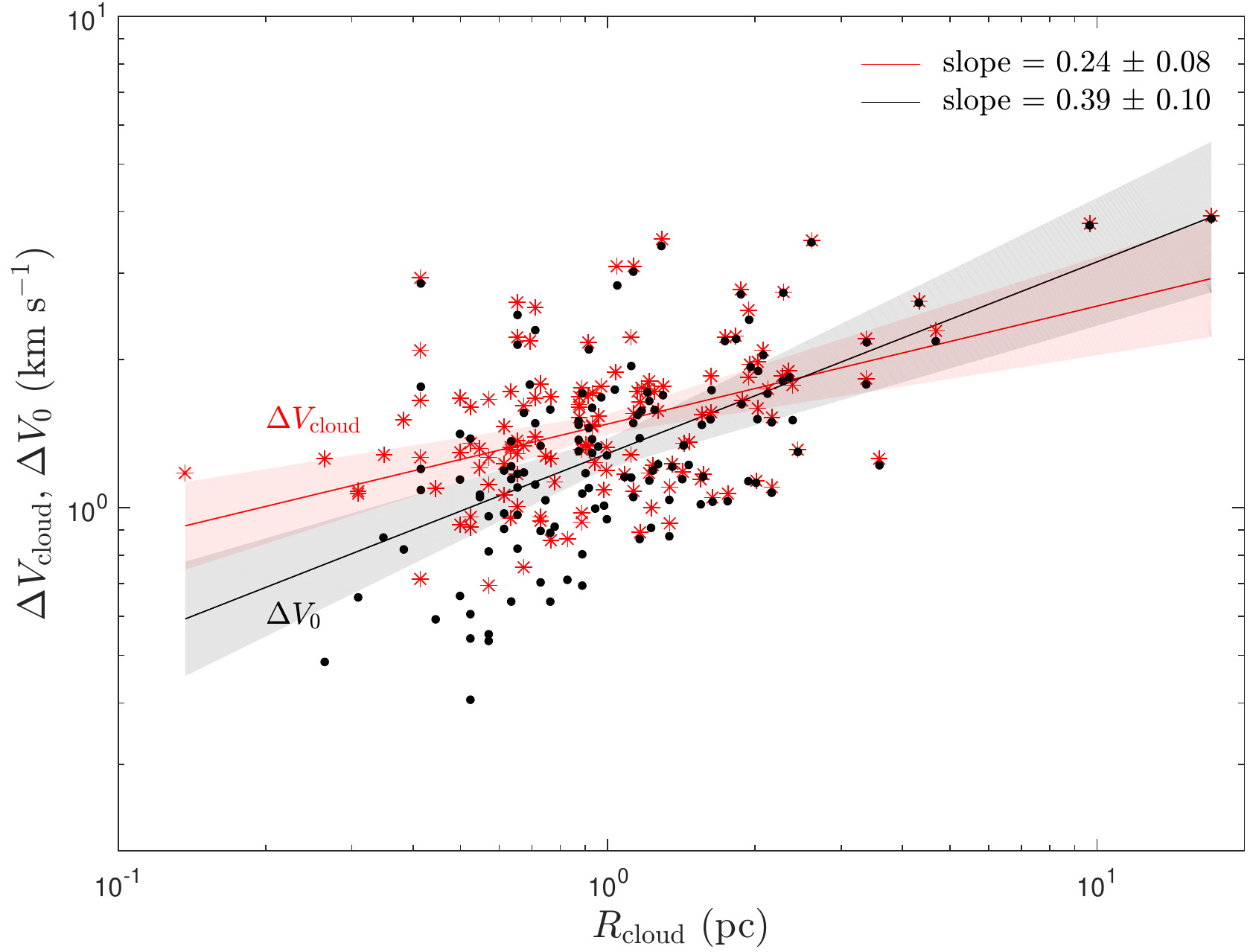}}
 \subfigure[The Local Arm]{\includegraphics[width=0.49\textwidth]{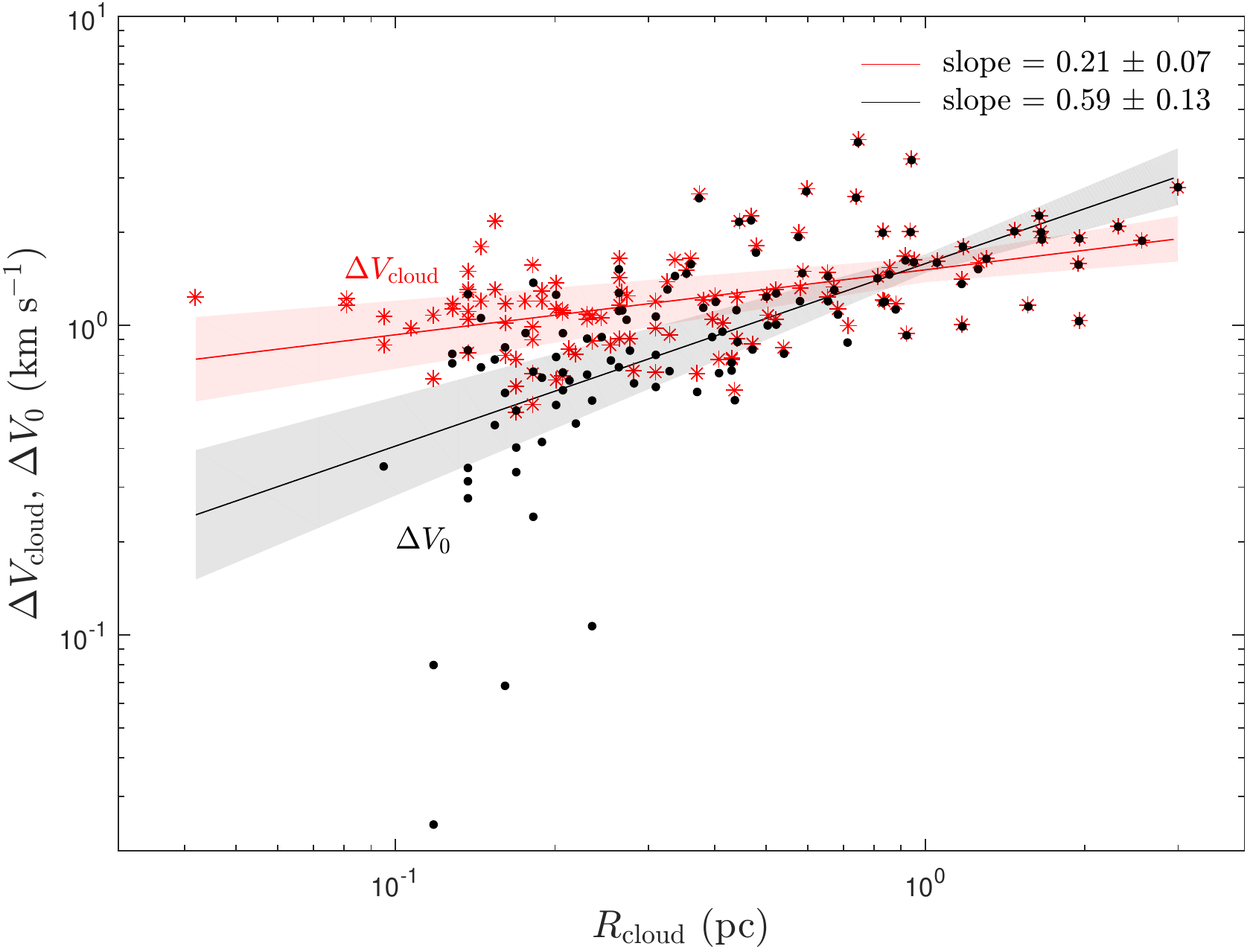}}
\caption{Linewidth-size relations for clouds in (a) the Perseus arm and (b) the Local arm, respectively. Red stars and black points represent $\Delta V_\mathrm{cloud}$ and $\Delta V_0$ (see the descriptions in the text) against $R_\mathrm{cloud}$, respectively. The solid lines with corresponding color denote the corresponding fitted linewidth-size relations.}
\label{Fig:larson}
\end{figure}

The linewidth, $\Delta V_\mathrm{cloud}$, as a function of cloud radius, $R_\mathrm{cloud}$, is shown in Figure \ref{Fig:larson}. The best-fitting linewidth-size relations for the clouds in the Perseus arm and the Local arm are
\begin{subequations}\label{equ:larson0}
\begin{align}
  \log (\Delta V_\mathrm{cloud}) = (0.24 \pm 0.08) \log R_\mathrm{cloud} + 0.17 \pm 0.02, c.c. = 0.46, \label{equ:larson1}\\
  \log (\Delta V_\mathrm{cloud}) = (0.21 \pm 0.07) \log R_\mathrm{cloud} + 0.18 \pm 0.04, c.c. = 0.48, \label{equ:larson2}
\end{align}
\end{subequations}
respectively. These two slopes are much shallower than the slope of 0.38 found by \citet{L1981} and 0.5 determined from large-scale observations \citep{PN2002, MO2007, TDE2018}. However, they are similar to the results from high-mass star-forming regions (e.g., $0.21 \pm 0.03$ found in \citealp{CM1995} and 0.3 in \citealp{SEY2003}). It is interesting to see whether outflow activity could play a role in some way.

To confirm the possibility of outflow feedback on the slopes reported in Equation (\ref{equ:larson0}), we compared $\Delta V_\mathrm{cloud}$ with the linewidth, $\Delta V_0$, which was used to attempt to subtract the contribution of the momentum injected by the outflow activity to the cloud turbulence. \footnote{There are two reasons for considering the momentum here: momentum is a conserved quantity, and the number of excluded samples with $\Delta V_0 \leqslant 0$ is small (see panel (b) of Figure \ref{Fig:ret}). Values of $\Delta V_0 \leqslant 0$ indicate that outflow candidates can eject momentum that is no less than the turbulent momentum of the cloud.}$\Delta V_0$ was measured by
\begin{equation}
  \Delta V_0 = \Delta V_\mathrm{cloud}(1-P_{\mathrm{flow}}/P_{\mathrm{turb}})= \frac{2\sqrt{2\ln 2}P_\mathrm{turb}}{\sqrt{3}M_\mathrm{cloud}}(1-P_{\mathrm{flow}}/P_{\mathrm{turb}}),
\end{equation}
where $P_{\mathrm{flow}}/P_{\mathrm{turb}}$ (see Table \ref{Table:ratio}) as a function of $R_\mathrm{cloud}$ (see Table \ref{Table:parameter}) is plotted in panel (b) of Figure \ref{Fig:ret}. The fitted linewidth-size relations of $\Delta V_0$ against $R_\mathrm{cloud}$ for the clouds in the Perseus arm and the Local arm are
\begin{subequations}\label{equ:larson7}
\begin{align}
  \log (\Delta V_0) = (0.39 \pm 0.10) \log R_\mathrm{cloud} + 0.11 \pm 0.03, c.c. = 0.58, \label{equ:larson5}\\
  \log (\Delta V_0) = (0.59 \pm 0.13) \log R_\mathrm{cloud} + 0.20 \pm 0.07, c.c. = 0.64, \label{equ:larson6}
\end{align}
\end{subequations}
respectively. Relative to the case of $\Delta V_\mathrm{cloud}$, the slopes in Equations (\ref{equ:larson7}) are steeper and closer to the slope found from large-scale observations (i.e., 0.5; see above). This fact supports the idea that feedback from outflow activity may potentially sculpt the linewidth-size relation. 

Going further, the slope reported in Equation (\ref{equ:larson5}) is shallower than the slope found for the large-scale observations mentioned above, likely indicating that additional feedback from star formation activities plays a role in further altering the linewidth-size relation. For instance, such additional feedback might be radiative feedback \citep{MKC2013} or could originate from turbulence regenerated through gravitational collapse \citep[e.g.,][]{FBK2008, KH2010, RG2012}.

\subsubsection{Uncertainty of Distance and Effect of Multi-component}\label{sec:velo-dis}

As discussed in Section \ref{sec:effect}, the effect of beam dilution is probably coupled with the effect of distance. In the constructed simulated cloud samples (the number of clouds is 136 and 124 for clouds in the Perseus arm and the Local arm, respectively), the distance is the same as that in Section \ref{sec:effect}.  For a single cloud, the linewidth is generated according to the similar probability density function with the expected value, standard deviation and the number of elements being $\Delta V_\mathrm{cloud}$, $\Delta V_\mathrm{err}$ and one thousand (it is enough to stabilize the result), respectively. In each test (totally one thousand tests), we take one of the generated linewith for a single cloud. The result shows that the slope is $0.24 \pm 0.01$ and $0.19\pm0.01$ (``the average value'' $\pm$ ``the standard deviation'' of the fits of one thousand tests) for clouds in the Perseus arm and the Local arm, respectively.

\begin{figure}[!ht]
\vspace{0.2cm}
%\figurenum{A. 5}
\centering
 \subfigure[The Perseus Arm]{\includegraphics[width=0.49\textwidth,trim={50 10 180 78},clip]{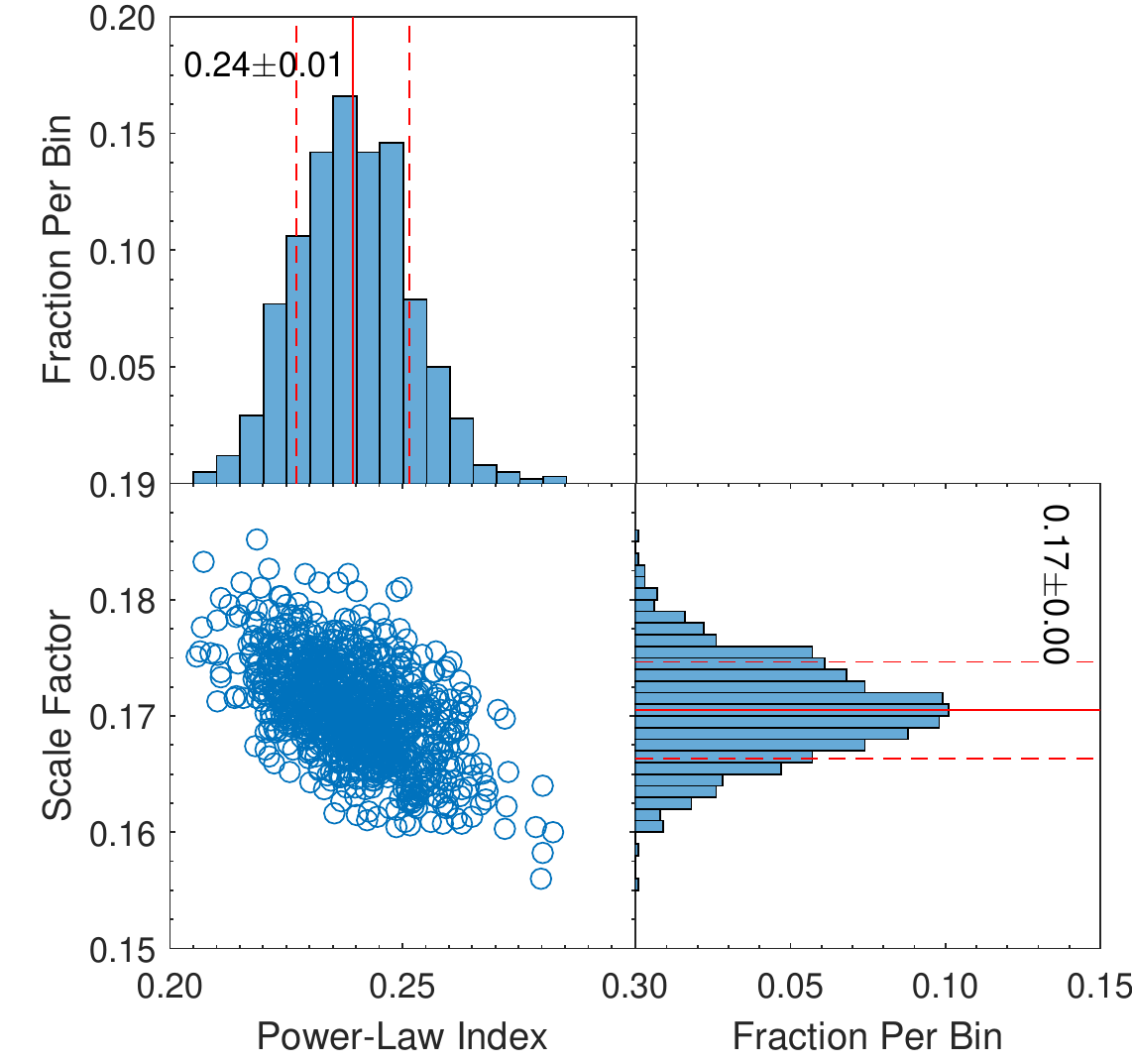}}
 \subfigure[The Local Arm]{\includegraphics[width=0.49\textwidth,trim={50 10 180 78},clip]{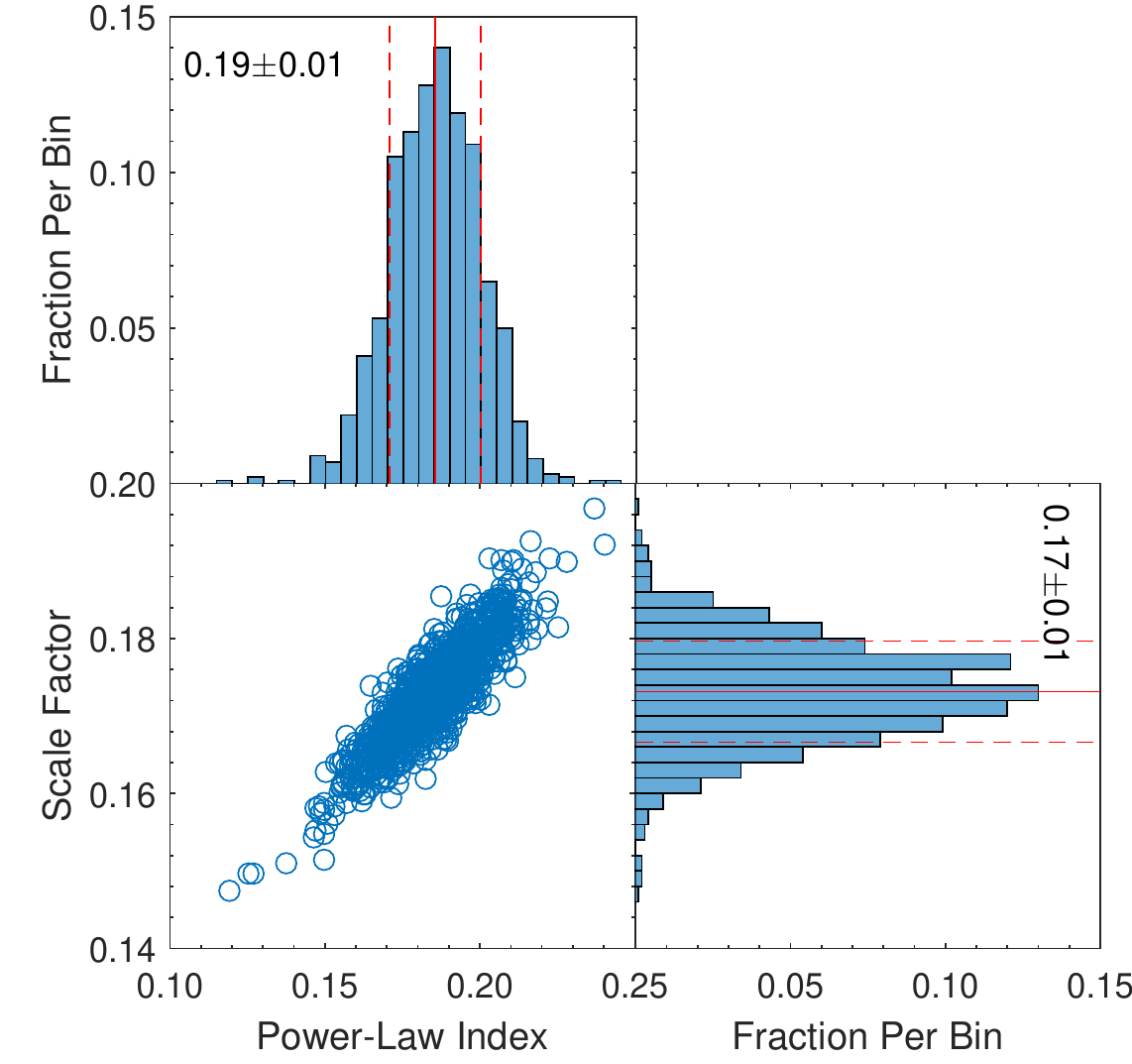}}
\caption{The effect of uncertainty of distance on the linewidth-size relations for clouds (a) in the Perseus arm and (b) the Local arm, respectively. }
\label{Fig:larson-dis}
\end{figure}

Similar to the treatment of the effect of multi-component in Section \ref{sec:emc} for the case of $L_{\mathrm{flow}}/L_{\mathrm{turb}}$ or $M_{\mathrm{esc}}/M_{\mathrm{flow}}$, clouds are merged to construct new samples. In this process, the linewidth of a new cloud is the average value of the corresponding old clouds weighted by the mass. The result shows that the slope is $0.24\pm0.01$ and $0.21\pm0.01$ for clouds in the Perseus arm and the Local arm, respectively.

\begin{figure}[!ht]
\vspace{0.2cm}
%\figurenum{A. 5}
\centering
 \subfigure[The Perseus Arm]{\includegraphics[width=0.49\textwidth,trim={50 10 180 78},clip]{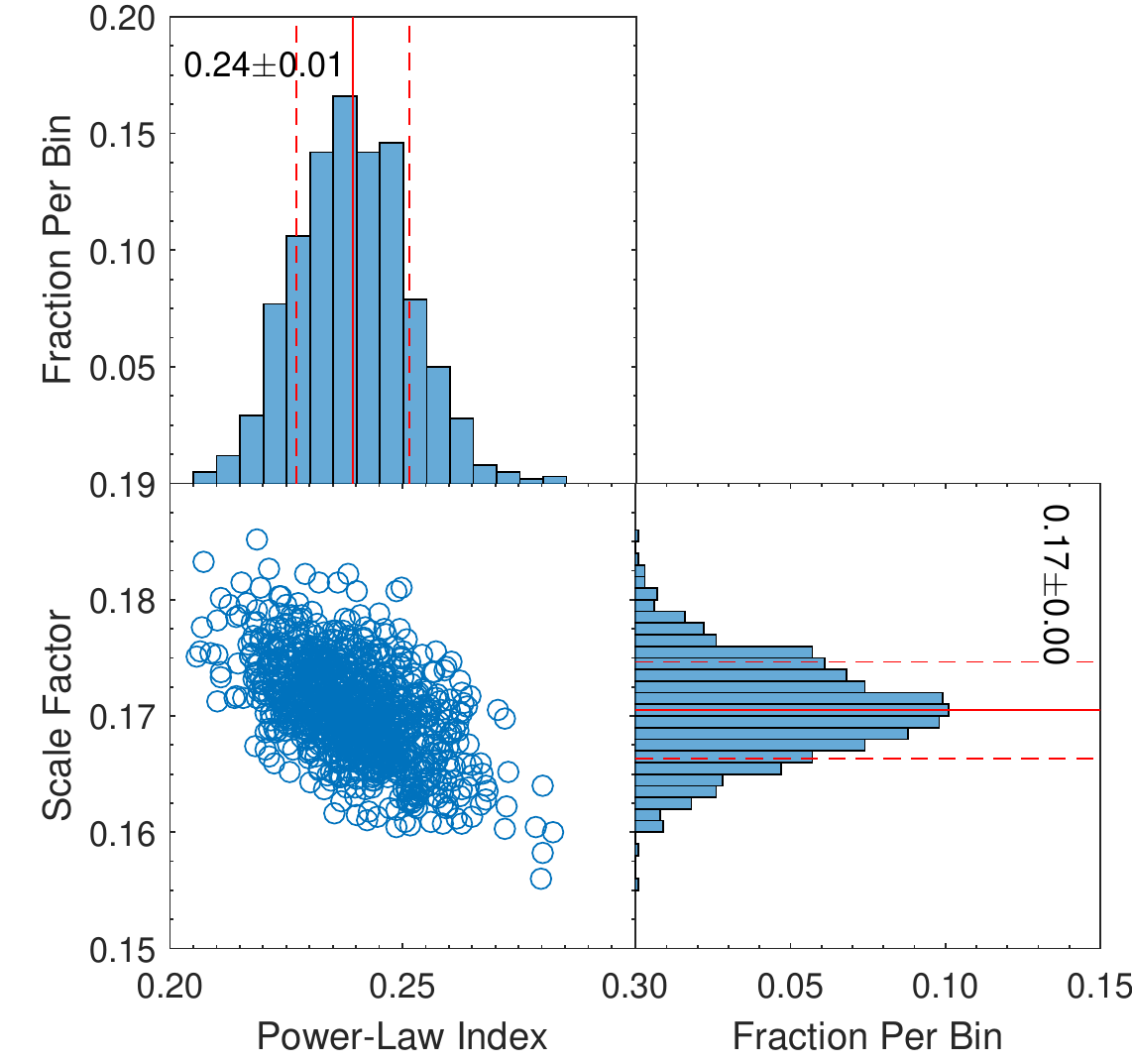}}
 \subfigure[The Local Arm]{\includegraphics[width=0.49\textwidth,trim={50 10 180 78},clip]{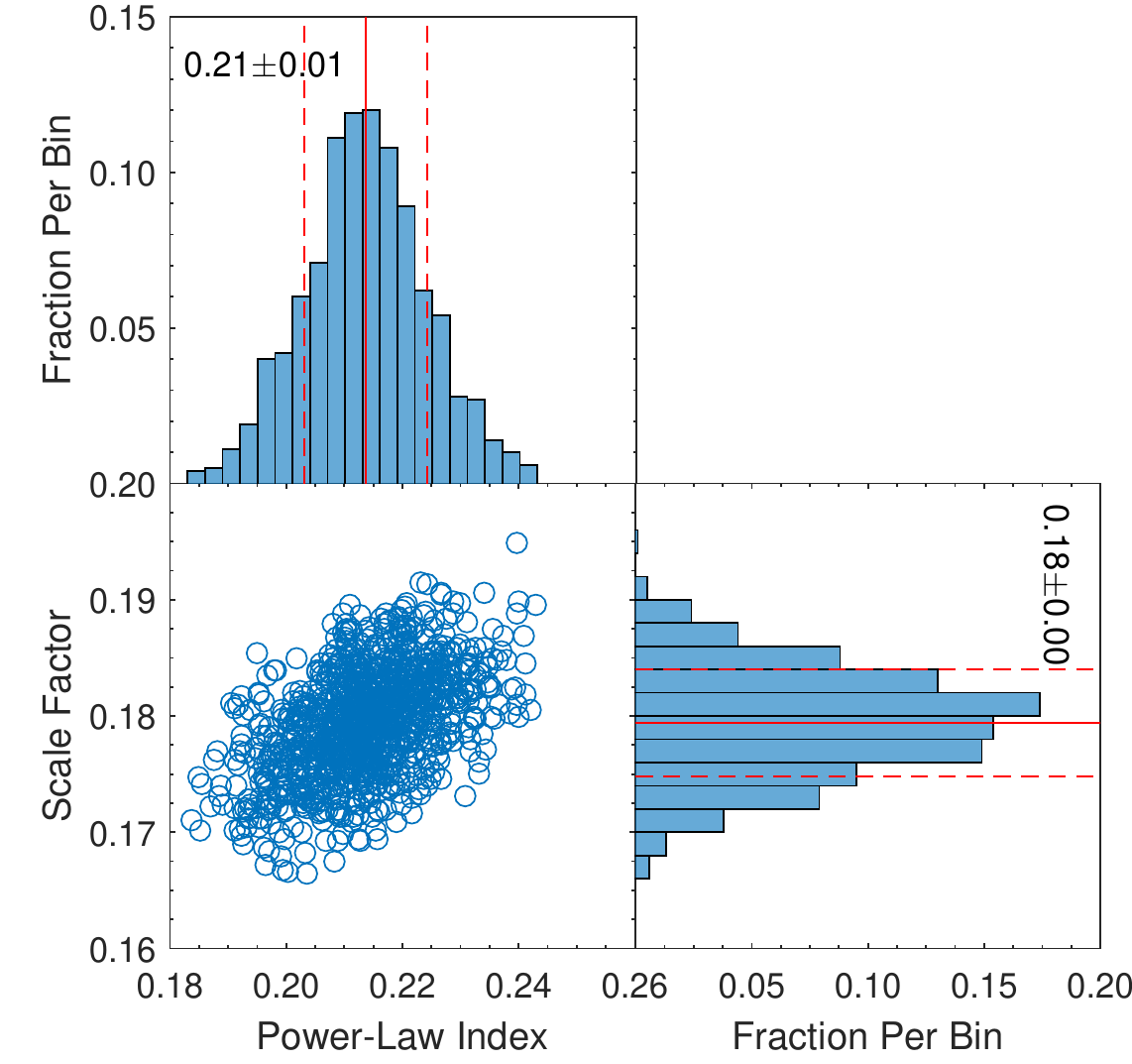}}
\caption{The effect of multi-component on the linewidth-size relations for clouds in (a) the Perseus arm and (b) the Local arm, respectively. }
\label{Fig:larson-mul}
\end{figure}

From the above results, beam dilution, uncertainty of distance and multi-component has little effect on the slope reported in Equation (\ref{equ:larson0}). This implies that the slope does change after the contribution of the momentum injected by the outflow activity to the cloud turbulence being subtracted, indicating that outflow feedback has the potential to alter the linewidth-size relation.

Based on the discussion above, especially the comparison of the properties of $V_\mathrm{cloud}$ and $V_0$, the feedback from outflow activity may potentially affect the linewidth-size relation, so that the slope becomes shallower. This scenario may also apply to the cases presented by \citet{CM1995} and \citet{SEY2003}, where other feedbacks related to star-forming activities may also have played a role.

%\restoregeometry
\section{Summary and Conclusions}{\label{summary}}

A large scale survey of \xco clouds which were associated with outflow candidates was conducted toward the W3/4/5 complex and its surroundings ($\sim$ 110 deg$^2$, within the Galactic coordinates of  $129\fdg75\leq l\leq140\fdg25$, and $-5\fdg25\leq b\leq5\fdg25$). A total of 265 clouds were identified, with radii ranging from $\sim 0.04$ to $\sim 17.12$ pc. Our main conclusions are as follows:

\begin{enumerate}
  \item The outflow activity was enough to maintain the turbulence in a minority of clouds if the outflows can couple to the dense gas where stars are forming, but in some cases it could potentially disrupt the entire cloud. It is likely that the scale break is $\gtrsim 4.7$ pc for both turbulence support and potential disruptive effect.
  \item Larger clouds might possess stronger ability to resisting feedback from outflow activities than small ones, because negative PLIs were obtained when fitting $M_{\mathrm{esc}}/M_{\mathrm{flow}}$ as a function of $R_\mathrm{cloud}$.
  \item The feedback of outflow activity has the potential to shape the linewidth-size relation, causing the profile to become shallower, because steeper power-law profiles were found when the contribution of momentum injected by outflow activity to the cloud turbulence was subtracted.
  \item For turbulent support, three ratios (i.e., $E_{\mathrm{flow}}/E_{\mathrm{turb}}$, $P_{\mathrm{flow}}/P_{\mathrm{turb}}$ and $L_{\mathrm{flow}}/L_{\mathrm{turb}}$) as functions of $R_\mathrm{cloud}$ all represented negative PLIs. From the analyses of the first two ratios, the CRRs were probably $\sim$ 0.2 -- 0.4 pc and $\sim$ 0.1 -- 0.2 pc for clouds in the Perseus arm and the Local arm, respectively, as derived from the nodes of the best-fitting power-law profiles and the ratios of unity. Outflow activity or other star-forming activities may affect the value of CRR.
  \item For potential disruptive effect, the power-law profiles of two ratios ($\eta_\mathrm{out} = 2E_{\mathrm{flow}}/E_{\mathrm{grav}}$ and $M_{\mathrm{esc}}/M_{\mathrm{cloud}}$) as functions of $R_\mathrm{cloud}$ had negative indices. The nodes of the best-fitting power-law profiles and the ratios of unity indicate that the possible CRR$_\mathrm{GE}$ is $\sim$ 1.0 and $\sim$ 0.6 pc respectively for clouds in the Perseus arm and the Local arm, and the corresponding values of CRR$_\mathrm{MC}$ are $\sim$ 0.3 and $\sim$ 0.2 pc. Similar to the case of turbulent support, outflow activity or other star-forming activities may affect the value of CRR.
\end{enumerate}

Note that all results of this work are based on the short segments of the Perseus arm, the Local arm and interarm 1 only over an interval of 11$\degr$ in Galactic longitude. Therefore, the conclusions reached here might not necessarily apply to the regions across the Perseus arm, the Local arm and interarms in the entire Milkey Way.

\acknowledgments

This work is part of the Milky Way Image Scroll Painting (MWISP) project, which is based on observations made with the PMO 13.7 m telescope at Delingha. We would like to thank all the staff members of Qinghai Radio Observing Station at Delingha for their help during the observations. We would like to thank the referee for reviewing the paper carefully and the constructive comments that improves this manuscript. This work was sponsored by the MOST under Grand No. 2017YFA0402701, Key Research Program of Frontier Sciences, CAS (grant No. QYZDJ-SSW-SLH047), the NSFC Grand NO. 11933011, 11873019, 11673066, 11773077 and 11503033, and the Key Laboratory for Radio Astronomy, CAS, and also partially funded by the ERC Advanced Investigator Grant GLOSTAR (247078).

\facility{PMO 13.7m}

\end{CJK*}

\begin{thebibliography}{}
\bibitem[Arce et al.(2010)]{ABG2010} Arce, H.~G., Borkin, M.~A., Goodman, A.~A., Pineda, J.~E., \& Halle, M.~W.\ 2010, \apj, 715, 1170
\bibitem[Arce \& Goodman(2002)]{AG2002} Arce, H.~G., \& Goodman, A.~A.\ 2002, \apj, 575, 911
\bibitem[Bally(2016)]{B2016} Bally, J.\ 2016, \araa, 54, 491
\bibitem[Beuther et al.(2002)]{BSS2002} Beuther, H., Schilke, P., Sridharan, T.~K., et al.\ 2002, \aap, 383, 892
\bibitem[Brunt(2010)]{B2010} Brunt, C.~M.\ 2010, \aap, 513, A67
\bibitem[Brunt et al.(2009)]{BHM2009} Brunt, C.~M., Heyer, M.~H., \& Mac Low, M.-M.\ 2009, \aap, 504, 883
\bibitem[Carroll et al.(2009)]{CFB2009} Carroll, J.~J., Frank, A., Blackman, E.~G., Cunningham, A.~J., \& Quillen, A.~C.\ 2009, \apj, 695, 1376
\bibitem[Caselli \& Myers(1995)]{CM1995} Caselli, P., \& Myers, P.~C.\ 1995, \apj, 446, 665
\bibitem[Cunningham et al.(2009)]{CFC2009} Cunningham, A.~J., Frank, A., Carroll, J., Blackman, E.~G., \& Quillen, A.~C.\ 2009, \apj, 692, 816
\bibitem[De Colle \& Raga(2005)]{DR2005} De Colle, F., \& Raga, A.~C.\ 2005, \mnras, 359, 164
%\bibitem[Downes \& Cabrit(2007)]{DC2007} Downes, T.~P., \& Cabrit, S.\ 2007, \aap, 471, 873
\bibitem[Drabek-Maunder et al.(2016)]{DHB2016} Drabek-Maunder, E., Hatchell, J., Buckle, J.~V., Di Francesco, J., \& Richer, J.\ 2016, \mnras, 457, L84
\bibitem[Du et al.(2017)]{DXY2017} Du, X., Xu, Y., Yang, J., \& Sun, Y.\ 2017, \apjs, 229, 24
\bibitem[Elmegreen(2007)]{E2007} Elmegreen, B.~G.\ 2007, \apj, 668, 1064
\bibitem[Elmegreen \& Scalo(2004)]{ES2004} Elmegreen, B.~G., \& Scalo, J.\ 2004, \araa, 42, 211
\bibitem[Evans et al.(2009)]{EDJ2009} Evans, N.~J., Dunham, M.~M., J{\o}rgensen, J.~K., et al.\ 2009, \apjs, 181, 321
\bibitem[Field et al.(2008)]{FBK2008} Field, G.~B., Blackman, E.~G., \& Keto, E.~R.\ 2008, \mnras, 385, 181
\bibitem[Frank et al.(2014)]{FRC2014} Frank, A., Ray, T. P., Cabrit, S., et al. 2014, in Protostars and Planets VI, ed.
H. Beuther et al. (Tucson, AZ: Univ. Arizona Press), 451
\bibitem[Frerking et al.(1982)]{FLW1982} Frerking, M.~A., Langer, W.~D., \& Wilson, R.~W.\ 1982, \apj, 262, 590
\bibitem[Goldsmith \& Langer(1999)]{GL1999} Goldsmith, P.~F., \& Langer, W.~D.\ 1999, \apj, 517, 209
\bibitem[Graves et al.(2010)]{GRB2010} Graves, S.~F., Richer, J.~S., Buckle, J.~V., et al.\ 2010, \mnras, 409, 1412
\bibitem[Hartmann \& Burkert(2007)]{HB2007} Hartmann, L., \& Burkert, A.\ 2007, \apj, 654, 988
\bibitem[Heyer \& Terebey(1998)]{HT1998} Heyer, M.~H., \& Terebey, S.\ 1998, \apj, 502, 265
\bibitem[Kawamura et al.(1998)]{KOY1998} Kawamura, A., Onishi, T., Yonekura, Y., et al.\ 1998, \apjs, 117, 387
\bibitem[Klessen \& Hennebelle(2010)]{KH2010} Klessen, R.~S., \& Hennebelle, P.\ 2010, \aap, 520, A17
\bibitem[Krumholz et al.(2014)]{KBA2014} Krumholz, M.~R., Bate, M.~R., Arce, H.~G., et al.\ 2014, in Protostars and Planets VI, ed. H. Beuther et al. (Tucson, AZ: Univ. Arizona Press), 243
\bibitem[Krumholz et al.(2011)]{KKM2011} Krumholz, M.~R., Klein, R.~I., \& McKee, C.~F.\ 2011, \apj, 740, 74
\bibitem[Ladd et al.(1994)]{LMG1994} Ladd, E.~F., Myers, P.~C., \& Goodman, A.~A.\ 1994, \apj, 433, 117
\bibitem[Larson(1981)]{L1981} Larson, R.~B.\ 1981, \mnras, 194, 809
\bibitem[Li et al.(2015)]{LLQ2015} Li, H., Li, D., Qian, L., et al.\ 2015, \apjs, 219, 20
\bibitem[Li et al.(2018)]{LLX2018} Li, Y., Li, F.-C., Xu, Y., et al.\ 2018, \apjs, 235, 15
\bibitem[Li et al.(2019)]{LXS2019} Li, Y., Xu, Y., Sun, Y., et al.\ 2019, \apjs, 242, 19(Paper I)
\bibitem[Li \& Nakamura(2006)]{LN2006} Li, Z.-Y., \& Nakamura, F.\ 2006, \apjl, 640, L187
\bibitem[Mac Low(1999)]{M1999} Mac Low, M.-M.\ 1999, \apj, 524, 169
\bibitem[Matzner(2007)]{M2007} Matzner, C.~D.\ 2007, \apj, 659, 1394
\bibitem[Matzner \& Jumper(2015)]{MJ2015} Matzner, C.~D., \& Jumper, P.~H.\ 2015, \apj, 815, 68
\bibitem[Maury et al.(2009)]{MAL2009} Maury, A.~J., Andr{\'e}, P., \& Li, Z.-Y.\ 2009, \aap, 499, 175
\bibitem[McKee(1989)]{Mc1989} McKee, C.~F.\ 1989, \apj, 345, 782
\bibitem[McKee \& Ostriker(2007)]{MO2007} McKee, C.~F., \& Ostriker, E.~C.\ 2007, \araa, 45, 565
\bibitem[Myers et al.(2014)]{MKK2014} Myers, A.~T., Klein, R.~I., Krumholz, M.~R., \& McKee, C.~F.\ 2014, \mnras, 439, 3420
\bibitem[Myers et al.(2013)]{MKC2013} Myers, A.~T., McKee, C.~F., Cunningham, A.~J., Klein, R.~I., \& Krumholz, M.~R.\ 2013, \apj, 766, 97
\bibitem[Nagahama et al.(1998)]{NMO1998} Nagahama, T., Mizuno, A., Ogawa, H., \& Fukui, Y.\ 1998, \aj, 116, 336
\bibitem[Nakamura et al.(2011a)]{NKK2011} Nakamura, F., Kamada, Y., Kamazaki, T., et al.\ 2011, \apj, 726, 46
\bibitem[Nakamura \& Li(2007)]{NL2007} Nakamura, F., \& Li, Z.-Y.\ 2007, \apj, 662, 395
\bibitem[Nakamura \& Li(2014)]{NL2014} Nakamura, F., \& Li, Z.-Y.\ 2014, \apj, 783, 115
\bibitem[Nakamura et al.(2012)]{NMK2012} Nakamura, F., Miura, T., Kitamura, Y., et al.\ 2012, \apj, 746, 25
\bibitem[Nakamura et al.(2011b)]{NSS2011} Nakamura, F., Sugitani, K., Shimajiri, Y., et al.\ 2011, \apj, 737, 56
\bibitem[Norman \& Silk(1980)]{NS1980} Norman, C., \& Silk, J.\ 1980, \apj, 238, 158
\bibitem[Offner \& Chaban(2017)]{OC2017} Offner, S.~S.~R., \& Chaban, J.\ 2017, \apj, 847, 104
\bibitem[Padoan \& Nordlund(2002)]{PN2002} Padoan, P., \& Nordlund, {\AA}.\ 2002, \apj, 576, 870
\bibitem[Pety(2005)]{P2005} Pety, J.\ 2005, SF2A-2005: Semaine de l'Astrophysique Francaise, 721
%\bibitem[Plunkett et al.(2013)]{PAC2013} Plunkett, A.~L., Arce, H.~G., Corder, S.~A., et al.\ 2013, \apj, 774, 22
\bibitem[Robertson \& Goldreich(2012)]{RG2012} Robertson, B., \& Goldreich, P.\ 2012, \apjl, 750, L31
\bibitem[Shan et al.(2012)]{SYS2012} Shan, W., Yang, J., Shi, S., et al.\ 2012, IEEE Transactions on Terahertz Science and Technology, 2, 593
\bibitem[Shirley et al.(2003)]{SEY2003} Shirley, Y.~L., Evans, N.~J., II, Young, K.~E., Knez, C., \& Jaffe, D.~T.\ 2003, \apjs, 149, 375
\bibitem[Su et al.(2019)]{SYZ2018} Su, Y., Yang, J., Zhang, S., et al.\ 2019, \apjs, 240, 9
\bibitem[Sun et al.(2020)]{SYX2020} Sun, Y., Yang, J., Xu, Y., et al.\ 2019, \apjs, 246, 7
\bibitem[Tan et al.(2006)]{TKM2006} Tan, J.~C., Krumholz, M.~R., \& McKee, C.~F.\ 2006, \apjl, 641, L121
\bibitem[Tan \& McKee(2002)]{TM2002} Tan, J.~C., \& McKee, C.~F.\ 2002, Hot Star Workshop III: The Earliest Phases of Massive Star Birth, 267, 267
\bibitem[Traficante et al.(2018)]{TDE2018} Traficante, A., Duarte-Cabral, A., Elia, D., et al.\ 2018, \mnras, 477, 2220
\bibitem[Wang et al.(2017)]{WYX2017} Wang, C., Yang, J., Xu, Y., et al.\ 2017, \apjs, 230, 5
\bibitem[Westerhout(1958)]{W1958} Westerhout, G.\ 1958, \bain, 14, 215
\bibitem[Zuo et al.(2011)]{ZLS2011} Zuo, Y.~X., Li, Y., Sun, J.~X., et al.\ 2011, Acta Astronomica Sinica, 52, 152
\end{thebibliography}
\end{document}